\documentclass[pdflatex,sn-mathphys-num]{sn-jnl}

\usepackage{graphicx}%
\usepackage{multirow}%
\usepackage{amsmath,amssymb,amsfonts}%
\usepackage{amsthm}%
\usepackage{mathrsfs}%
\usepackage[title]{appendix}%
\usepackage{xcolor}%
\usepackage{textcomp}%
\usepackage{manyfoot}%
\usepackage{booktabs}%
\usepackage{algorithm}%
\usepackage{algorithmicx}%
\usepackage{algpseudocode}%
\usepackage{listings}%

\theoremstyle{thmstyleone}%
\newtheorem{theorem}{Theorem}
\newtheorem{proposition}[theorem]{Proposition}%

\theoremstyle{thmstyletwo}%
\newtheorem{example}{Example}%
\newtheorem{remark}{Remark}%

\theoremstyle{thmstylethree}%
\newtheorem{definition}{Definition}%

\begin{document}

\title[A universal critical accretion rate for black hole jet formation]{
A universal critical accretion rate for black hole jet formation}


\author*[1]{\fnm{Adelle J.} \sur{Goodwin}}\email{ajgoodwin.astro@gmail.com}
\equalcont{These authors contributed equally to this work.}

\author*[2]{\fnm{Andrew} \sur{Mummery}}\email{amummery@ias.edu}
\equalcont{These authors contributed equally to this work.}

\affil[1]{\orgdiv{International Centre for Radio Astronomy Research}, \orgname{Curtin University}, \orgaddress{\street{GPO Box U1987}, \city{Perth}, \postcode{6845}, \state{WA}, \country{Australia}}}

\affil[2]{\orgdiv{School of Natural Sciences}, \orgname{Institute for Advanced Study}, \orgaddress{\street{ 1 Einstein Drive}, \city{Princeton}, \postcode{08540}, \state{NJ}, \country{USA}}}







\abstract{
It has long been suspected that black hole accretion-outflow coupling is invariant from the stellar to supermassive scales \cite{Rees1982}. Stellar mass black hole accretion flows are known to launch jets and outflows as they transition through critical accretion rate thresholds, with values well constrained observationally owing to their short evolutionary timescales. In contrast, accretion flows in typical supermassive black hole (SMBH)  systems (those in active galactic nuclei) evolve over thousands of years, making the critical transitions at which jets are launched impossible to constrain in individual systems.
Tidal disruption events (TDEs) provide the unique opportunity to witness the birth and evolution of an accretion flow onto a SMBH which evolves on timescales of years \citep{Rees88}. 
Here we show that TDEs launch outflows during a super-Eddington accretion phase and a second, physically distinct outflow, at a critical accretion rate of $L_{\rm crit} \approx0.02$\,$L_{\rm Edd}$, the same as the critical accretion rate for state transitions observed in accreting stellar mass black holes \citep{Maccarone2003,Vahdat2019}.
This work naturally explains the mechanism, observed properties, and detection rate for prompt and delayed outflows observed in TDEs, which until now have been open problems. 
More broadly, we demonstrate that SMBHs exhibit the same accretion-outflow coupling as stellar mass black holes and that the critical low accretion rate threshold for jet formation in black holes is scale invariant. 
}




\maketitle

\section{Main Text}\label{sec1}

The accretion of material onto black holes powers some of the most energetic outflows observed in the Universe \citep{Fender01,Feruglio2010,King2015}. It has long been assumed that the coupling between the outflows launched from black hole accretion flows and the state of these flows themselves should be scale (i.e., black hole mass $M_\bullet$) invariant.  At low (stellar $M_\bullet \simeq 10 M_\odot$) black hole masses, the coupling between outflows and the state of the accretion flow is well-studied and categorized via X-ray binary systems (XRBs). XRBs exhibit clearly defined accretion disk states, and are observed to launch powerful radio-emitting outflows or jets during transitions between these states \citep{Fender04}. At high black hole masses ($M_\bullet \gtrsim 10^6 M_\odot$), active galactic nuclei (AGN) can power galaxy-sized outflows and jets. Studies have shown that a strong correlation exists in AGN between the radio and X-ray luminosities \citep{Merloni03,Falcke2004}, suggesting some form of a disk-jet connection, but the jet launching and accretion flow states in AGN disks are poorly understood due to the extremely long timescales on which these systems evolve \citep{Maccarone2003_agn}. In this work, we use a population of significantly more rapidly evolving SMBH disks, those that form following a tidal disruption event, to constrain the physical properties of SMBH disks at the time at which outflows and jets are launched. We demonstrate that indeed accretion-outflow coupling is scale invariant, and that the critical accretion rate for jet formation is the same in SMBH accretion flows as stellar mass accretion flows.

A natural assumption, if accretion is indeed scale invariant, would be that the same accretion-outflow coupling and critical transitional scales seen in XRBs would be present in accreting supermassive black hole systems. While this is certainly a well motivated assumption it has not, however, been tested rigorously by modelling sources from across the full spectrum of black hole masses present in accreting systems $M_\bullet \sim 10-10^8 M_\odot$. The reason for this is practical: the most numerous super-massive accreting systems (AGN) evolve on the viscous timescale associated with their disk outer edge \cite{Pringle81}, an extremely long timescale $t_{\rm visc, AGN} \sim \alpha^{-1}\theta^{-2} \sqrt{r_{\rm out}^3/ GM_\bullet} \sim 500,000 (M_\bullet/10^8M_\odot) \, {\rm yr}$  (where we have taken  the usual \cite{SS73} turbulence parameter $\alpha = 0.1$, $\theta = h/r = 0.1$ is the disk aspect ratio, and we have assumed a typical outer radius scale of $r_{\rm out} = 10^5 GM_\bullet/c^2$). Limited by observational campaigns that probe $\sim 10^{-5}$ of the natural evolutionary timescale of the system, it will never be possible to observe a single AGN transition through different classical accretion states. 

There exists, however, another population of accreting super-massive black hole systems which are much more amenable to a detailed (individual source level) analysis. These are those disks which form following a tidal disruption event (TDE) -- a transient event initiated by a star being scattered sufficiently close to a black hole that it is ripped apart by tidal forces, initiating a (relatively) rapidly evolving accretion flow \citep{Rees88, mummery2024fitted, Guolo25c}. The disks which form from these events are extremely compact, with initial radial extent set roughly (up to order unity constants) by the tidal radius (the radius at which the star is disrupted) $r_{\rm out} \simeq 2 R_T = 2R_\star(M_\bullet/M_\star)^{1/3}$ \cite{Rees88}. The viscous timescale associated with tidal disruption event disks is therefore {\it independent} of black hole properties (to leading order), $t_{\rm visc, TDE} \sim \alpha^{-1}\theta^{-2} \sqrt{r_{\rm out}^3/ GM_\bullet} \sim \alpha^{-1}\theta^{-2} \sqrt{8R_\star^3/ GM_\star} \sim  50r_\star^{3/2}m_\star^{-1/2} \, {\rm d} $ and depends only on the density of the disrupted star (we have defined $r_\star \equiv R_\star/R_\odot$ and $m_\star \equiv M_\star/M_\odot$ with all other parameters set to the same values as above).   The exact same scaling holds for the viscous timescale for the outer edge of the disks in low mass X-ray binaries fed by Roche-Lobe overflow, as the Roche-overflow radius is $r_{\rm out} \simeq R_{L} \approx  R_\star(M_\bullet/8M_\star)^{1/3}$ \cite{Eggleton83}, meaning $t_{\rm visc, LMXRB} \sim \alpha^{-1}\theta^{-2} \sqrt{R_{L}^3/ GM_\bullet} \sim \alpha^{-1}\theta^{-2} \sqrt{R_\star^3/ 8GM_\star} \sim  6r_\star^{3/2}m_\star^{-1/2}\, {\rm d} $.  While LMXRBs typically evolve somewhat quicker than this (as their outbursts are initiated within the disk, not at the outer disk edge), we see that these two systems (despite ranging in black hole mass by plausibly $\sim 7$ orders of magnitude) are closely analogous, and can be probed on similar observational timescales. 

Tidal disruption events therefore represent a unique probe of time-evolving accretion onto supermassive black holes, and are the only systems in which the scale-invariance of black hole accretion-outflow coupling can be tested on human-observable timescales. We utilise this simple yet profound result in this work. 


With the advent of all-sky optical and X-ray surveys, the number of known TDEs has drastically increased over the last decade to some $\sim200$ events currently known \citep{vanVelzen19_ZTF,Hammerstein23,Yao24,Grotova25}. Long-term optical and UV monitoring has revealed long-lived accretion disk emission following TDEs, characterised by a slowly evolving plateau in excess of the host galaxy emission \citep{vanVelzen19, MumBalb20a, Mummery_et_al_2024}. 
At radio frequencies, observations of TDEs have revealed the switching on of jets and outflows from SMBHs \citep[see][for a review]{Alexander2020}. 
The underlying physical mechanism which produces these outflows  has been a subject of much debate, with possibilities including the unbound debris stream \citep{Krolik2016}, material ejected during debris collision shocks \citep{Lu2020}, or accretion disk driven winds or jets \citep{Alexander16,vanVelzen16,Stein21}. To complicate the picture, recent long-term monitoring observations of TDEs have revealed a population that show rising radio emission hundreds to thousands of days after the TDE occurred \citep{Cendes2024}, and in some cases TDEs have shown two distinct radio flares \citep{Horesh2021_15oi,Goodwin25}. The mechanism that powers these late-time (re)flares has also been a subject of much debate \cite{Cendes2022_hyz,Horesh2021_15oi,Matsumoto2023,Zhuang2025}. 

A natural resolution of both problems (the scale-invariance of accretion and the origin of TDE radio emission) would be if these two distinct populations of TDE radio flares   corresponded to two well-understood accretion rate scales associated with radio emission in XRBs. Namely, $\dot m\sim 1$ (an outflow launched by super-Eddington accretion) producing prompt TDE radio-emission, and a drop through $\dot m\sim 0.02$ and the switching on of a steady compact jet producing the second, delayed, radio flares at late times. Clearly the only way to test this hypothesis is to constrain the properties of TDE accretion flows at the time of radio outflow launch. 

\begin{figure}
    \centering
    \includegraphics[width=0.48\linewidth]{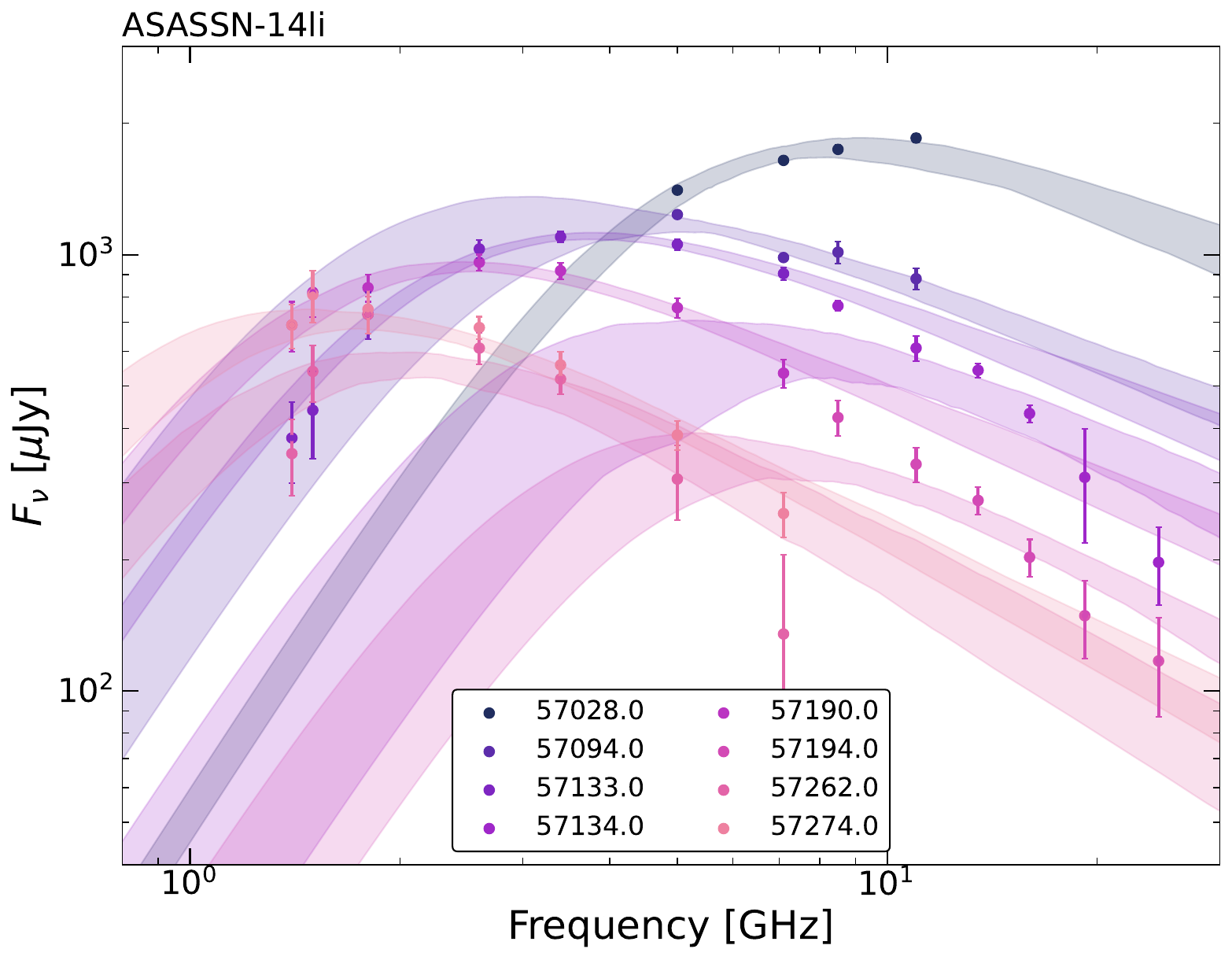}
    \includegraphics[width=0.48\linewidth]{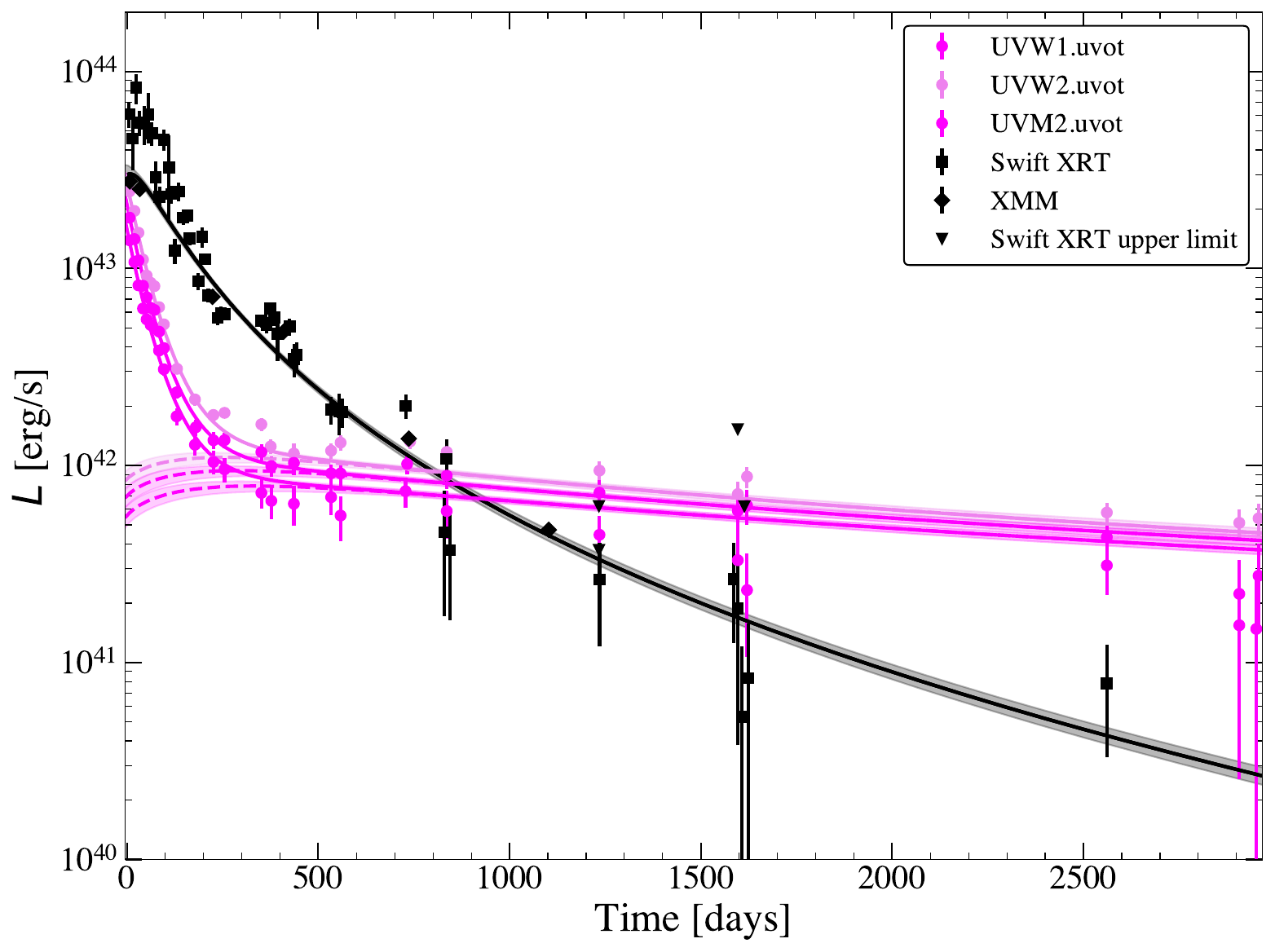}
    \includegraphics[width=0.48\linewidth]{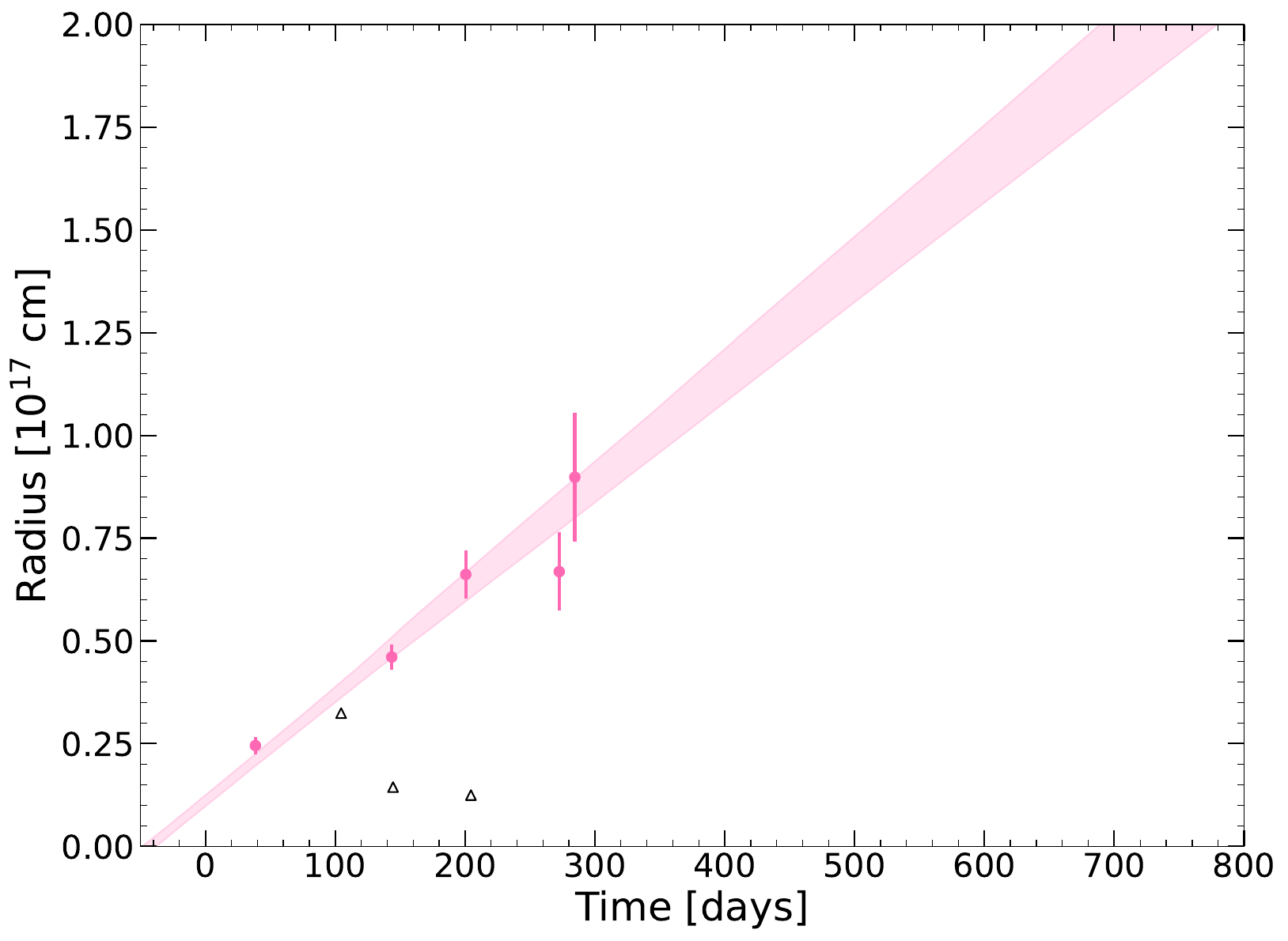}
    \includegraphics[width=0.48\linewidth]{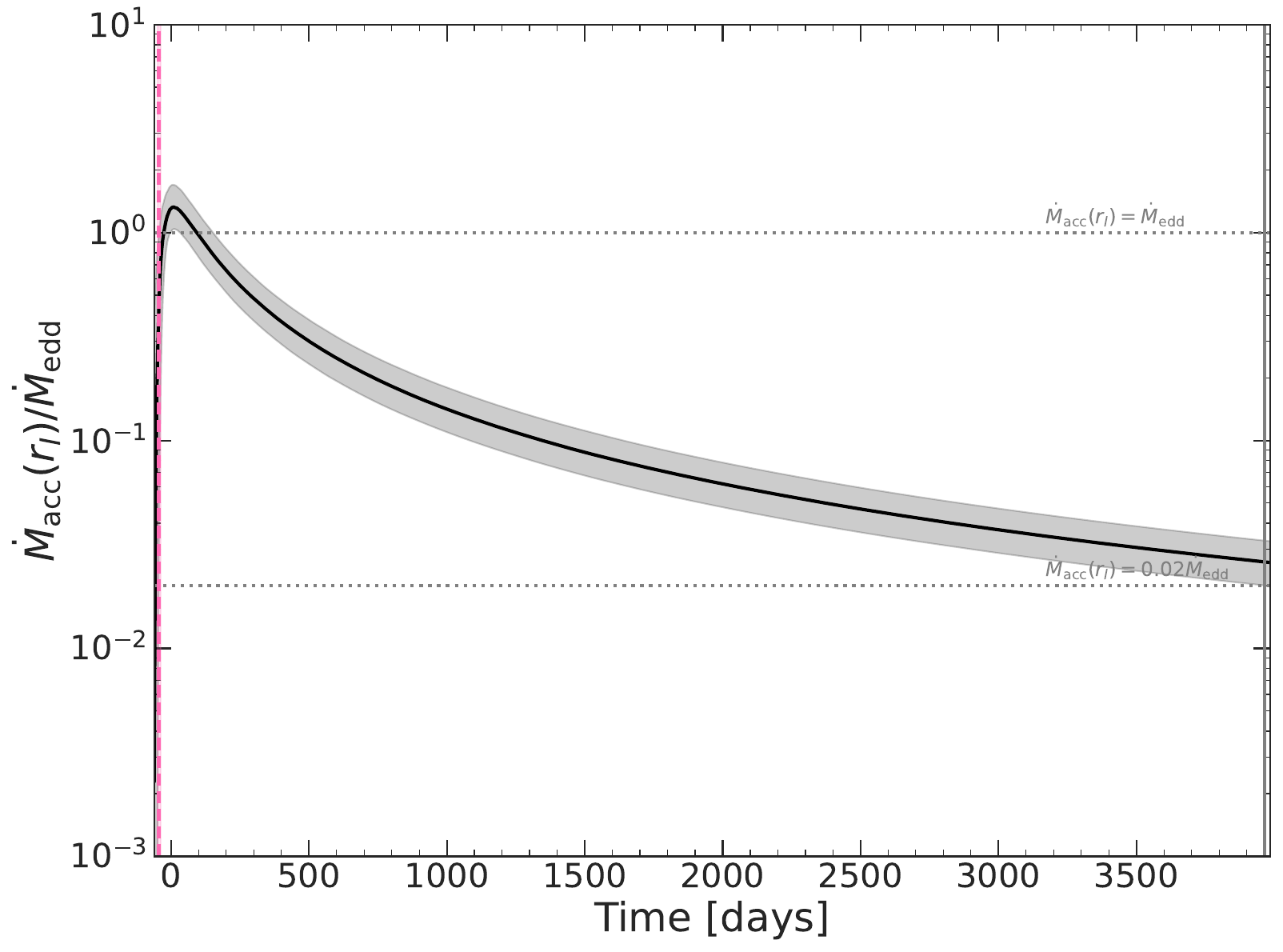}
    \includegraphics[width=0.7\linewidth]{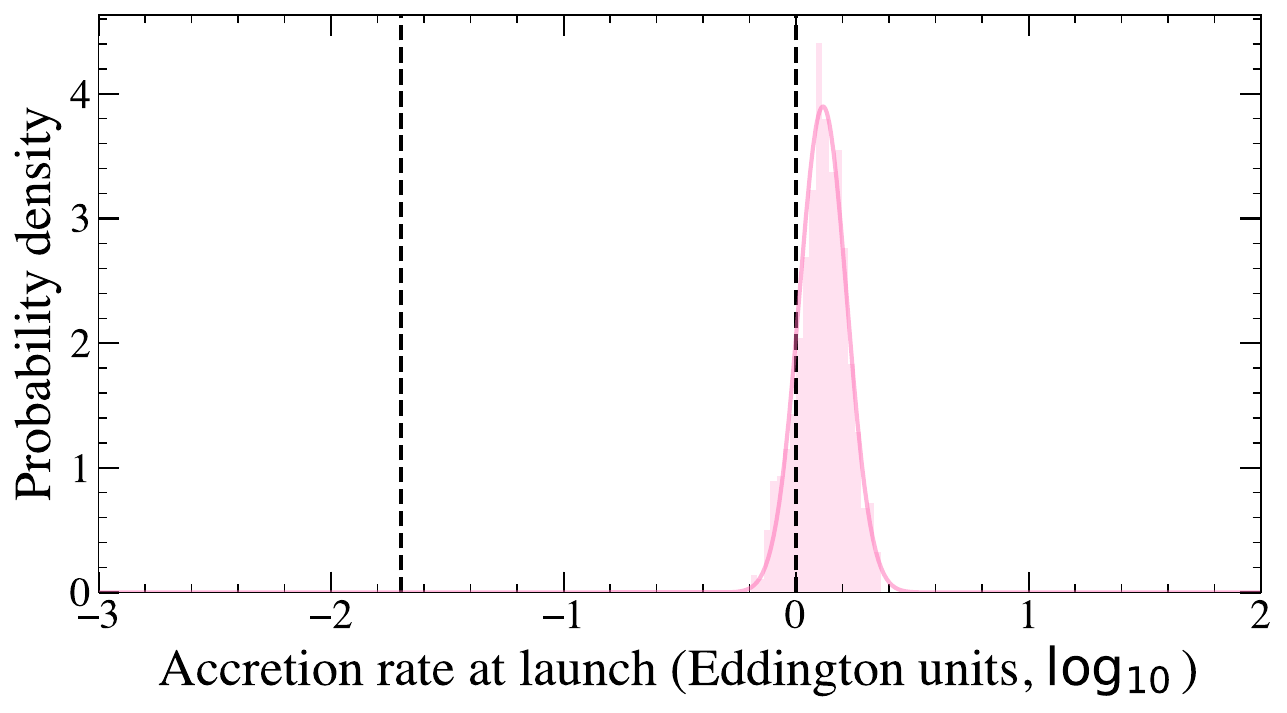}
    \caption{An example of the four steps in our analysis which constrain the accretion rate at outflow launch. Displayed is the TDE ASASSN-14li, all other sources are shown in the methods. Firstly, the radio spectra at multiple epochs are modelled to constrain the peak frequency, flux density, and optically-thin spectral slope (top left). Next, the constrained radio spectral properties are used to infer the outflow radius at each epoch - excluding epochs where the peak of the spectrum was unable to be constrained (black triangles) - which is then modelled to determine the outflow launch time (middle left). The optical, UV, and X-ray lightcurves of the TDE are modelled to constrain the disk properties over time (top right), from which the accretion rate evolution with time can be extracted (middle right). The accretion rate posterior is then sampled at the time of the outflow launch, and a posterior distribution for the accretion rate at the time of the outflow launch is obtained (bottom panel). The shaded regions on each plot indicate $1\sigma$ confidence intervals. }
    \label{fig:models_14li}
\end{figure}

To probe the accretion-outflow coupling in TDE systems we collated a sample of all 20 TDEs with well-sampled optical, UV, X-ray, and radio observations publicly available (see \ref{sec:sample_selection} for detailed sample selection). The radio observations allow robust inference of the outflow launch date (Section \ref{sec:radio}) and the optical, UV, and X-ray observations allow robust inference of the accretion rate at the time of outflow launch when modelled as an accretion disk using fully time-dependent relativistic disk theory \citep[e.g.,][]{MumBalb20a, MumBalb20b, Goodwin22, Nicholl24,  Goodwin25, Chakraborty25, mummery2024fitted, Guolo25time} (Section \ref{sec:disk}).  
An example of the modelling steps that allow us to infer the accretion rate at the time of the outflow launch is shown in Figure \ref{fig:models_14li}, and model fits for all well-constrained TDEs are shown in Section \ref{sec:radio}. 
For 3 of the sources the late-time optical/UV data was of insufficient quality to constrain the disk, and these sources were entirely removed from the sample. For eight further sources the disk is only well constrained at late times (when  no outflow is inferred to have been launched), but not at early times, meaning these sources were also removed from our sample. This leaves 10 sources, and 11 radio flares, at which we have robust constraints on the properties of the disk at outflow  launch time (\ref{sec:disk}). This is comparable to the sample sizes of modern XRB studies \cite{Vahdat2019}. 

\begin{figure}
    \centering
    \includegraphics[width=0.95\linewidth]{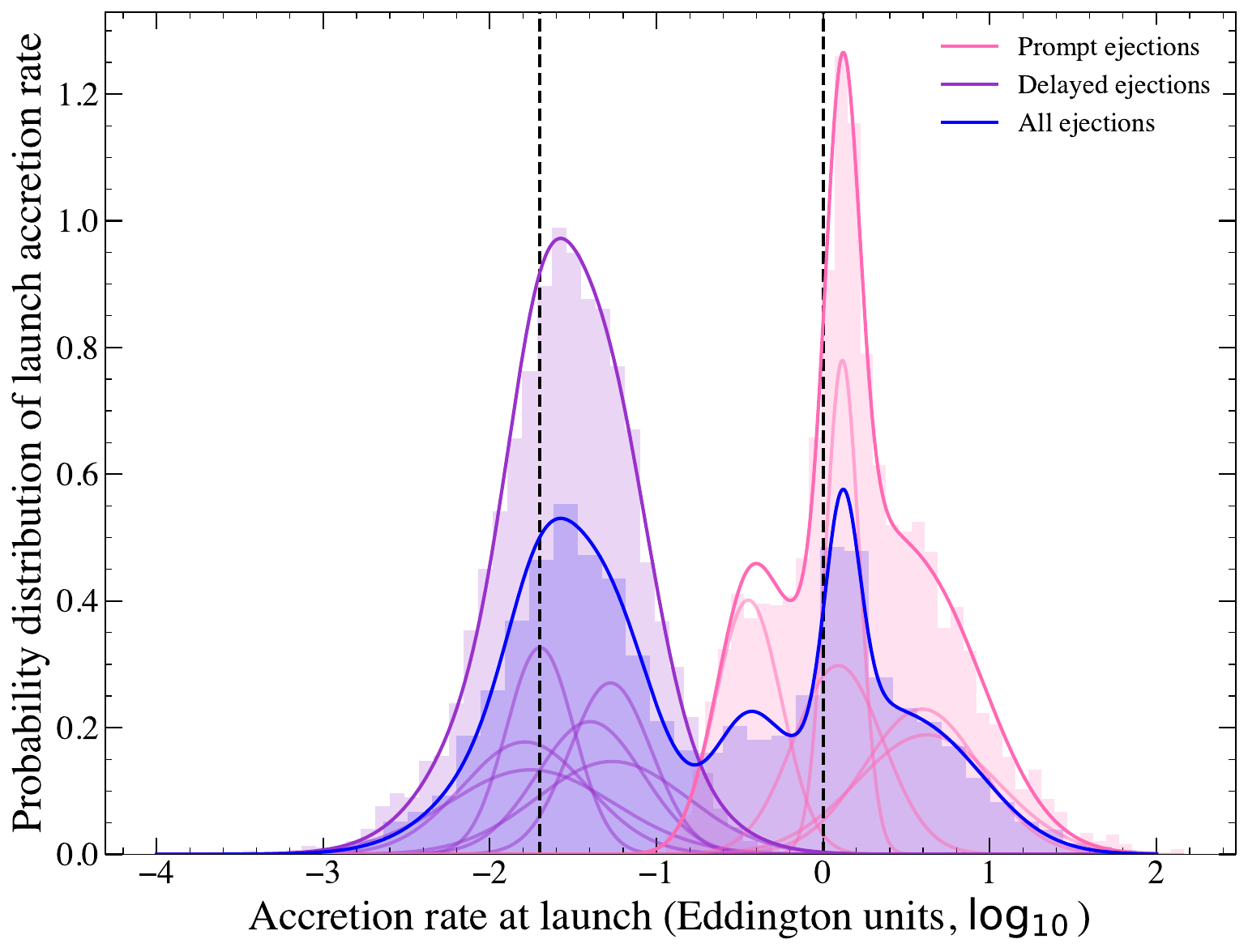}
    \caption{The accretion rate of TDE accretion disks, normalised by the Eddington and presented on a log scale, at the time at which outflows were launched from the 10 TDE systems studied in this work. We split the outflows by prompt (pink) and delayed (purple). The total distribution is shown in blue, and is distinctly bi-modal. Every prompt flare is consistent with having been launched above $\dot m \sim 1$, while all delayed flares are consistent with being launched at $\dot m \sim 0.02$. Individual flares are shown by smoothed Gaussian profiles, while the full (numerical) distributions are shown by histograms. }
    \label{fig:solved}
\end{figure}

We find that the accretion rate at the time of the radio outflow launch is a distinctly bi-modal distribution (Figure \ref{fig:solved}). In Figure \ref{fig:solved} we colour delayed radio outflows by purple curves (with individual sources shown by smoothed Gaussian profiles), while prompt flares are coloured pink. All delayed flares are consistent with being launched at $\dot m \sim 0.02$ with a posterior average of $\dot m_{\rm launch, low} = 0.03^{+0.03}_{-0.02}$, and all prompt flares are consistent with being launched at (or above) $\dot m \simeq 1$ with a posterior average of $\dot m_{\rm launch, high} = 2.2^{+2.8}_{-1.1}$ (here $\dot m \equiv \dot M/\dot M_{\rm Edd}$ is the Eddington-normalised mass accretion rate, see \ref{sec:disk}). 
It is well-established that XRB disks undergo a state transition that switches on a compact radio-emitting jet at the critical low accretion rate threshold of $L_{\rm Bol} \approx0.02L_{\rm Edd}$ \citep{Maccarone2003,Kalemci2013,Vahdat2019}, and also launch outflows when accreting at high (super) Eddington rates. The state transition is characterised by a change in the X-ray spectrum (from predominately soft X-ray emission to predominately hard X-ray emission), and the switching on of a compact persistent radio jet \citep{Fender04}. It is suggested that the hard X-rays are produced by a corona that forms when the inner accretion disk extends closer to the black hole, providing the conditions necessary for a radio jet to form \citep{Fender04}. X-ray observations of the TDEs in our sample which have launched delayed radio flares show emission consistent with a corona: three of the TDEs in our sample have shown evidence of hard X-ray emission $>3$\,yr after the TDE \citep{Guolo2024,Jonker20}, and in all cases are consistent with a corona switching on at the time of the delayed radio outflow launch (see Section \ref{sec:discussion}). 

\begin{figure}
    \centering
    \includegraphics[width=0.8\linewidth]{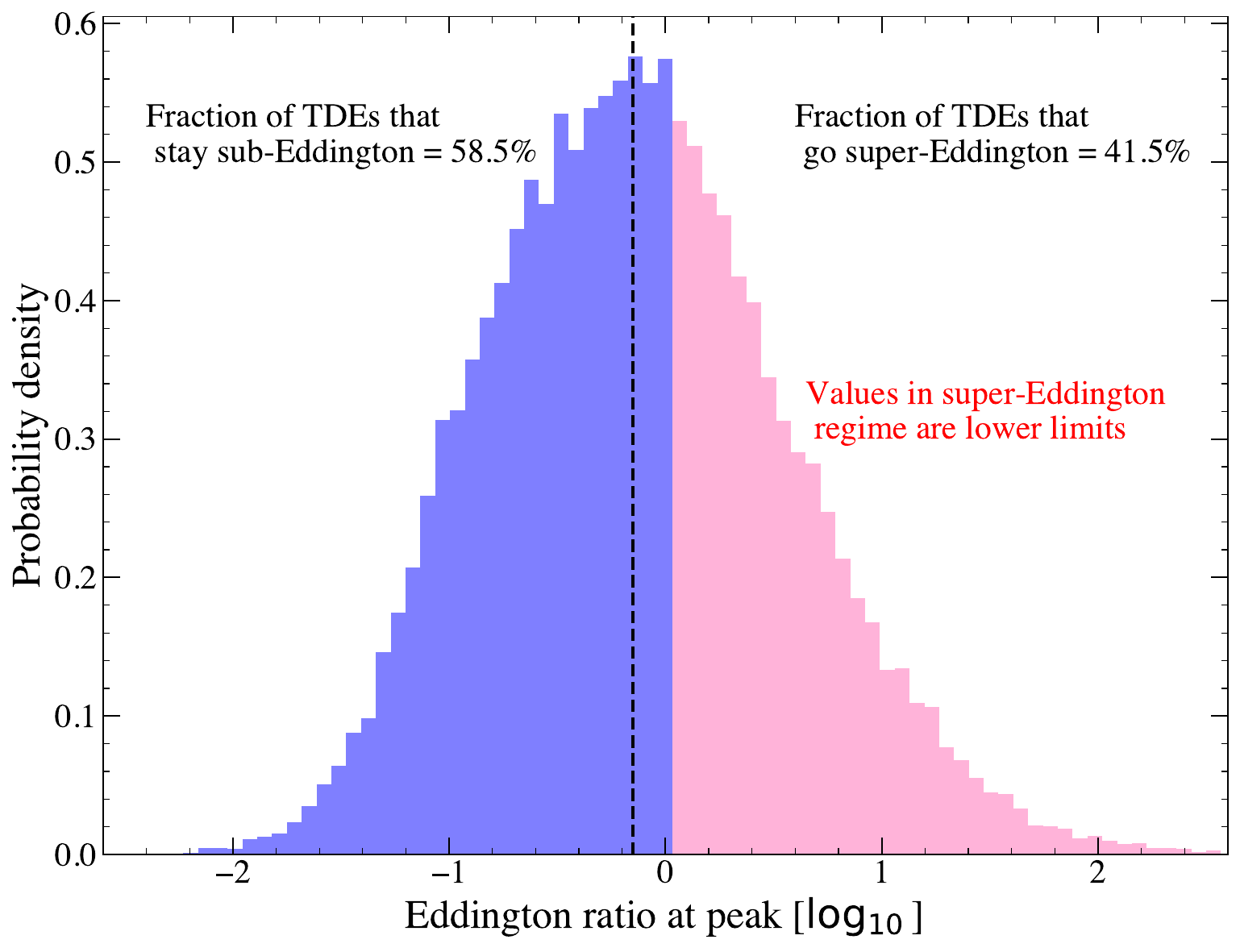}
    \includegraphics[width=0.8\linewidth]{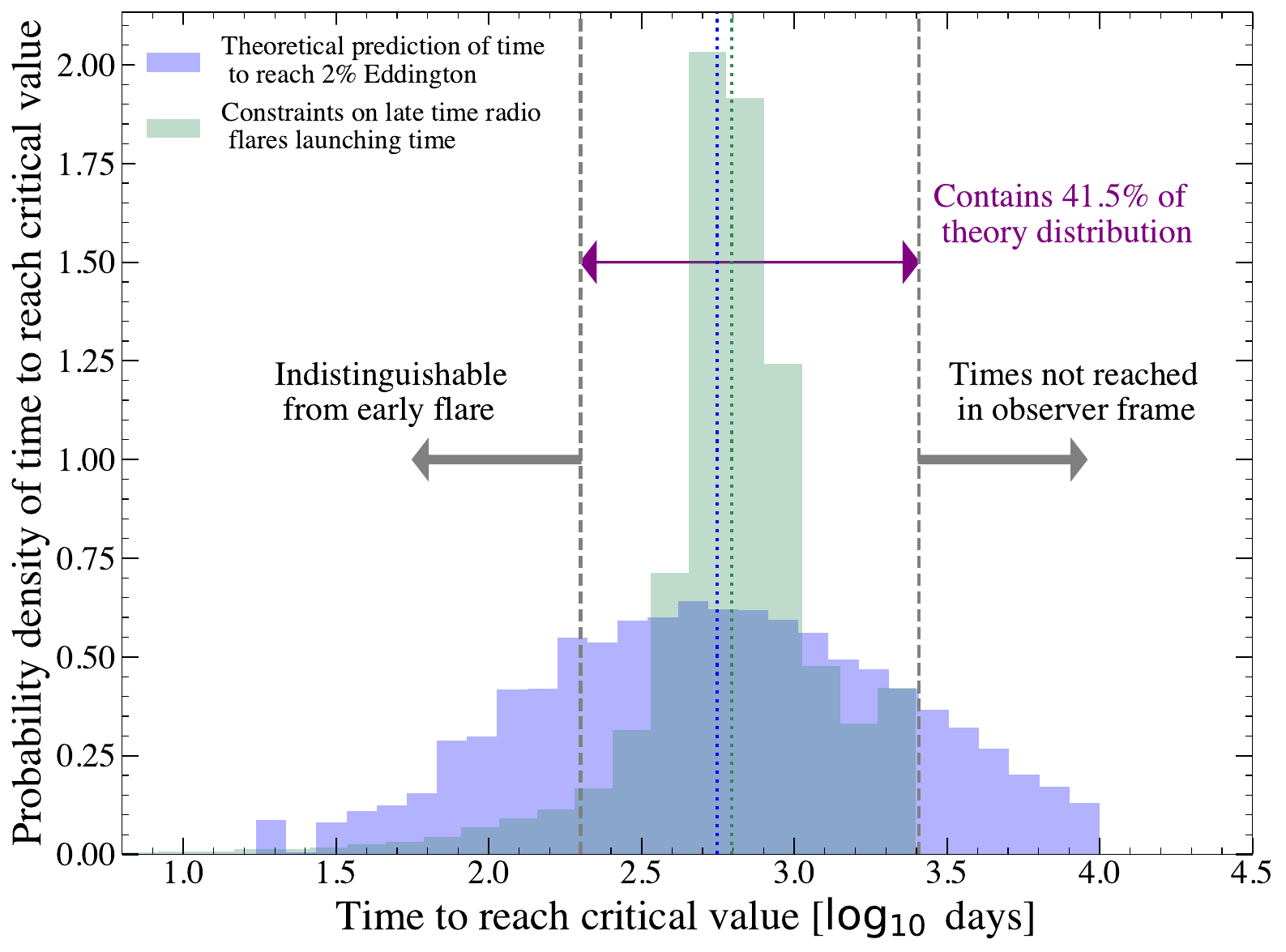}
    \caption{Population synthesis of TDEs, computed from a simulation of $N=10^6$ TDEs using the observationally constrained TDE black hole mass function. The upper panel shows the peak Eddington ratio reached in TDE disks, where $\simeq 42\%$ of TDE disks go super-Eddington (and can therefore produce prompt radio emission), comparable to the observed prevalence of $\sim 30-50\%$ (\citep{Alexander2020, Anumarlapudi2024, Somalwar2025, Goodwin2025_eros}) of prompt radio flares. The lower panel shows the time for TDE disks to reach $2\%$ Eddington (in blue), for which $\simeq 42\%$ to do within the first $7$ years (bounding the observational window of real TDEs). This is comparable with the observed prevalence of delayed flares $\sim 40\%$ (\cite{Cendes2024}). In green we show the observed distribution of delayed flare launch times (computed in this work), which has near identical distribution median to that computed by  population synthesis  (vertical dotted lines).  }
    \label{fig:populations}
\end{figure}

We have demonstrated that the critical accretion scales $\dot m\sim 0.02$ and $\dot m \sim 1$ are {\it necessary} conditions for launching an outflow from a TDE. This begs the question of whether this is simply a necessary condition, or whether this is necessary {\it and sufficient} for radio-bright emission to be produced by a TDE.  In effect we wish to determine whether the population of radio non-detected TDEs are consistent with our paradigm. To do this we perform population synthesis (with procedure spelt out in \ref{sec:disk}), using the recently constrained TDE black hole mass function \cite{MummeryVV25}. We sample $N = 10^6$ TDE disk systems, and for each system record both the peak accretion rate $\dot m_{\rm peak}$ reached during the disks evolution, and the time it takes for that disk to drop to the critical scale $\dot m(t_{\rm tr}) = 0.02$. The distributions of $\dot m_{\rm peak}$ (upper) and $t_{\rm tr}$ (lower) are displayed in Figure \ref{fig:populations}. 

We find that only $\approx 40\%$ of TDE disks should have transitioned through $\dot m=0.02$ within the time-frame $200\, {\rm d}<t_{\rm tr}< 7\, {\rm yr}$ (broadly bounding the observable window to  for delayed flares in TDEs), agreeing remarkably well with the observational estimate of \cite{Cendes2024}, who also infer that $\approx 40\%$ of TDEs launch delayed radio flares. Interestingly, the posterior median for the transition time predicted in this population synthesis ($\bar t_{\rm tr} \approx 560\, {\rm d}$) is very close to the posterior median of that observed in our sample ($\bar t_{\rm late} \approx 625\, {\rm d}$). 
Similarly, only $\approx 42\%$ of TDE disks are expected to undergo a super-Eddington accretion phase, which compares favourably with current rates of prompt radio outflow detection in TDEs of 30--50$\%$ \citep{Alexander2020, Anumarlapudi2024, Somalwar2025, Goodwin2025_eros}.  This population synthesis allows us also to perform a null hypothesis test, by computing the accretion rates that would be inferred at launch  across a population of TDEs if the launch time knew nothing about the state of the accretion flow and was instead log-uniformly distributed between 10 days and 7 years (broadly bounding the observing windows of TDEs). In this case, the distribution would be well described by a single Gaussian peaking at $\dot m \sim 0.2$, far removed from our observations (see section \ref{sec:stats}). 

%

Our findings reveal that TDEs launch radio-emitting outflows at two distinct evolutionary times, with different mechanisms producing these two outflows (a super-Eddington wind and a compact state transitional jet).
A natural prediction of this paradigm is that the observed radio properties of prompt and delayed outflows in TDEs should be distinct. 
Analyses of the few TDEs that have shown both prompt and delayed outflow behaviour to date have found different energetics in the radio properties of the two flares \citep{Hajela2025,Goodwin25,Christy2025}. Modelling the radio lightcurve evolution of the TDEs in our sample, overall we find the rise indices of the prompt and delayed outflow populations are bi-modal, with prompt outflows rising with indices 1--2.5, whereas delayed outflows rise with indices $>4$ (see Section \ref{sec:radio_lcs}). The prompt outflow rise indices are well-explained by a  ballistically expanding outflow sweeping up material from a circumnuclear medium with little continuous energy injection into the outflow. Conversely, the steep rise indices of the delayed flares require energy injection, consistent with the presentation of a compact jet formed from the inner accretion disk as the low-hard  state is entered.

The individual aspects of our analysis are subject to some physical uncertainty, but not in a manner which biases our results. By computing the synchrotron equipartition radius we are inferring a formal lower bound on the emitting radius at a given epoch. However, our inference of {\it launch dates} are robust to assumptions regarding the outflow geometry and shock microphysics (e.g. relative equipartition fraction), as these assumptions merely scale the absolute value of the inferred radius, but cannot change the location of its temporal zero (i.e., the launch date, see \ref{sec:radio}). More consequential is our assumption that the velocity of the outflow is constant with time. Whilst for accretion disk winds observed during the rise to radio peak this assumption is likely valid, for compact radio jets there may be some variability in the jet velocity depending on the dynamics of the accretion flow and jet properties. We believe it unlikely that this would significantly shift the inferred launch dates (which already have moderate observational uncertainty which we propagate through our analysis). Additionally, the relativistic disk model used in this work will begin to break down as the super-Eddington accretion regime is entered. However, the bias here is that our model should {\it under-predict} the accretion rate in these states, meaning that our inference of a super-Eddington outflow launching would only be strengthened by a disk model which incorporates additional high $\dot m$ physics.

This work presents compelling evidence that the critical accretion rate threshold for the launching of radio jets in the low-hard disk state around SMBHs is the same as XRBs.
While the ubiquity of compact jets in black holes accreting at low rates at all mass scales has been known since the discovery of the fundamental plane correlation between radio luminosity, X-ray luminosity, and black hole mass \citep{Merloni03}, the critical accretion rate threshold for  compact jet launching was unconstrained in supermassive black holes prior to this work. This observed mass invariance has theoretical support, as \cite{Rees1982} argued that a critical accretion rate threshold should exist for a black hole accretion flow which supports the launching of a radio jet that is dependent on the scale height of the disk, $h$, and the viscosity parameter, $\alpha$, but crucially not the BH mass.  In demonstrating that the critical accretion rate that initiates the low-hard state transition is the same in stellar and supermassive black holes, we have confirmed this long-held conjecture.

Our results naturally explain the timing and detection rate of both prompt and delayed outflows observed in TDEs, elucidating the underlying mechanisms that produce radio bright outflows in TDEs at all times during their evolution. Our analysis was carried out with the full sample of 10 TDE systems in which publicly available (and constraining) multiwavelength data were available. As we enter the era of LSST and future large ($N_{\rm TDE}\sim 10^4-10^5$) samples of TDEs \cite{Bricman20}, the framework put forward in this paper will facilitate the detailed modelling of jet launching conditions in SMBH systems. The unique constraining power of rapidly evolving TDE disks, which we have made use of in this work, will likely lead to much further insight into the complex physics of disk-jet coupling at supermassive black hole scales.

\bmhead{Acknowledgements}

The authors would like to thank Muryel Guolo for helpful discussions and James Miller-Jones for extremely useful comments on an earlier draft of the manuscript. 
AJG is grateful for support from the Forrest Research Foundation. 
A.M. acknowledges support from the Ambrose Monell Foundation, the W.M. Keck Foundation and the John N. Bahcall Fellowship Fund at the Institute for Advanced Study.  
This research benefited from discussions at the Kavli Institute for Theoretical Physics (KITP) program on tidal disruption events in April 2024 that was supported in part by the National Science Foundation under PHY- 1748958, and conversations at the conference Tidal Disruption Events and Nuclear Transients: Entering the Data-Rich Era in Crete in 2024, and the conference Tidal Disruption Events and Nuclear Transients: Entering the Data-Rich Era in Madrid in 2025.

\section*{Author contributions}


Both authors collated the public multiwavelength data used in this work. AJG selected the TDEs in the sample. AJG carried out the radio data modelling and AM carried out the disk modelling. AM carried out the statistical tests. Both authors contributed to the analysis and interpretation of the results, and the writing of the manuscript. 




\section{Methods}

The X-ray luminosity observed from an XRB provides a robust estimate of the bolometric luminosity of its disk, owing to the high temperatures of a typical XRB disk in the soft state ($T\sim 1-2$ keV) which puts the bulk of the disk emission in the X-ray band. TDE disks, however, are significantly cooler $T \sim 50$ eV \citep[e.g.,][]{Guolo2024, Mummery_Wevers_23}, a simple result which can be understood via classical disk theory at fixed Eddington ratio $T\sim M_\bullet^{-1/4}$ and the higher black hole masses of TDEs. This means that the X-ray luminosity observed from a TDE is a poor tracer for the accretion rate (the X-ray to bolometric correction is an exponential function of $\dot m$), and many TDEs near the assumed critical transition scale $\dot m\sim 0.02$  indeed show no detectable X-ray emission at these times.  Fortunately, while it is known that the early time optical/UV luminosity in a TDE does not trace the accretion rate \cite{Guolo25c}, on longer timescales ($\Delta t \gtrsim 1\, {\rm yr}$) TDE disks enter an observable ``plateau-phase''  at optical/UV frequencies \cite{vanVelzen19, MumBalb20a, Mummery_et_al_2024}, which provides robust constraints on the disk properties even in the absence of X-ray emission. 

The only way to infer the properties of a TDE accretion flow  is therefore to model the full spectral energy distribution (SED) of the system from optical to X-ray energies in the disk-dominated phase. This can be done on an epoch-by-epoch basis \citep[e.g.,][]{GuoloMum24,Guolo25, Wevers25}, but the more powerful technique is to dynamically model the evolving disk using fully time-dependent relativistic disk theory \citep[e.g.,][]{MumBalb20a, MumBalb20b, Goodwin22, Nicholl24,  Goodwin25, Chakraborty25, mummery2024fitted, Guolo25time}, which also allows the accretion rate to be inferred during epochs when no observations were taken (essential for providing constraints at outflow launch times). We apply this latter technique in this paper, while verifying that our results are consistent with those inferred at individual epochs where such fitting has been performed \cite{Guolo25c}. 

To probe the accretion-outflow coupling in TDE systems we collated a sample of all 20 TDEs with well-sampled optical, UV, X-ray, and radio observations publicly available (see \ref{sec:sample_selection} for detailed sample selection). 
We first modelled the radio spectra for each source in order to infer the synchrotron-emitting region size as a function of time, using a standard equipartition analysis. The  spectral and equipartition models used are described in Section (\ref{sec:radio}). For each TDE, we were then able to infer a launch date of each observed outflow by backwards-modelling the measured outflow radius with time (\ref{sec:radio_equi}). Consistent with current literature, we identified two distinct populations of outflows: those launched within $100$\,d of the optical flare (``prompt") and those launched hundreds to thousands of days after the optical flare (``delayed''). We additionally modelled the radio lightcurves of each of the TDEs in the sample, employing a broken power-law model appropriate for the evolution of a synchrotron-emitting source in which the peak of the emission is associated with synchrotron self-absorption \citep{Chevalier1998}. We find that these two outflow types are physically distinct, with delayed flares rising much faster than prompt flares (Section \ref{sec:radio_lcs}). 

With the outflow launch times constrained for each outflow, we then modelled the optical, UV, and X-ray lightcurves of each of the 20 sources in the sample, using the {\tt FitTeD} relativistic disk code \cite{mummery2024fitted}, see section (\ref{sec:disk}). For 2 of the sources the late-time optical/UV data was of insufficient quality to constrain the disk, and these sources were entirely removed from the sample. For eight further sources the disk is only well constrained at late times (when  no outflow is inferred to have been launched), but not at early times, meaning these sources were also removed from our sample. This leaves 10 sources, and 11 radio flares, at which we have robust constraints on the properties of the disk at outflow  launch time (\ref{sec:disk}). This is comparable to the sample sizes of modern XRB studies \cite{Vahdat2019}.

With posterior distributions of both the launch time $p(t_{\rm launch})$ and the disk parameters $p_{\rm disk}$ constrained from observations, we sampled simultaneously from both, computing the inferred distribution of the (Eddington normalised) accretion rate $\dot m\equiv \dot M_{\rm acc}/\dot M_{\rm Edd}$ at launch, which we denote $p(\dot m|t_{\rm launch})$. The choice of accretion rate rather than bolometric luminosity makes no difference to the statistical significance or inferred critical values of these results (see \ref{sec:stats}).

\subsection{Sample selection}\label{sec:sample_selection}
We manually searched the literature for all TDEs that satisfied the following selection criteria:

\begin{itemize}
    \item Classified as a bona-fide TDE based on typical optical spectral features, optical lightcurve, and quiescent host galaxy
    \item Radio-detected and at least one broadband radio spectrum (with $>3$ frequencies sampled) publicly available
    \item Localised to the nucleus of the host galaxy\footnote{The radio properties of synchrotron emitting outflows are influenced by the density of the ambient medium the outflow propagates through. To maintain a consistent sample, we therefore limit our sample to nuclear SMBHs which are more likely to have similar ambient density environments than black holes elsewhere in the host galaxy.}
    \item UV and optically-detected with lightcurves publicly available
\end{itemize}
This search resulted in 20 TDEs, with discovery papers listed in Table \ref{tab:discovery} (note that this is the reference for original TDE discovery, not necessarily the reference at which radio emission was first discovered). We exclude the relativistic (jetted) TDEs from this work as their multi-wavelength emission is likely dominated by the relativistic jet rather than an accretion disk, preventing us from determining a quantitative description of their accretion flows. 

\begin{table}
    \centering
    \begin{tabular}{p{80pt} p{145pt} }
    Discovery reference & Tidal disruption event  \\
    \hline
\cite{Holoien14} & ASASSN-14ae  \\
\cite{Miller15,Holoien16} & ASASSN-14li  \\
\cite{Holoien16b} & ASASSN-15oi \\
\cite{Blagorodnova17} & iPTF-16fnl \\
\cite{vanVelzen19b,Holoien18a} & AT2018zr \\
\cite{Holoien19} & AT2019ahk \\
\cite{Leloudas19} & AT2018dyb \\
\cite{Short20,vanVelzen20} & AT2018hyz \\
\cite{Nicholl19,vanVelzen20} & AT2019qiz \\
\cite{vanVelzen20} & AT2018hco, \mbox{AT2019dsg}, \mbox{AT2019ehz}  \\
\cite{vanVelzen20,Hinkle21,Liu22} & AT2019azh \\
\cite{vanVelzen21} & AT2019eve  \\  
\cite{Wevers22,Hammerstein23} & AT2020zso \\
\cite{Hammerstein23} &  \mbox{AT2019teq}, \mbox{AT2020opy} \\
\cite{Goodwin23,Yao23} & AT2020vwl \\
\cite{Yao23} & \mbox{AT2021sdu} \\ 
\cite{Homan2023_eJ2344} & eRASSt J234402.9-352640 = ``eJ2344'' \\ \\
    \end{tabular}
    \caption{Literature references (``discovery papers") for the TDEs used in work. We use AT names where possible, and for simplicity shall refer to the tidal disruption event eRASSt J234402.9-352640 as eJ2344 for the remainder of this paper.  }
    \label{tab:discovery}
\end{table}

For these 20 TDEs we then gathered all publicly available multi-wavelength data. For optical/UV and X-ray emission this is rather simple, as compilations already exist. We make use of the {\tt manyTDE} database \cite{Mummery_et_al_2024} for the optical/UV emission for all TDEs except eJ2344 (which was discovered after the database was produced), and AT2019ahk (ASASSN-19bt). For the X-ray emission, if it was detected (in a thermal-dominated state) we use the compilation of \cite{Guolo24} (again except for eJ2344 and AT2019ahk). For X-ray non-detected sources (which have not been compiled into a database), we return to the original discovery papers. For the radio detections of TDEs, we took data for each source from individual papers.  The references for our multi-wavelength data sets are presented in Table \ref{tab:data}. 

\begin{table}
    \centering
    \begin{tabular}{p{90pt} p{60pt} p{80pt} p{80pt} }
    Tidal disruption event & Optical/UV data source & X-ray data source & Radio data source   \\
    \hline
ASASSN-14ae  & \cite{Mummery_et_al_2024} & \cite{Holoien14}   & \cite{Cendes2024}\\
ASASSN-14li  &\cite{Mummery_et_al_2024} & \cite{Guolo24} & \cite{Alexander16}\\
ASASSN-15oi &\cite{Mummery_et_al_2024} & \cite{Guolo24} & \cite{Horesh2021_15oi,Hajela2025}\\
iPTF-16fnl &\cite{Mummery_et_al_2024}& \cite{Blagorodnova17}& \cite{Horesh2021_16fnl}\\
AT2018zr &\cite{Mummery_et_al_2024} & \cite{Guolo24} & \cite{Cendes2024}\\
AT2019ahk 
&\cite{Mummery_et_al_2024}& \cite{Holoien19} &\cite{Christy2024} \\
AT2018dyb &\cite{Mummery_et_al_2024}& \cite{Leloudas19} &\cite{Cendes2024} \\
AT2018hyz &\cite{Mummery_et_al_2024} & \cite{Guolo24} & \cite{Cendes2022_hyz, Cendes2025_hyz}\\
AT2019qiz &\cite{Mummery_et_al_2024}&\cite{Guolo24,Nicholl24}& \cite{Alexander2025}\\
AT2018hco &\cite{Mummery_et_al_2024}& \cite{vanVelzen20} &\cite{Cendes2024} \\
AT2019dsg &\cite{Mummery_et_al_2024} & \cite{Guolo24} &\cite{Cendes2021_dsg} \\
AT2019ehz &\cite{Mummery_et_al_2024}& \cite{Guolo24} & \cite{Cendes2024} \\
AT2019azh &\cite{Mummery_et_al_2024} & \cite{Guolo24} & \cite{Goodwin2022,Burn2025} \\
AT2019eve &\cite{Mummery_et_al_2024}&\cite{vanVelzen21}& \cite{Cendes2024} \\  
AT2020zso &\cite{Mummery_et_al_2024}&\cite{Hammerstein23}& \cite{Christy2025}\\
AT2019teq &\cite{Mummery_et_al_2024}&\cite{Guolo24} &\cite{Cendes2024} \\ 
AT2020opy &\cite{Mummery_et_al_2024}&\cite{Goodwin2023_opy}&\cite{Goodwin2023_opy}\\
AT2020vwl &\cite{Mummery_et_al_2024}&\cite{Goodwin25} & \cite{Goodwin23,Goodwin25}\\
AT2021sdu &\cite{Mummery_et_al_2024}&\cite{Yao23}&\cite{Christy2025} \\ 
eJ2344 &\cite{Homan2023_eJ2344}&\cite{Homan2023_eJ2344}&\cite{Goodwin2024_eJ2344} \\ \\
    \end{tabular}
    \caption{The references from which we obtain the publicly available multi-wavelength data used in this study. The reference \cite{Mummery_et_al_2024} collated all publicly available optical/UV data of TDEs (at the time) into the {\tt manyTDE} database, which we make primary use of. All X-ray detections come from the compilation presented in \cite{Guolo24}, except for the more recently discovered eJ2344 source. X-ray upper limits are typically taken from the discovery papers.  For TDE radio data we took observations for each source from individual papers.    }
    \label{tab:data}
\end{table}

\begin{table}
    \centering
    \begin{tabular}{|p{62pt} | p{47.5pt} p{47.5pt}| p{20pt} | p{47.5pt} p{47.5pt}| p{20pt}|}
    \hline 
    Tidal disruption event & Early Radio Constraints? & Early Disk Constraints? & Use Early? & Late Radio Constraints? & Late Disk Constraints?  & Use Late?  \\
    \hline
    {\bf ASASSN-14ae} & No & No & No & Yes & Yes & {\bf Yes} \\
    {\bf ASASSN-14li} & Yes & Yes & {\bf Yes} & No & Yes & No \\ 
    {\bf ASASSN-15oi} & Yes & Yes & {\bf Yes} & Yes & Yes & {\bf Yes} \\ 
    AT2018dyb & No & No & No & No & Yes & No \\ 
    AT2018hco & No & No & No & No & Yes & No \\
    {\bf AT2018hyz} & No & No & No & Yes & Yes & {\bf Yes} \\ 
    AT2018zr & No & No & No & No & Yes & No \\
    {\bf AT2019ahk} & Yes & No & No & Yes & Yes & {\bf Yes} \\
    {\bf AT2019azh} & Yes & Yes & {\bf Yes} & No & Yes & No \\
    {\bf AT2019dsg} & Yes & Yes & {\bf Yes} & No & Yes & No \\
    AT2019ehz & No & No & No & Yes & No & No \\
    AT2019eve & No & No & No & No & No & No \\
    AT2019qiz & Yes & No & No & No & Yes & No \\
    AT2019teq & No & No & No & No & No & No \\
    AT2020opy & Yes & No & No & No & Yes & No \\
    {\bf AT2020vwl} & Yes & No & No & Yes & Yes & {\bf Yes} \\ 
    {\bf AT2020zso} & Yes & No & No & Yes & Yes & {\bf Yes} \\
    AT2021sdu & Yes & No & No & No & Yes & No \\
    {\bf eJ2344} & Yes & Yes & {\bf Yes} & No & Yes & No \\
    iPTF-16fnl & No & No & No & No & Yes & No \\
    \hline 
\end{tabular}
\caption{The data quality cuts made in this work. Sources with names in bold have useable epochs with which radio properties can be meaningfully compared to disk properties, this ``gold'' sample totals 10 sources and 11 individual radio flares. While all sources were of course detected (at least once) in the radio, many only had one epoch at either early or late times. This is insufficient to meaningfully constrain a radio launch time. Sources are considered to have a well constrained disk at early times if there are numerous X-ray detections of the inner disk at early epochs, and are considered to have a well constrained disk at late times if the optical/UV disk flux at late times exceeds that of the decaying early flare. See text for more details.   }\label{tab:constrained_tdes}
\end{table}

\subsection{Radio modelling}\label{sec:radio}
The purpose of our analysis is to constrain the physical properties of the outflows observed in our sample of TDEs in order to infer the outflow launch time. We do this by fitting each observed radio spectrum of each source with a physically motivated synchrotron emission model to constrain the synchrotron self absorption flux and frequency. We then apply an equipartition analysis to constrain the minimum radius of the emitting region over time for each TDE. Finally we fit the radius over time to infer the outflow launch time and velocity. 

\subsubsection{Radio spectral modelling}\label{sec:radio_spec}
The radio emission from TDEs is well-described by a peaked synchrotron spectrum produced by outflows ejected during the stellar disruption or subsequent accretion of material onto the SMBH (see \cite{Alexander2020} for a review). Ambient electrons are accelerated into a power-law distribution by the blastwave from the outflow, $N(\gamma)\propto \gamma^{-p}$, where $\gamma$ is the electron Lorentz factor ($\gamma\geq \gamma_{\rm m}$, where $\gamma_{\rm m}$ is the minimum Lorentz factor) and $p$ is the synchrotron energy index. The resulting synchrotron spectrum has characteristic break frequencies corresponding to the minimum frequency ($\nu_m$), self-absorption frequency ($\nu_a$), and the cooling frequency ($\nu_c$). TDEs with well-constrained radio spectral observations are most commonly observed to be in the regime $\nu_m < \nu_a < \nu_c$, where the radio spectral peak is associated with synchrotron self-absorption \citep[e.g][]{Alexander16,Goodwin23}. Occasionally, and where sufficient spectral coverage is available, the cooling break corresponding to $\nu_c$ has been detected at frequencies $>15$\,GHz \citep{Cendes2021_dsg,Christy2024}. In this work, the radio data available for each source primarily span 0.65--15\,GHz, with varying levels of spectral coverage within those frequency ranges, so we do not include the cooling break in our spectral model.

Since the synchrotron spectral peak is associated with self-absorption, we fit each observed spectrum with the following equation from \citep{Granot2002}, appropriate for a source in the regime $\nu_m < \nu_a < \nu_c$ where the spectral peak is due to synchrotron self-absorption 

\begin{equation}
    \label{eq:Fnua}
    \begin{aligned}
        F_{\nu, \mathrm{synch}} = F_{\nu,\mathrm{a}} \left[
        \left(\frac{\nu}{\nu_{\rm a}}\right)^{-s\beta_1} +  \left(
        \frac{\nu}{\nu_{\rm a}}\right)^{-s\beta_2
        }\right]^{-1/s}
        \end{aligned}
    \end{equation}
where $\nu$ is the observed frequency, $F_{\nu,\mathrm{a}}$ is the normalisation (corresponding to the self-absorption flux density), $s = 5/4-9p/50$, $\beta_1 = {5}/{2}$, and $\beta_2 = {(1-p)}/{2}$.  

We fit each observed spectrum using a Python implementation of MCMC, \texttt{emcee} \citep{EMCEE}, implementing a Gaussian likelihood function where the variance is underestimated by some fractional amount $f$ of the form
\begin{equation}
    \ln {\cal L}(\Theta) = - {1 \over 2}\sum_{{\nu}} \left[ \frac{\left(O_{\nu} - M_{\nu}\right)^2  }{E_{\nu}^2} + \ln{(2\pi E_{\nu}^2)} \right] ,
\end{equation}
where the model parameters are $\Theta = [F_{\nu,a}, \nu_a, p]$,  the observed flux density at frequency $\nu$ is denoted  $O_{\nu}$, and $M_{\nu}$ is the modelled flux density at frequency $\nu$. The parameter $E_{\nu}$ is equal to $E^2_\nu = O_{\nu, {\rm err}}^2 + M_{\nu}^2f^2$ where $O_{\nu, {\rm err}}$ is the uncertainty of the observed flux density at frequency $\nu$. 

We assume flat prior distributions for all parameters, allowing  $10^{-6} < F_{\nu,\mathrm{a}}$\,(mJy)$ < 2F_{\nu,{\rm obs,max}}$, $0.1 < \nu_a$\,(GHz)$ < 20 $, and $2 < p < 3.5$. The prior on the flux amplitude is flat in log space, with all other priors flat in linear space. We run each chain for 2000 steps with 400 walkers, discarding the first 1000 steps for burnin. We allow individual frequencies of observations to be combined into one epoch if taken within 20\,d and we only include epochs for which three or more frequency points are available (excepting AT2019qiz for which we fit spectra with 2 points, but necessarily exclude this source from the ``gold" sample). All sources in the sample were undetected in archival radio survey data prior to the TDE, indicating weak/no host radio emission, except ASASSN-14li. For this source, we subtracted a host emission component as described in \cite{Alexander16} before carrying out the spectral fitting. 
The individual radio spectral fits for each of the ``gold" TDEs are shown in Figures \ref{fig:models_1}--\ref{fig:models_9}.

\subsubsection{Outflow radius and launch time modelling}\label{sec:radio_equi}
The synchrotron self-absorption flux and frequency provide tight constraints on the minimum emitting region size of a synchrotron-emitting source using the equipartition method \citep{BarniolDuran2013}. An equipartition radius can be estimated by taking advantage of the fact that the total energy of the system is minimised at some radius when the  electrons and magnetic field are roughly at equipartition. The equipartition method has been extensively used to interpret radio observations of TDEs in order to constrain physical outflow properties such as the radius, energy, magnetic field strength, velocity, and ambient density \citep[e.g.][]{vanVelzen16,Alexander16,Horesh2021_15oi,Cendes2021_dsg,Goodwin2022,Cendes2022_hyz,Christy2024,Goodwin23,Goodwin2024_eJ2344,Goodwin25,Christy2025}. Numerous studies of radio-emitting outflows from non-relativistic TDEs have demonstrated that these outflows are consistent with sub-relativistic velocities $\beta\approx0.1$ \citep{Alexander16,Goodwin2022,Cendes2021_dsg}. Therefore, in this work we assume the Newtonian limit in the derivation of the equipartition radius, using the following equation from \citep{BarniolDuran2013} to estimate the equipartition radius for each observed spectrum  
\begin{equation}\label{eq:R_eq}
\begin{aligned}
\begin{split}
    R_{\mathrm{eq}} = 1\times10^{17} (21.8 (525^{(p-1)})^{\frac{1}{13+2p}}
    \chi_{\rm e}^{\frac{2-p}{13+2p}}
    F_{\mathrm{\nu,a}}^{\frac{6+p}{13+2p}} \left(\frac{d}{10^{28}\,\rm{cm}}\right)^{\frac{2(p+6)}{13+2p}}\\ \left(\frac{\nu_{\mathrm{a}}}{10\,\rm{GHz}}\right)^{-1}
    (1+z)^{-\frac{19+3p}{13+2p}}
    f_{\mathrm{A}}^{-\frac{5+p}{13+2p}} f_{\mathrm{V}}^{-\frac{1}{13+2p}} 4^{\frac{1}{13+2p}} \xi^{\frac{1}{13+2p}}\quad\rm{cm}. 
\end{split}
\end{aligned}
\end{equation}
where $d$ is the luminosity distance from the observer, $z$ is the redshift, $\chi_{\rm e} = \left(\frac{p-2}{p-1}\right) \epsilon_{\rm e} \frac{m_{\rm p}}{m_{\rm e}}$ ($m_{\rm e}$ is the electron mass and $m_{\rm p}$ is the proton mass), and $\xi = 1 + \frac{1}{\epsilon_{\rm e}}$. We assume the fraction of the total energy in the electrons is 10$\%$  ($\epsilon_e=0.1$, \citep[e.g.][]{Alexander16}).
This equipartition radius is appropriate for a source in which $\nu_m<\nu_a$, as assumed in our spectral fitting (and consistent with the spectral shape of most sources). The factor $4^{\frac{1}{13+2p}}$ arises due to a correction to the isotropic number of radiating electrons in the non-relativistic limit.
The two geometric factors $f_{\rm A}, f_{\rm v}$ are given by $f_{\mathrm{A}}=A/(\pi R^2/\Gamma^2)$ and $f_{\mathrm{v}} = V/(\pi R^3/\Gamma^4)$, for an outflow with area, $A$, volume, $V$, and distance from the origin of the outflow, $R$. All reported values in this work assume the emitting region is approximately spherical and the emitting region is a shell of thickness 0.1$R_{eq}$ ($f_{\mathrm{A}}=1$ and $f_{\mathrm{V}}=\frac{4}{3}(1-0.9^3)=0.36$), but we additionally ran our analysis assuming the emitting region is conical with ($f_{\mathrm{A}}=0.13$ and $f_{\mathrm{V}}=1.15$), corresponding to a mildly collimated outflow with a half-opening angle of 30\,degrees. We choose to assume a spherical outflow as it represents the simplest assumptions in the equipartition analysis, and to date there has been little evidence for significant beaming in radio observations of non-relativistic TDEs.  

The radius is related to the outflow velocity, $v$, where $\beta = v/c$, via
\begin{equation}\label{eq:vej}
    \delta t = \frac{R (1-\beta)(1+z)}{\beta c},
\end{equation}
where $\delta t = t_{\rm launch} - t_{\rm obs}$ is time since outflow launch \citep{BarniolDuran2013}. 

In order to constrain the launch time of each of outflow, we use Equation \ref{eq:vej} to fit the equipartition radius observed over time, leaving $\beta$ and $t_{\rm launch}$ as free parameters. We fit the radii observed for each TDE using \texttt{emcee}, again implementing a Gaussian likelihood function. We use flat prior distributions, assuming allowed ranges of $10^{-3} < \beta < 1$ and $-50 < t_{\rm launch} < 2500$\, d. We run each chain for 5000 steps with 32 walkers, discarding the first 3000 steps for burnin. Occasionally, the radio spectral coverage for individual epochs was insufficient to constrain the peak of the synchrotron spectrum, resulting in a lower limit on the radius obtained. We exclude these points from the radius fitting, which are indicated by open triangles in Figures \ref{fig:models_1}--\ref{fig:models_9}. 

In order to test the dependence  of the outflow launch date on the assumed geometry factors $f_A, f_V$ and assumptions about the shock microphysics encoded in $\epsilon_e$, we also ran the radius fitting procedure for a conical geometry, and with $\epsilon_e$ reduced to $5\times10^{-3}$. Whilst changing these nuisance parameters scales the absolute values of the equipartition radius at each epoch, we found that all inferred launch dates were entirely consistent with the spherical geometry case. The lack of dependence of the outflow launch date to changes in the geometry and relative equipartition fractions can also be seen by noting the power-law dependence of radius on these parameters in Equation \ref{eq:R_eq}. While the function $R_{\rm eq}(t)$ can be rescaled, e.g., as $R_{\rm eq}(t) \to R_{\rm eq}(t) \, f_A^{-(5+p)/(13+2p)}$, we are searching for the zero of this function. A simple rescaling of a function of course cannot change the location of its zero. As a final test of the inferred outflow launch dates, we compared all inferred launch dates with values available in the literature from independent analyses which in some cases employ slightly different spectral shape and equipartition assumptions. For all sources where a literature inferred launch date was available (18/20 TDEs), our inferred launch dates agree within error. 

We therefore emphasise that the outflow launch dates inferred for this analysis are robust, and are not sensitive to changes in the assumed outflow geometry or relative equipartition fraction. 

All sources in our sample have at least one radius measurement (by definition of the sample selection). However, without assuming a velocity, a single radius measurement does not provide a strong constraint on the outflow launch time. Therefore, for the disk model comparisons outlined in subsequent sections, we only include sources in which the outflow launch date was able to be robustly constrained with two or more radii measurements (see Table \ref{tab:constrained_tdes}). The inferred radio outflow launch dates for these constrained flares are listed in Table \ref{tab:launchtimes}.

\begin{table}
    \centering
    \begin{tabular}{p{85pt} p{40pt} p{60pt} p{40pt} p{60pt} }
    Tidal disruption event & $t_0$ (MJD) & $t_{\rm{launch}} - t_0$ (d) & $\dot{m}_{\rm{launch}}$ & $t_{0.02\rm{Edd}}-t_0$ (d)  \\
    \hline
    Prompt flares  & & & \\
    \hline
    ASASSN-14li &56989 & $-44^{+8}_{-4}$& $1.3^{+0.3}_{-0.2}$ & $5200^{+1300}_{-1200}$ \\
    ASASSN-15oi &57248 &  $9^{+63}_{-41}$ & $0.4^{+0.3}_{-0.1}$ & - \\
    AT2019azh  &58566 & $-35^{+16}_{-10}$& $3.9^{+4.6}_{-2.0}$ & $1510^{+250}_{-280}$\\
    AT2019dsg  &58604 & $-37^{+15}_{-10}$& $4.2^{+8.0}_{-2.5}$ & $690^{+180}_{-180}$\\
    eJ2344  & 59159& $-15^{+35}_{-24}$& $1.3^{+1.1}_{-0.6}$ & $1310_{-290}^{+280}$\\
    \hline
    Delayed flares  & & & \\
    \hline    
    ASASSN-14ae & 56685& $1561^{+582}_{-911}$ & $0.02^{+0.03}_{-0.01}$ & - \\
    ASASSN-15oi &57248 &  $833^{+267}_{-433}$ & $0.04^{+0.05}_{-0.02}$&  -\\
    AT2018hyz  &58429 & $421^{+254}_{-281}$& $0.04^{+0.04}_{-0.02}$ & - \\
    AT2019ahk  & 58544& $527^{+132}_{-205}$&$0.02^{+0.01}_{-0.01}$ & -  \\
    AT2020vwl  &59170 & $535^{+60}_{-94}$&$0.02^{+0.02}_{-0.01}$ & -\\
    AT2020zso  &59192 & $720^{+110}_{-194}$& $0.05^{+0.07}_{-0.04}$ &- \\
    \hline
    \end{tabular}
    \caption{The modelled radio outflow launch times ($t_{\rm{launch}}$; measured from optical peak, $t_0$) and the modelled accretion rate at this time ($\dot{m}_{\rm{launch}}$) for each of the flares where these values could be robustly constrained. We also include the time that the modelled accretion rate passes through 0.02 Eddington, ($t_{0.02\rm{Edd}}$), allowing predictions for sources that are yet to show late-time flares.}
    \label{tab:launchtimes}
\end{table}

\subsubsection{Lightcurve modelling}\label{sec:radio_lcs}

The evolution of the radio lightcurves of TDEs is primarily driven by the total energy deposited in the outflow and the density of the ambient medium that the shock propagates through \citep{Alexander2020}. In this section we fit a broken power-law lightcurve model to the 5\,GHz radio lightcurves in our sample in order to compare the lightcurve rise indices, and therefore probe differences in the total shock energy for prompt and delayed outflows.

Following the \citet{Chevalier1998} prescription for modelling the time evolution of a self-absorbed synchrotron source, we assume the observed flux density at frequency $\nu$ is described by
\begin{equation}\label{eq:lcfit}
    F_{\nu} = 1.582 F_{\nu,tc} \left(\frac{t}{t_c}\right)^a \left[1 - \exp\left(-\left(\frac{t}{t_c}\right)^{-(a+b)}\right)\right]
\end{equation}
where $F_{\nu}$ is the flux density at frequency $\nu$, $F_{\nu,tc}$ is the flux density at the break time, $t_c$ and $a$ and $b$ the powerlaw slope either side of the break time. 

We select 5\,GHz to model as the majority of TDEs in our sample had 5\,GHz observations in most epochs, and the lightcurve evolution at different frequencies is also affected by the location of the self-absorption break, so it is important to model the lightcurves at the same observing frequency to compare the lightcurve evolution between TDEs. Due to paucity of data during the rising portion of the radio lightcurves, only 9 TDEs in our sample have sufficient data to constrain the powerlaw index of the rise at the selected frequency of 5\,GHz. Four of these TDEs have inferred launch times consistent with a delayed outflow, three of which also showed a prompt radio flare. In order to model the TDEs with two radio flares, we assume two broken powerlaws such that 
\begin{equation}\label{eq:2broken_plaw}
    F_{\nu,\rm{total}} = F_{\nu,1} + F_{\nu,2}
\end{equation}
where $F_{\nu,1}$ and  $F_{\nu,2}$ (and associated $t_{c,1}$, $t_{c,2}$, $a_1$, $a_2$, etc.) are each given by Eq. \ref{eq:lcfit}. 

We fit each of the 9 TDE lightcurves using \texttt{emcee}, implementing a Gaussian likelihood function and flat log prior distributions. For each flare, we assume the allowed parameter ranges of $10<F_{\nu,tc} (\mu\rm{Jy}) < 10^{3}$ , $0 < a < 10$, $0 < b < 10$, $1 < t_c < 500$, and in the case of two flares, we additionally assume $t_c < t_{c2}  < 2000 (\rm{d})$ keeping the prior ranges of $F_{\nu,tc2}$, $a_2$, and $b_2$ the same as for the first flare. The prior on the flux amplitude is flat in log space, with all other priors flat in linear space. We notably increase the prior range of the peak flux density to  $10<F_{\nu,tc} (\mu\rm{Jy}) < 2F_{\nu,obs}$ for ASASSN-15oi and AT2018hyz due to the higher observed flux densities of these sources.  

\begin{figure}
    \centering
    \includegraphics[width=0.32\linewidth]{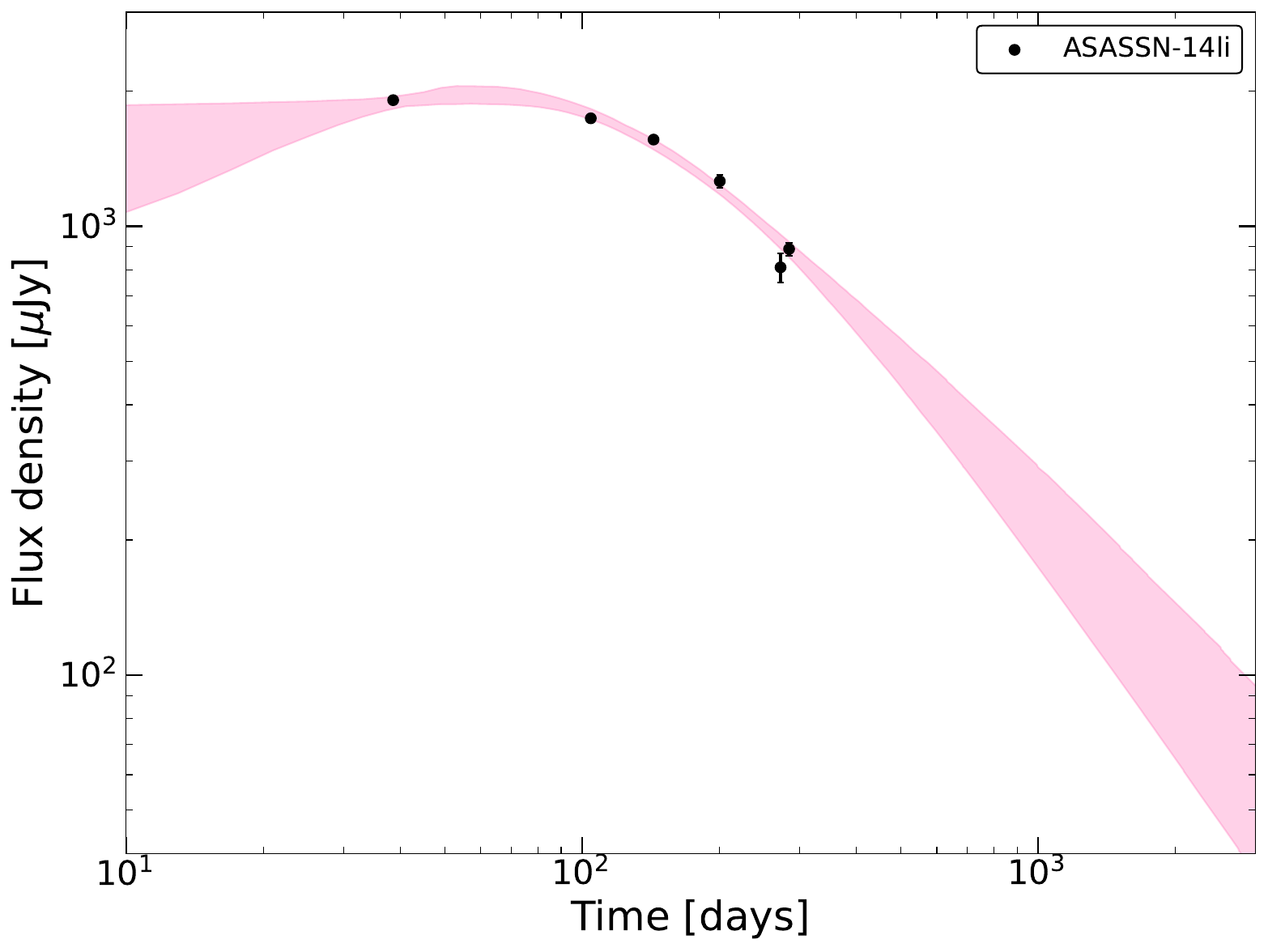}
    \includegraphics[width=0.32\linewidth]{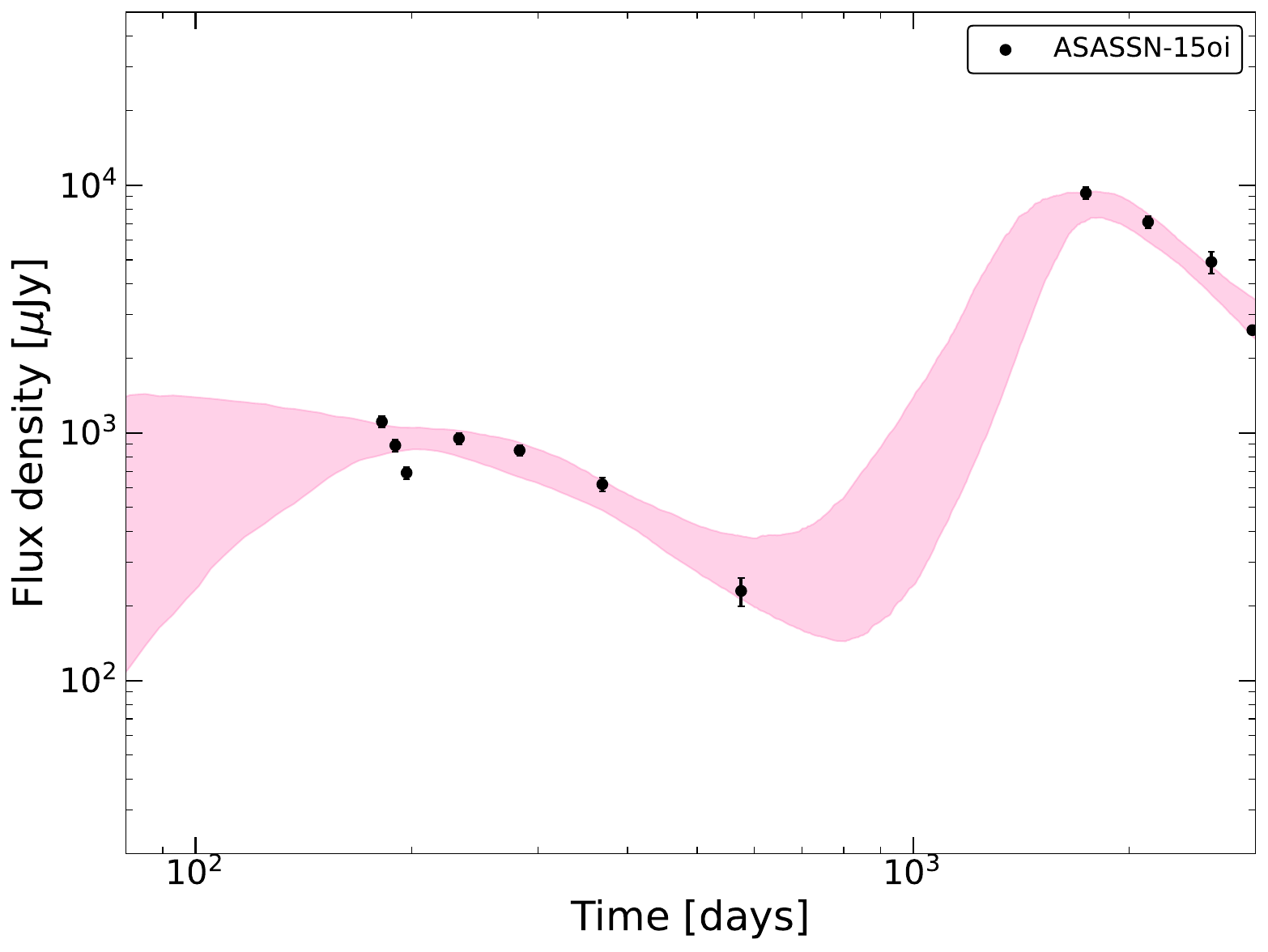}
    \includegraphics[width=0.33\linewidth]{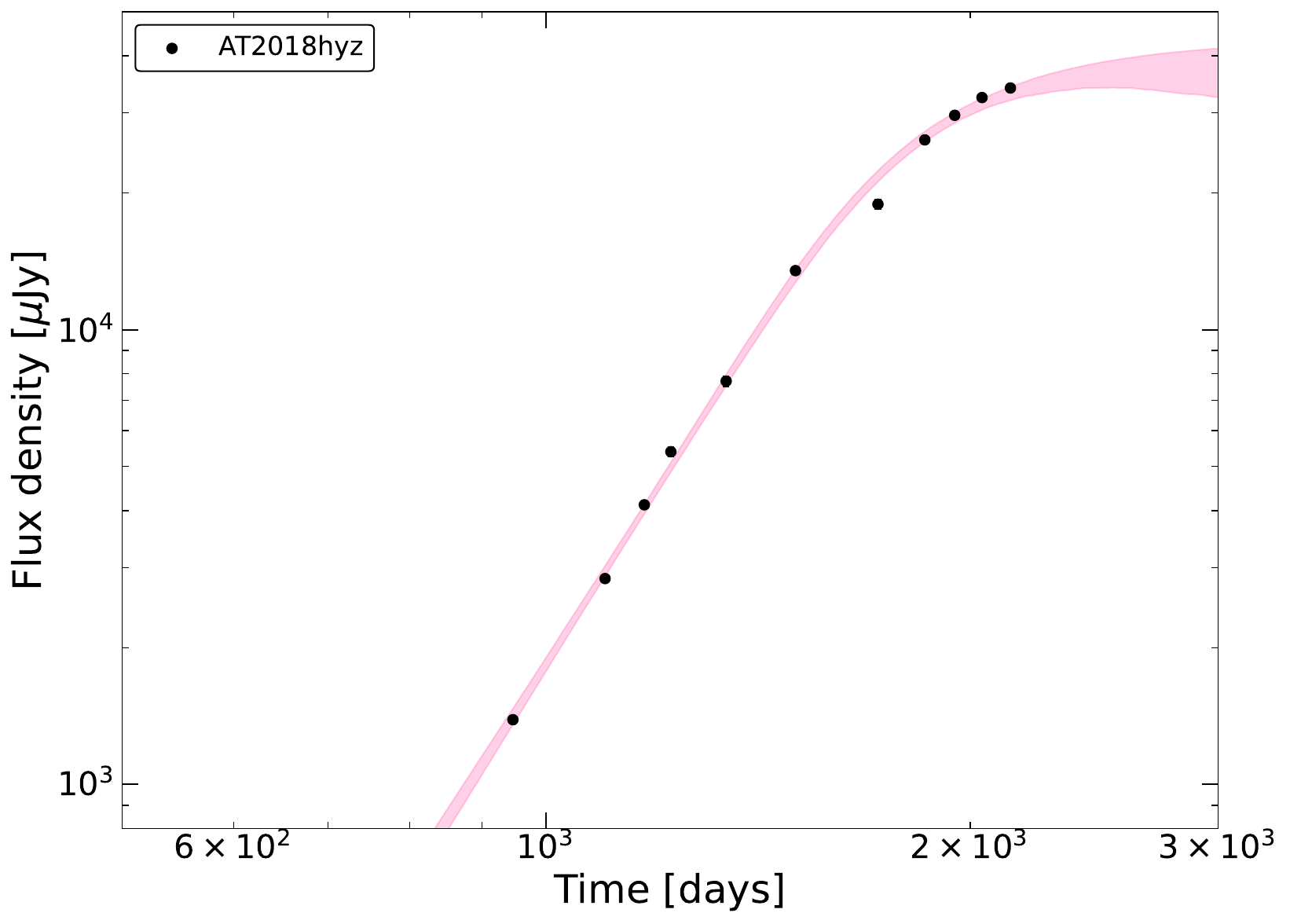}
    \includegraphics[width=0.32\linewidth]{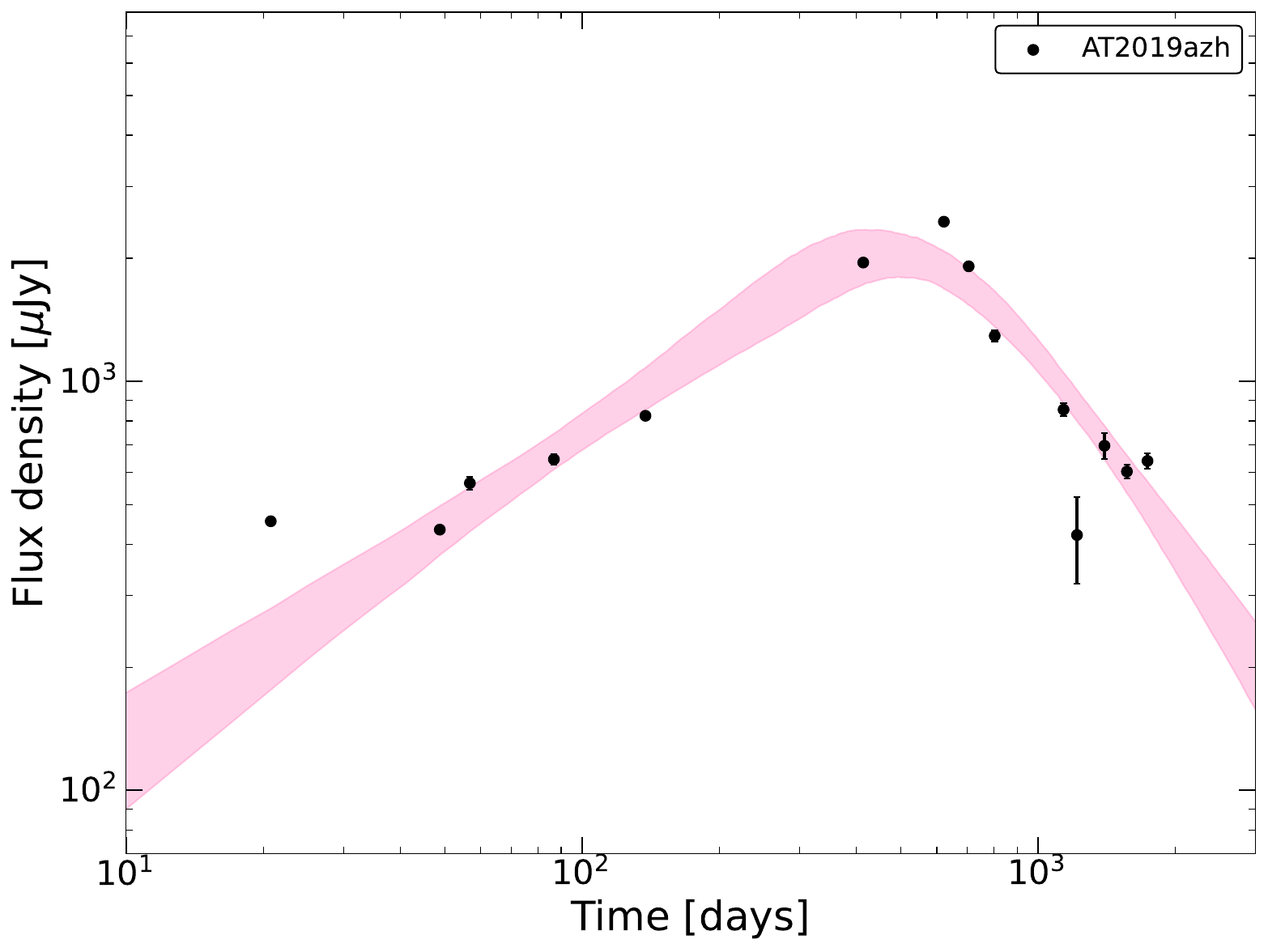}
    \includegraphics[width=0.32\linewidth]{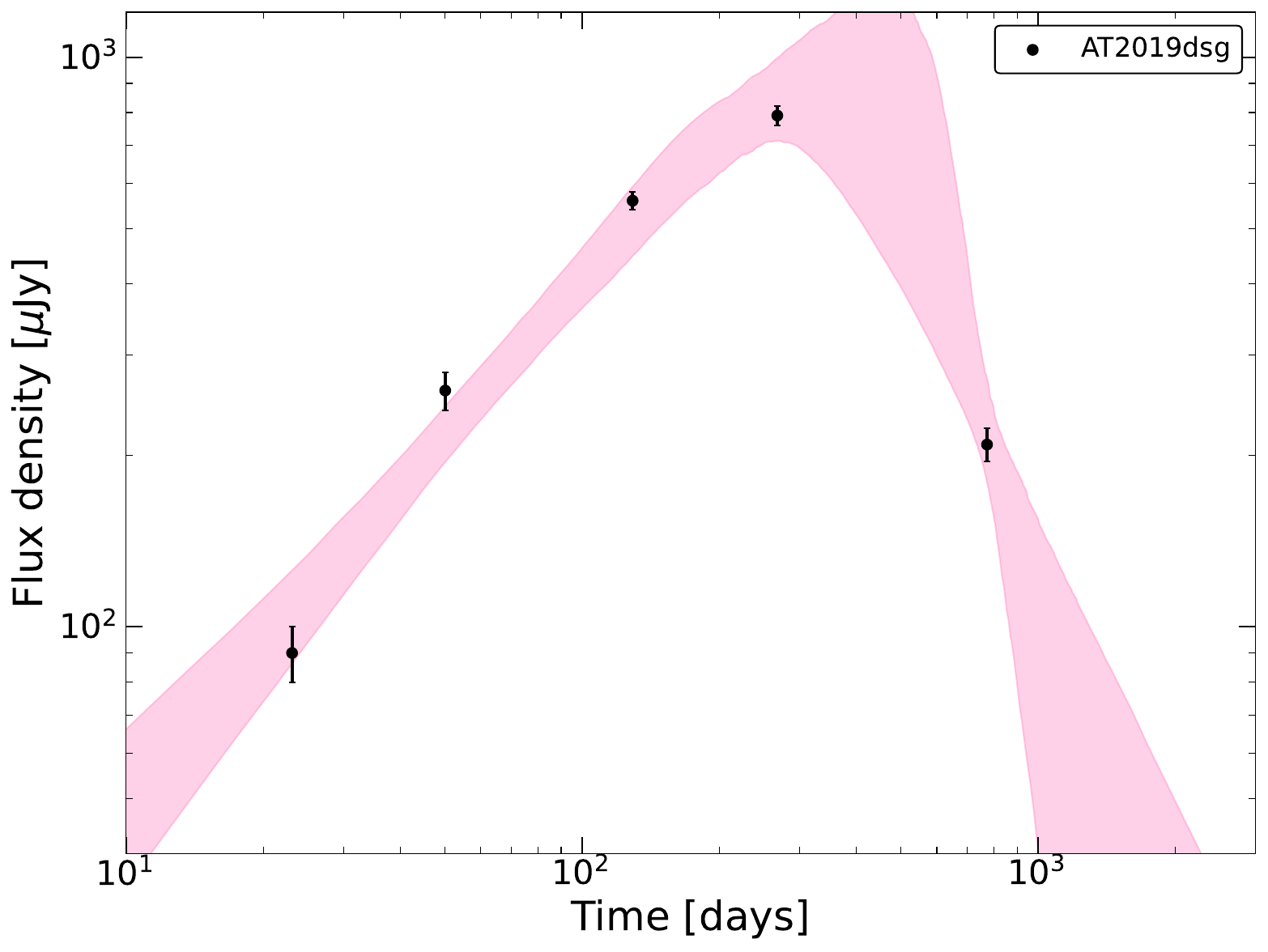}
    \includegraphics[width=0.32\linewidth]{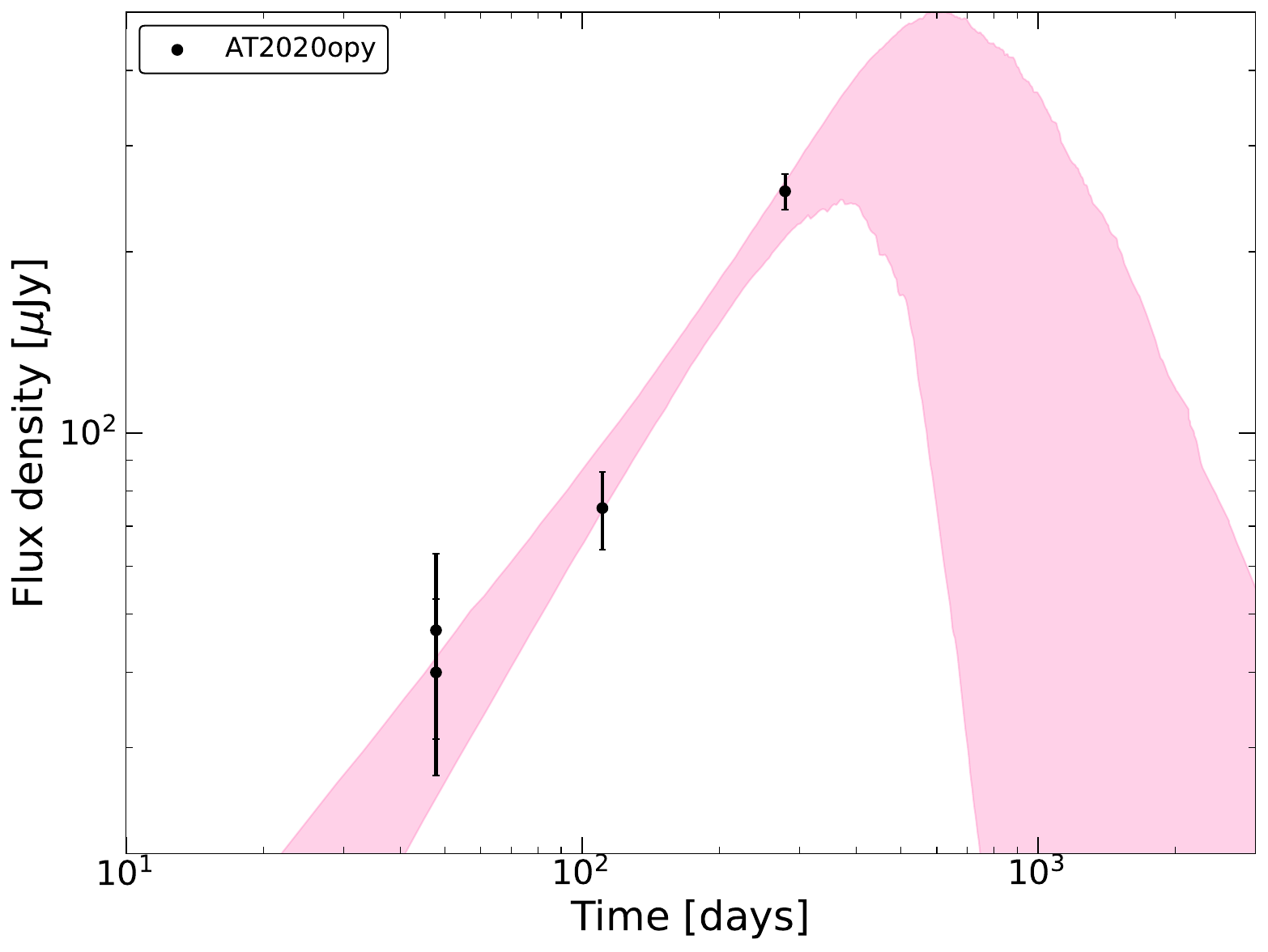}
    \includegraphics[width=0.32\linewidth]{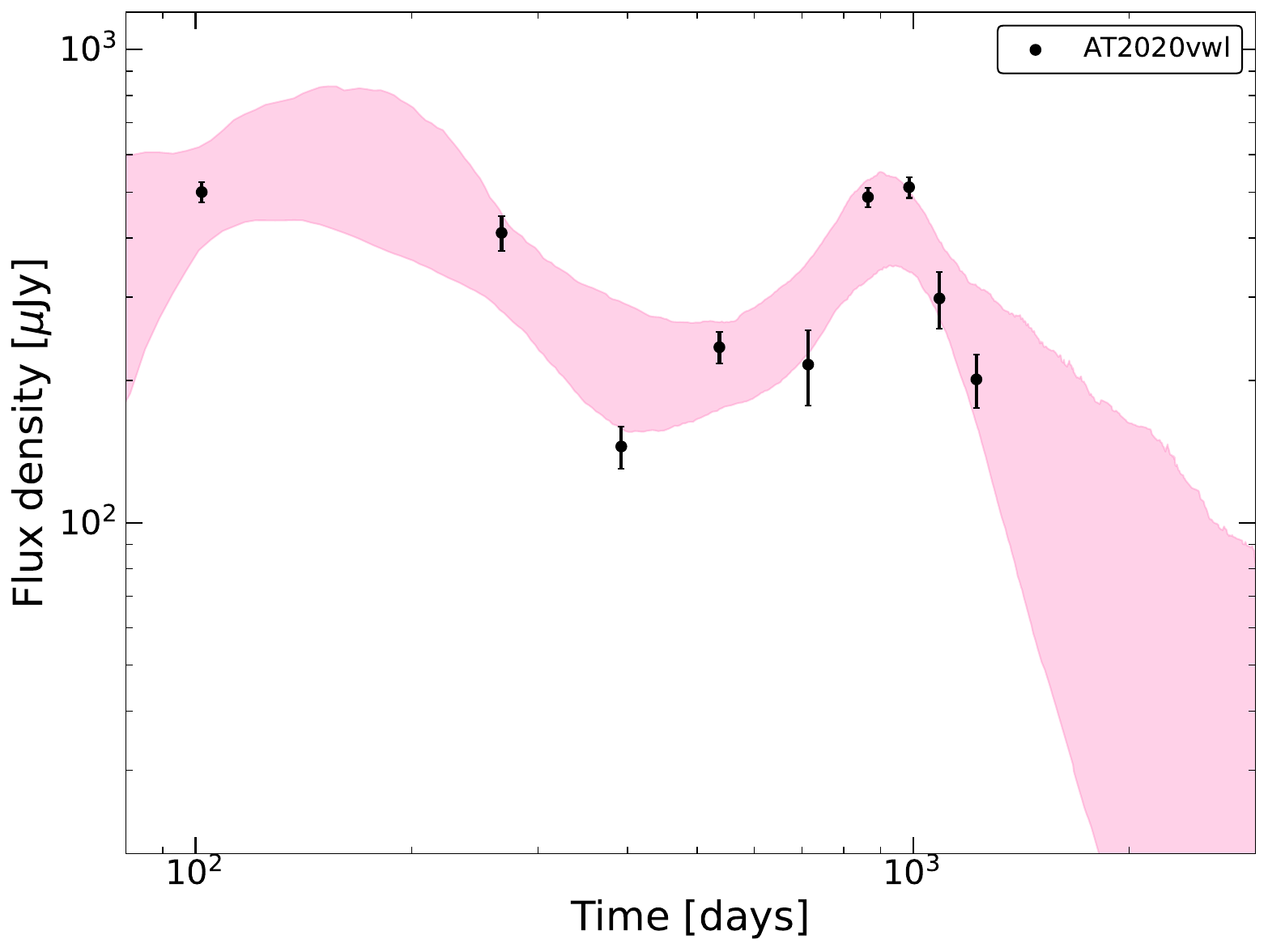}
    \includegraphics[width=0.32\linewidth]{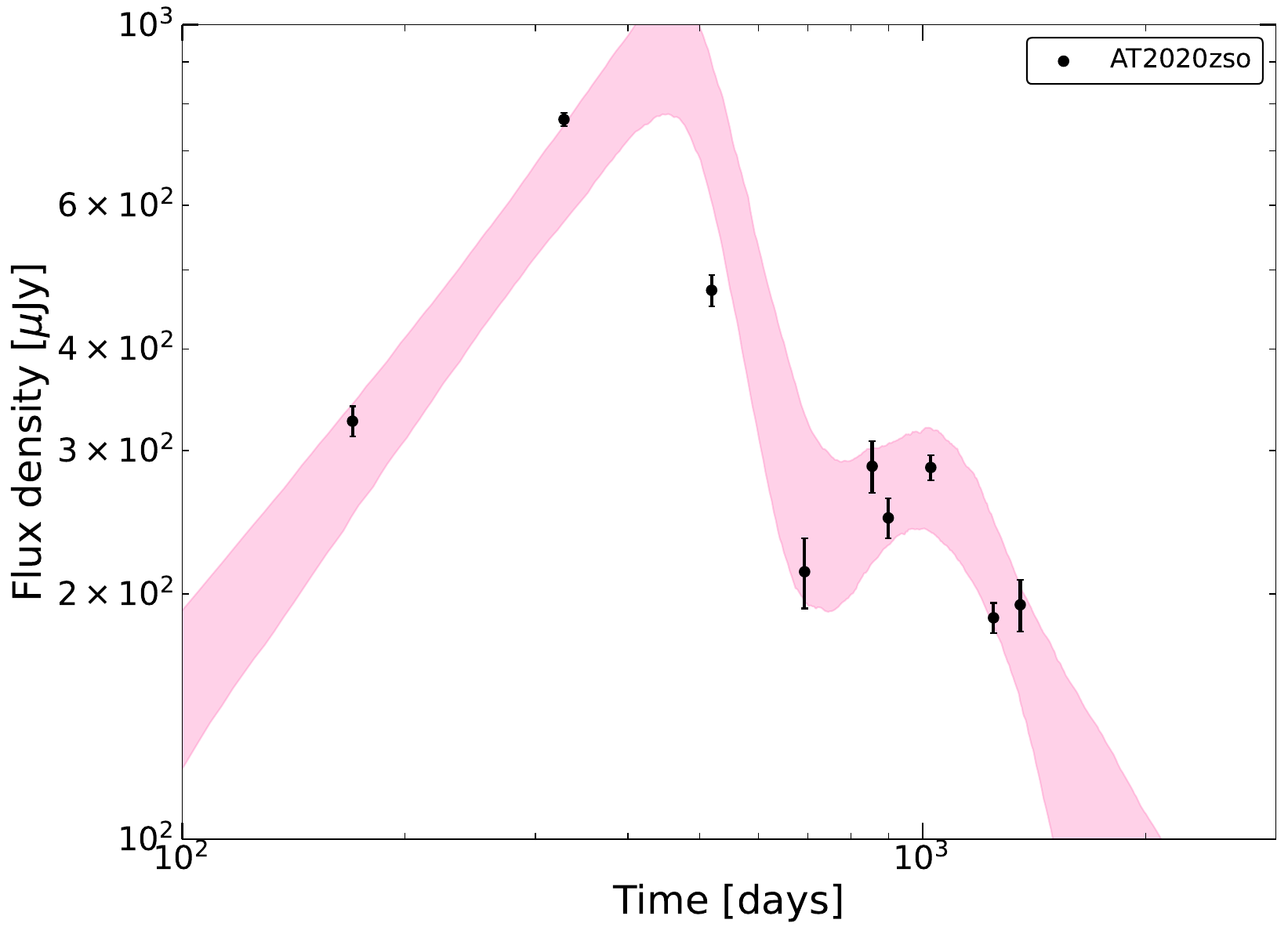}
    \includegraphics[width=0.32\linewidth]{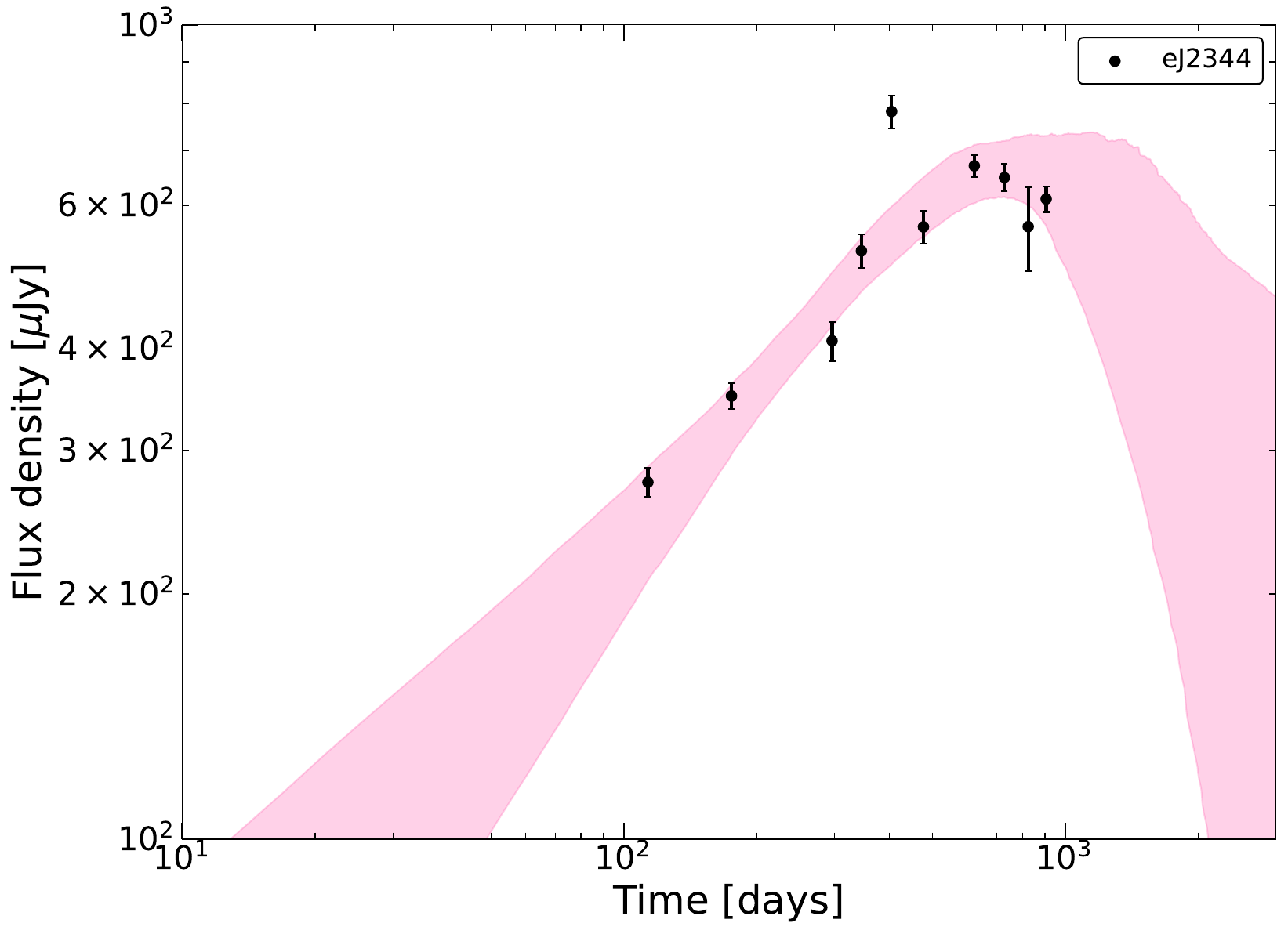}
    \caption{The 5-6\,GHz observed radio lightcurves (black points) and the corresponding broken power-law model fits (pink) for each of the TDEs in which the rise indices were able to be constrained. The shaded regions indicate $1\sigma$ confidence intervals.}
    \label{fig:radio_lcs}
\end{figure}

\begin{figure}
    \centering
    \includegraphics[width=0.45\linewidth]{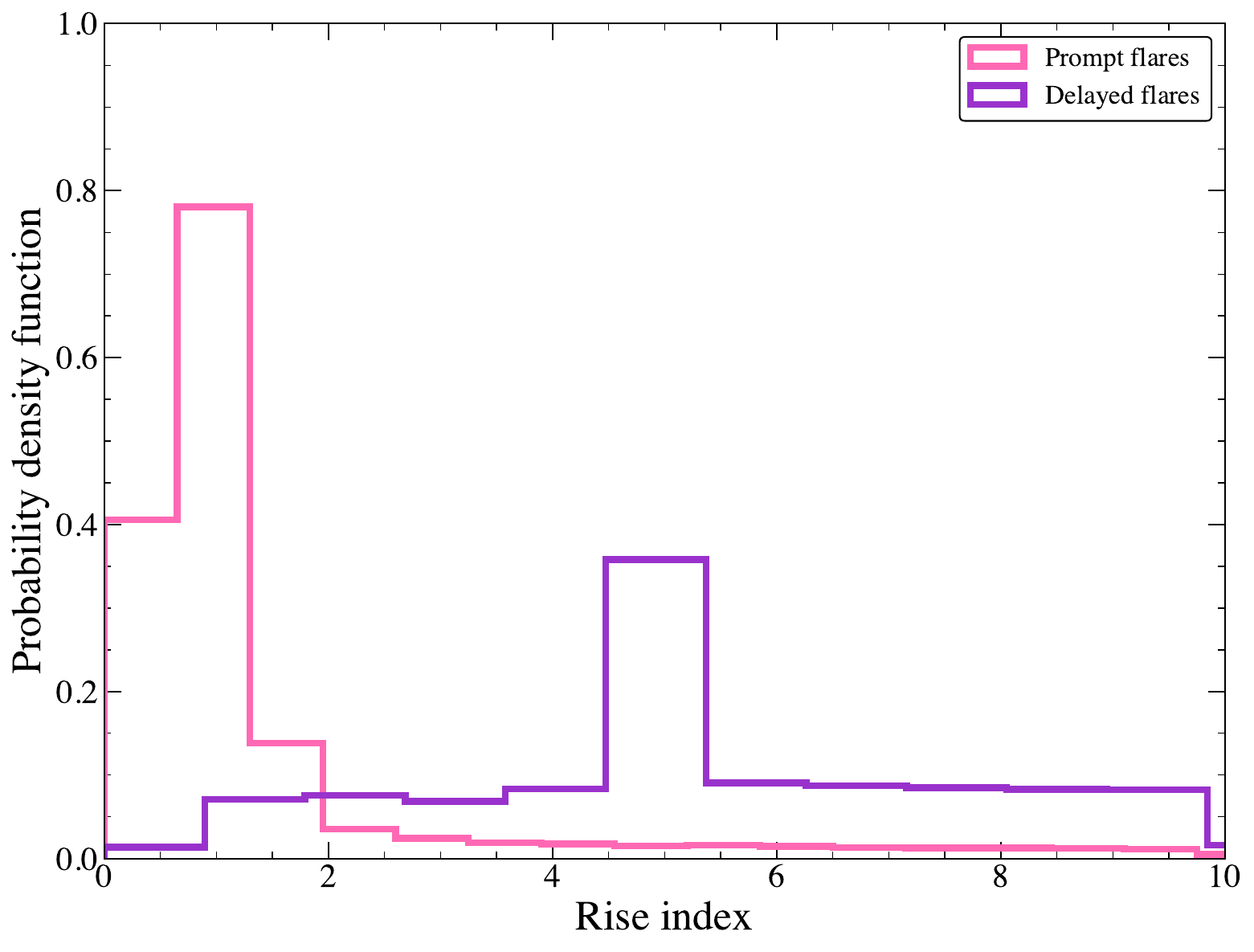}\includegraphics[width=0.45\linewidth]{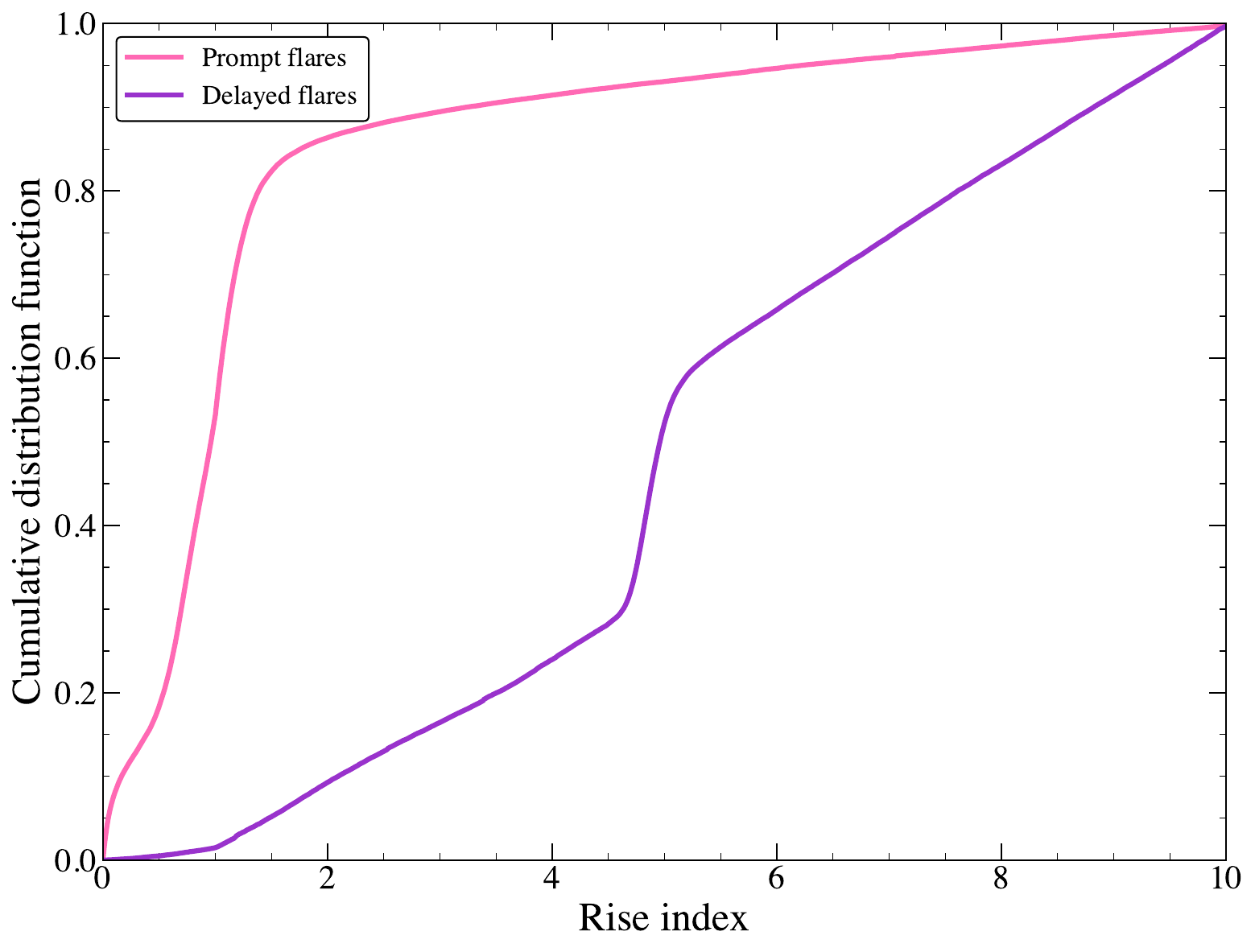}
    \caption{The distribution of power-law rise indices for the ``prompt" (pink) and ``delayed" (purple) flare radio lightcurves modelled in Figure \ref{fig:radio_lcs}. On the left we show the two probability density functions, while on the right we show the cumulative distribution functions, highlighting the clearly  distinct nature of the two distributions of rise index. This is strongly suggestive of two different physical mechanisms powering the two temporally distinct types of flares.  }
    \label{fig:rise_index}
\end{figure}

Of the 20 TDEs in our sample, at 5-6\,GHz, the rise indices for ASASSN-14ae, AT2019ahk, AT2018dyb, AT2018hco, AT2018zr, AT2019ehz, AT2019eve, AT2019qiz, and AT2021sdu were completely unconstrained by the model (i.e. the model returned the prior), so we excluded these sources from the lightcurve analysis. The lightcurve fits for the remaining 9 TDEs in which the rise was able to be constrained (to varying degrees of uncertainty) are plotted in Figure \ref{fig:radio_lcs}.
Interestingly, we find a distinct difference in the power-law index rise times of the prompt outflows compared to the delayed outflows. The prompt outflows rise slowly (with power-law indices in the range 0.5--2), whereas the delayed outflows rise significantly more quickly (with power-law indices $>4$). The bi-modal distribution of power-law indices can be seen in Figure \ref{fig:rise_index}, where we have plotted the combined posterior on the rise index for the flares observed. Such a bi-modal distribution provides evidence that the energetics, and thus the mechanism, of the prompt and delayed outflows are different. 

\subsection{Disk modelling}\label{sec:disk}
The purpose of our analysis is to constrain the physical properties of the accretion flows in TDEs at the time at which radio outflows are inferred to be launched from the system. We do this by modelling the multi-band light curves of each TDE using the {\tt FitTeD} code \cite{mummery2024fitted}. {\tt FitTeD} solves \cite{Mummery23a} the time-dependent relativistic disk equations \cite{Balbus17} to produce a time and radius dependent disk temperature profile $T(r, t)$, from which light curves can be generated at any given observing frequency (or across broad X-ray bands). Once fit to data, the {\tt FitTeD} code returns posterior distributions on the key black hole and disk parameters describing the system, namely the black hole mass $M_\bullet$ and (dimensionless) spin $a_\bullet$, the initial disk mass $M_{\rm disk}$, radial scale $r_0$, formation time $t_0$ and viscous timescale $t_{\rm visc}$, and the disk-observer inclination angle $i$.  Once the parameters of the system are constrained, the accretion rate and bolometric luminosity of the disk can be determined (with appropriate uncertainties) at all times. 

\subsubsection{Data sets used}
In this work we take the multi-band light curves of all twenty TDEs which have radio spectral data in the literature. We include all available optical, UV and X-ray data taken at all times during the evolution of the system. Not all of this data represents an observation of direct thermal emission from an accretion disk (which is what {\tt FitTeD} models). In particular, the early-time ($t \lesssim 1$ year) optical-UV flares observed in TDEs are known not to represent direct emission from the accretion flow, and instead result from some poorly-understood initial phase \cite{Roth20}, which is likely involved in the disk formation process. However, after typically $\Delta t \approx 1$ year post-peak, all optical/UV emission from a TDE results from a spreading and cooling accretion flow \cite{vanVelzen19b, MumBalb20a}, which is well described with time-dependent relativistic accretion theory \cite{Mummery_et_al_2024}.  While it does not originate directly from the accretion flow, it is important to include the data describing this early flare  as the decay of this initial phase still imprints some (fading) emission at late times, which can subtly impact the disk parameters. All optical/UV data is corrected for host emission (i.e., it is host-subtracted) and also corrected for absorption along the line of sight. See the description in \cite{Mummery_et_al_2024} for details. 

Not all TDE X-ray emission represents direct emission from a thermal disk either, and we ensure that every data point included in our light curve fitting had an X-ray spectrum consistent with an unobscured thermal accretion flow. This represents the vast majority of the TDE X-ray data in the literature, which are near universally observed in soft accretion states (in stark contrast with AGN). Our modelling necessitates not including coronal (power-law) dominated epochs in the fitting process, which we exclude. We also do not include early-time (first $\approx 200$ days) epochs for ASASSN-15oi and AT2019azh where it is known that the X-ray emission is suppressed by obscuration \cite{Guolo24}, with a poorly constrained correction factor. We deemed this obscuration correction too poorly constrained to be reliably included. For both AT2019azh and ASASSN-15oi we have verified that the quantitative accretion rate results presented in this work are not impacted by this later choice (indeed they are consistent within the error bars), and our solutions neglecting this early time emission are generally consistent with these early X-ray detections at the $\sim1-2\sigma$ level. When any non-detections were found with X-ray instruments, we produced $3\sigma$ upper luminosity limits assuming that the spectrum was that of a $T=75$ eV blackbody, a canonical temperature measured from a wide range of TDE X-ray spectra \citep[e.g.,][]{Miller15, Brown17, Mummery_Wevers_23, Guolo24}. All X-ray luminosities are corrected for absorption along the line of sight, see \cite{Guolo24} for details. Note that since the publication of a fit to AT2019dsg in \cite{mummery2024fitted}, further X-ray and late-time UV data has become available, and so we re-run the fit. This cause slight changes in the posterior accretion rate distribution at early times compared to \cite{mummery2024fitted}. No other parameters where notably changed. 

We process the data slightly, to aid the fitting process. This involves binning the optical/UV data to reduce noise, using a 5-day bin for optical/UV emission at early times ($t \leq 1$ year, this also reduces the number of early data points, to prevent over fitting to emission not resulting from a disk), and broader (20 day) bins at later times. The emission at later times is fainter, and this larger bin size aids in reducing the scatter in the light curves during this disk-dominated phase. For X-ray data we only binned the light curves if there was an extremely large amount of data available (only relevant for ASASSN-14li, ASASSN-15oi, and AT2019azh) when we used 7-day bin sizes (a choice made for numerical fitting feasibility), otherwise we included every upper limit and detection at X-ray energies. We verified (using ASASSN-14li) that our choices of bin size did not affect the results, and that all posterior distributions were consistent when the bin sizes were either doubled or halved.

\subsubsection{Fitting procedure}
With the reduced multi-band data available, we use the inbuilt {\tt emcee} \cite{EMCEE} Monte Carlo Markov Chain fitting algorithm in {\tt FitTeD} (see \cite{mummery2024fitted} for details) to produce posterior distributions of all parameters. We model the early (non-disk) optical/UV flare with a phenomenological Gaussian rise (except for ASASSN-14ae, ASASSN-14li, ASASSN-15oi ipTF-16fnl which were never observed in the rising phase) and power-law decay. 

Specifically, these phenomenological models have the functional form  
\begin{equation}
    L_{\rm rise}(\nu, t) = L_0 \times f_{\rm rise}(t) \times \frac{\nu B(\nu, T)}{\nu_0 B(\nu_0, T)} , 
\end{equation}
and 
\begin{equation}
    L_{\rm decay}(\nu, t) = L_0 \times f_{\rm decay}(t) \times \frac{\nu B(\nu, T)}{\nu_0 B(\nu_0, T)} , 
\end{equation}
where $B(\nu, T)$ is the Planck function, and $\nu_0 = 6 \times 10^{14}$ Hz is a reference frequency (approximately equal to the ZTF $g$-band). The amplitude $L_0$ and temperature $T$ are common to both the rise and decay models. The functions $f_{\rm rise}(t)$ and $f_{\rm decay}(t)$ are defined so as to have a maximum amplitude of unity, so that $L_0$ remains the physical peak amplitude. The rise model is that of a Gaussian rise
\begin{equation}\label{gauss_rise}
    f_{\rm rise}(t) = \exp\left( - {\left(t - t_{\rm peak}\right)^2 \over 2\sigma_{\rm rise}^2}\right) ,
\end{equation}
with fitting parameters $t_{\rm peak}$ and $\sigma_{\rm rise}$ (both with units of days). The decay is modelled with a power-law 
\begin{equation}\label{pl_decay}
    f_{\rm decay}(t) = \left(  {t - t_{\rm peak} + t_{\rm fb} \over t_{\rm fb}}\right)^{-p} ,
\end{equation}
with fitting parameters $t_{\rm peak}$, $t_{\rm fb}$ and $p$. The parameters $t_{\rm peak}$ and $t_{\rm fb}$ have  units of days, while the index $p$ is dimensionless. The parameter $t_{\rm peak}$ is by default assumed to be common between the rise and decay models, and if no rise model is specified the parameter is fixed to $t_{\rm peak} = 0$.  

We also compute the evolving accretion disk spectral luminosity $\nu L_\nu(\nu, t)$, and the total model luminosity is simply the addition of the two components. The early component has no impact on any X-ray emission from the source, as the inferred temperatures are all too cool ($T< 10^5$ Kelvin) to produce thermal X-rays. 

For each set of parameters $\left\{ \Theta \right\} = [M_\bullet, a_\bullet, M_{\rm disk}, ...]$, we compute the total log-likelihood assuming each measurement is chi-squared distributed 
\begin{equation}
    \ln {\cal L}(\Theta) = - {1 \over 2}\sum_{{\rm bands}, \, i} \sum_{{\rm data, \, j}} \frac{\left(O_{i, j} - M_{i, j}\right)^2  }{E_{i, j}^2} , 
\end{equation}
where $O_{i, j}$, $M_{i, j}$ and $E_{i, j}$ are the observed luminosity, model luminosity and luminosity uncertainty of the $j^{\rm th}$ data point in the $i^{\rm th}$ band respectively. The following log-likelihood is used for upper limits 
\begin{equation}
    \ln {\cal L}_{U}(\Theta) = - {1 \over 2}\sum_{{\rm bands}, \, i} \sum_{\substack{{\rm limits, \, j, } \\ {\, M_{i, j}>U_{i, j}}}} \frac{\left(U_{i, j} - M_{i, j}\right)^2  }{(U_{i, j}/N_{\sigma, i, j})^2} , 
\end{equation}
where $U_{i, j}$ is the upper limit luminosity, with statistical significance of $N_{\sigma, i, j}$ standard deviations (always 3 in this work). The model luminosity is again $M_{i, j}$, and only those model values with luminosities greater than the upper limit contribute to the likelihood. 

We include prior bounds on the parameters of the model with the following simple function 
\begin{equation}
    \ln {\cal P}(\Theta) = \sum_{{\rm parameters}, k} \ln \, \left[{\cal U}({\rm low}_k, {\rm high}_k, k) \right],
\end{equation}
where ${\cal U}(a, b, x)$ is the uniform distribution, which is equal to 1 if $x$ lies between ``$a$'' and ``$b$'',  and  equal to zero elsewhere. We use the default prior bounds $\{{\rm low}_k, {\rm high}_k\}$ for each parameter as summarised in \cite{mummery2024fitted}. This means we do not impose any prior knowledge of (e.g.,) the black hole masses in these sources from galaxy scaling relationships, allowing a population-level test of the reliability of the parameter inference. 

We then maximise, numerically, the total (log) probability
\begin{equation}
    \ln {\rm Prob}(\Theta) = \ln {\cal L}(\theta) +  \ln {\cal L}_U(\theta) + \ln {\cal P}(\theta) .
\end{equation}

\subsubsection{Fits to individual sources}
In this sub-section we provide an overview of the fitting parameters inferred in this analysis, and describe which sources have well constrained disk properties at which evolutionary stages. 

The {\tt FitTeD} code works best when there are observations available which span a broad range of observing frequencies, with X-ray emission probing the very innermost (hottest) disk regions and optical/UV emission generally probing the outermost (cooler) disk regions. Five sources, namely ASASSN-14li, ASASSN-15oi, AT2019dsg, AT2019azh and eJ2344 satisfied both observational constraints, and represent the posterior disk solutions which are robust at all evolutionary stages. The remaining 15 sources either had only one or two X-ray detections at early times (insufficient to properly constrain the disk over the first $\sim$ year), or were undetected at X-ray frequencies throughout their evolution. For these poorly X-ray sampled TDEs we are reliant on late time optical/UV emission to constrain the disk on timescales typically exceeding one year into the TDEs evolution. 

The vast majority of the sources (17/20) in this work show long-lived optical/UV emission far exceeding both the host emission and that which would be inferred from extrapolating the early-time light curve to these late times. This is entirely in keeping with previous studies of long-lived accretion disk emission in TDEs \cite[e.g.,][]{vanVelzen19, MumBalb20a, Mummery_et_al_2024, MummeryVV25}. The sources AT2019ehz, AT2019eve and AT2019teq  had insufficient late time detections of disk emission to reliably inform disk parameters (which are very hard to constrain, even in the presence of X-ray emission, without robust optical/UV data at late times). The {\tt FitTeD} code was unable to constrain the disk and black hole parameters in each of these cases, effectively just returning the priors, and we removed them from our sample entirely. 

For the remaining 12 sources (without prompt X-ray emission) the disk is well constrained on timescales exceeding the first year of evolution, owing to the long lived and well-sampled optical/UV emission sourced from the outer disk. Some sources also had X-ray upper limits at these late times, which aides in constraining the inner disk during these same epochs, which generally acts to shrink the uncertainty on the accretion rate/bolometric luminosity at these stages. For these 12 sources the early evolutionary stages of the disk were not well constrained, as this is only possible when well-sampled X-ray light curves are available. A good proxy for whether or not the early disk behaviour was constrained was the posterior distribution of the disk formation radius $r_0$. For the five sources with well sampled X-ray emission at early times, $r_0$ typically converged to a reasonably precise value well within the bounds of the imposed prior. For those sources only with constraints from late-time optical/UV emission and X-ray upper limits we generally found that $r_0$ was very poorly constrained, returning a posterior representing a sizeable fraction of the uniform prior. In these cases we do not deem the disk to be well constrained prior to the late evolutionary stage. 

So as to be quantitive about when the disk was deemed well constrained, for these 12 sources we only analyse epochs at which the disk was dominating the flux (i.e., $L_{\rm disk}>L_{\rm decay}$) across all optical and UV bands. For each TDE in our sample this was at some point within the window $\Delta t = (130, 475)$ days post peak. 

When we cross reference the seventeen TDEs with disks well constrained at early (5/17) and late (17/17) times with those TDE systems with well constrained radio emission at those same stages we are left with 10 sources (and 11 distinct radio flares). This means that we lose 7 sources which have a disk which is well constrained at late times, but have insufficient radio coverage to determine properties of any radio flares. 

\begin{figure}
    \centering
    \includegraphics[width=0.45\linewidth]{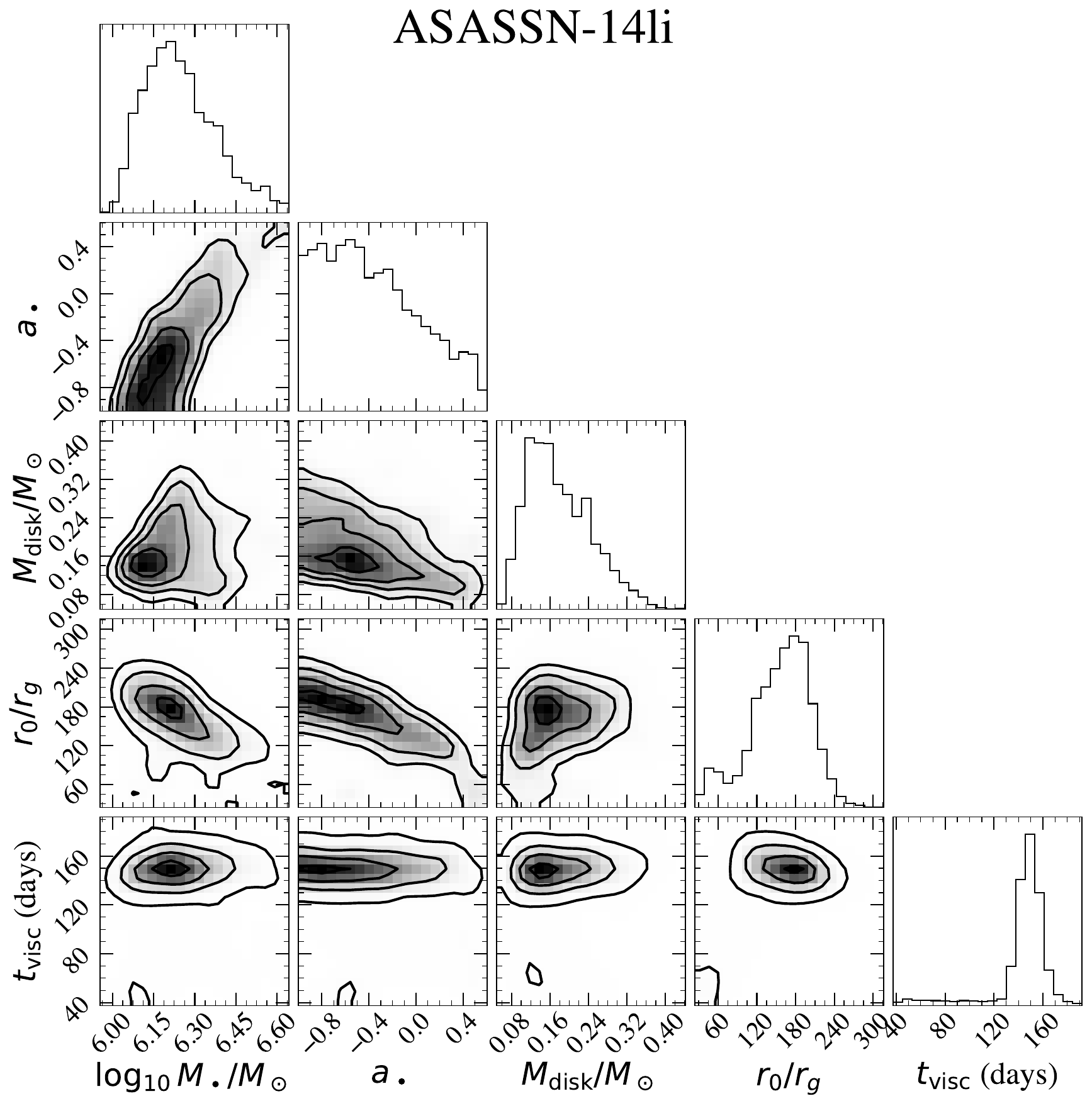}
    \includegraphics[width=0.45\linewidth]{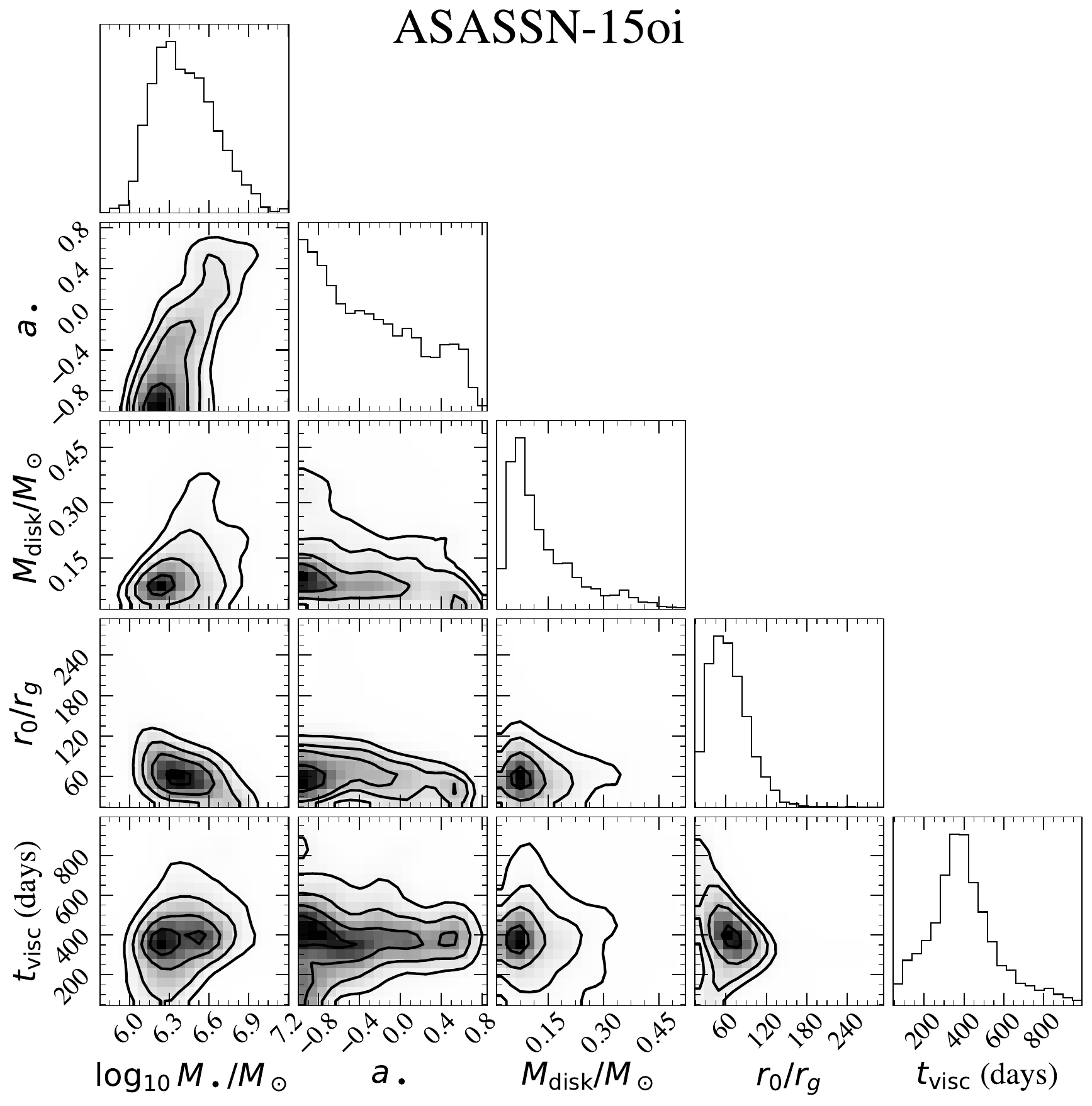}
    \includegraphics[width=0.45\linewidth]{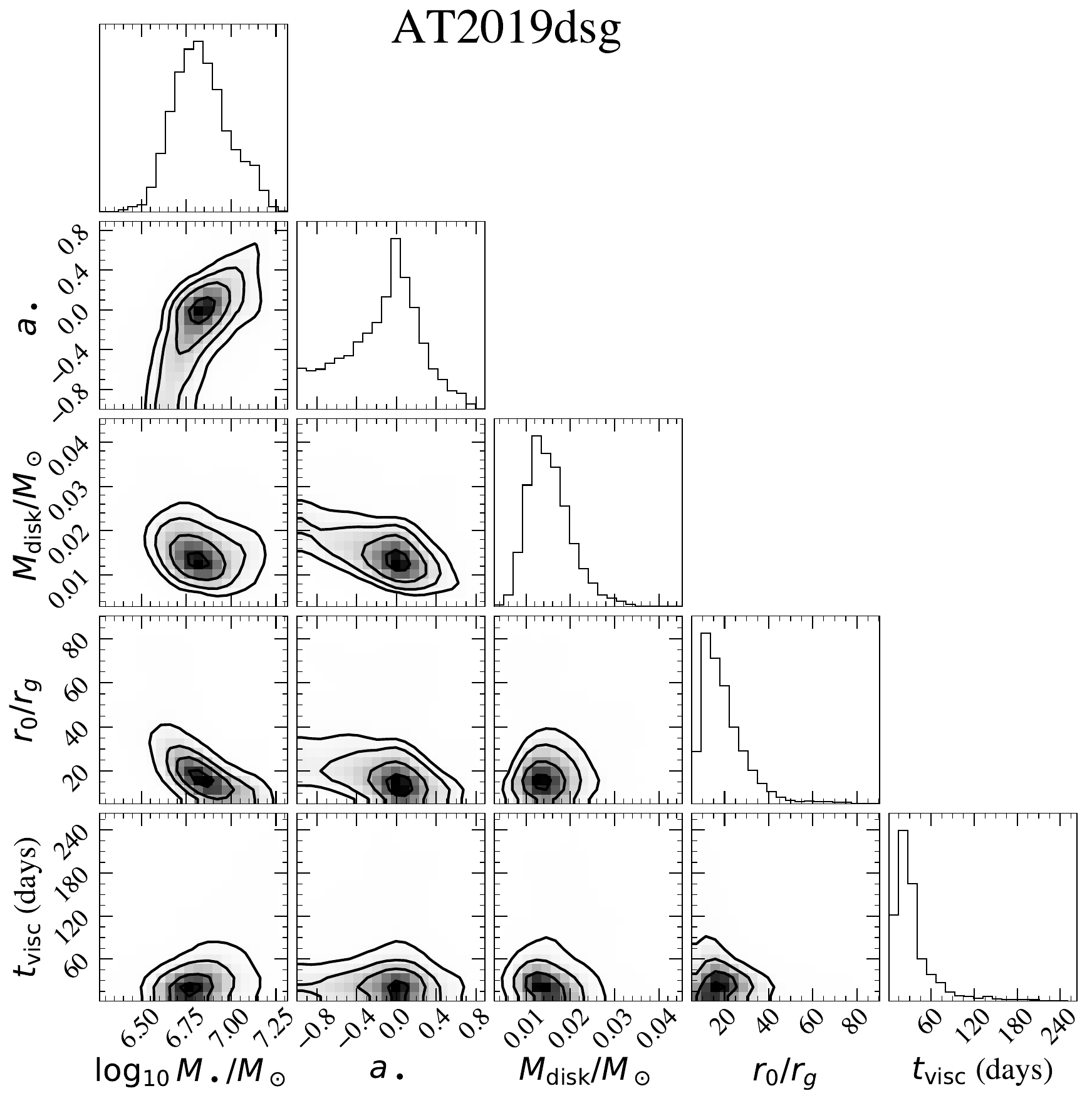}
    \includegraphics[width=0.45\linewidth]{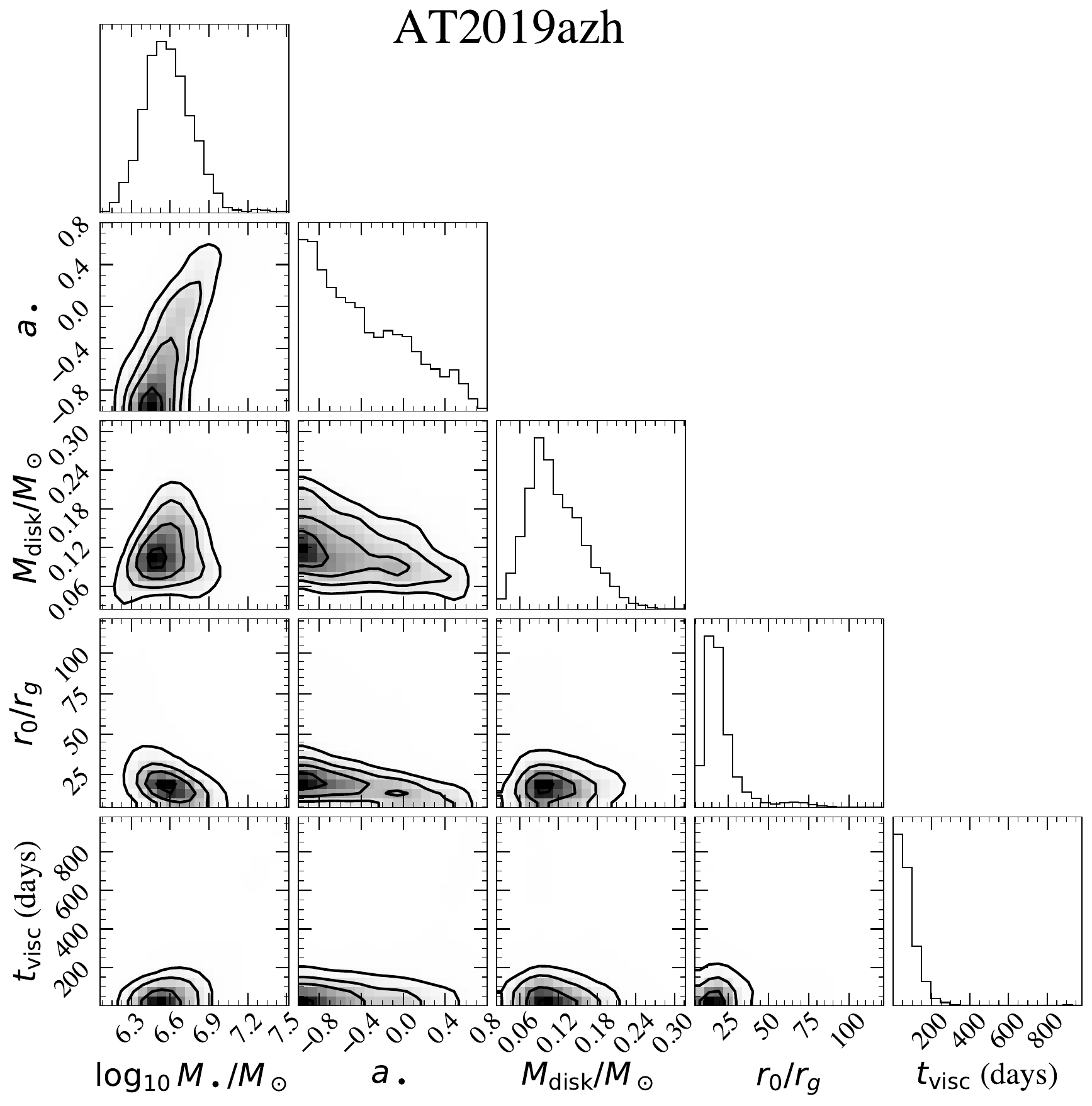}
    \includegraphics[width=0.45\linewidth]{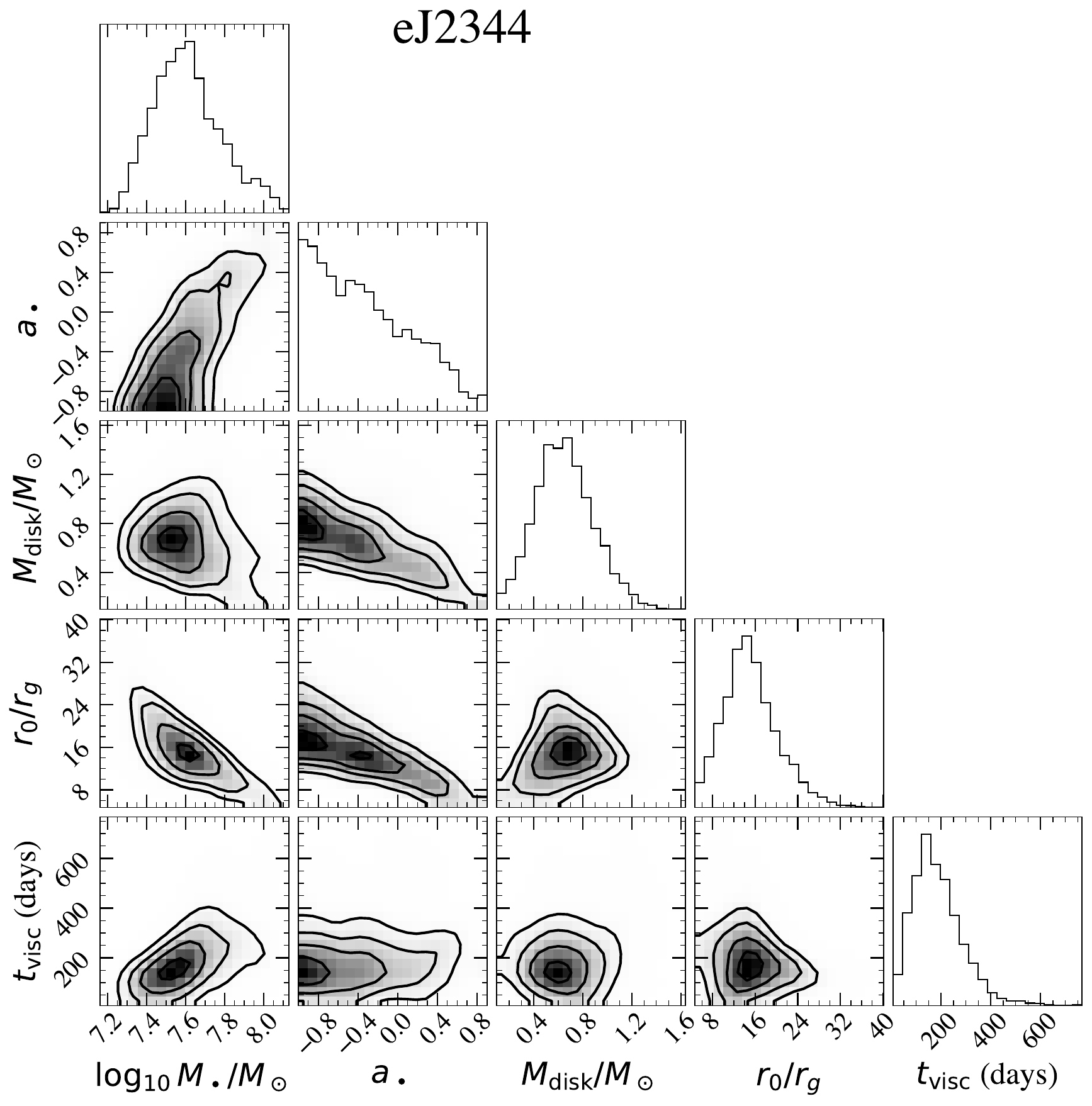}
    \caption{The posterior distributions of the parameter values which set the accretion rate of the disk at early and late times, for the five TDE systems from our gold sample with well constrained early X-ray emission. The name of each TDE system is displayed on each corner plot. }
    \label{fig:corners1}
\end{figure}

\begin{figure}
    \centering
    \includegraphics[width=0.45\linewidth]{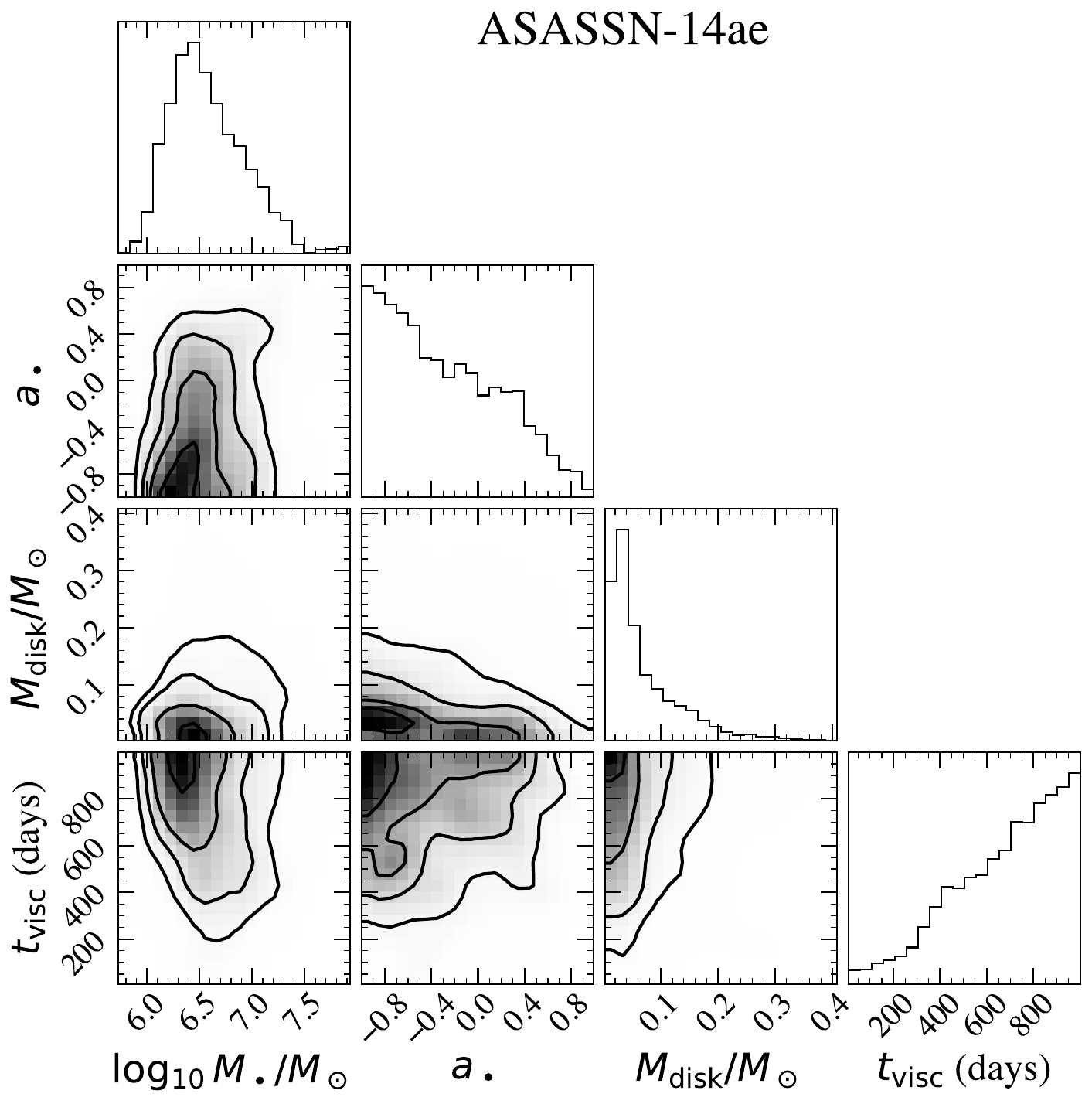}
    \includegraphics[width=0.45\linewidth]{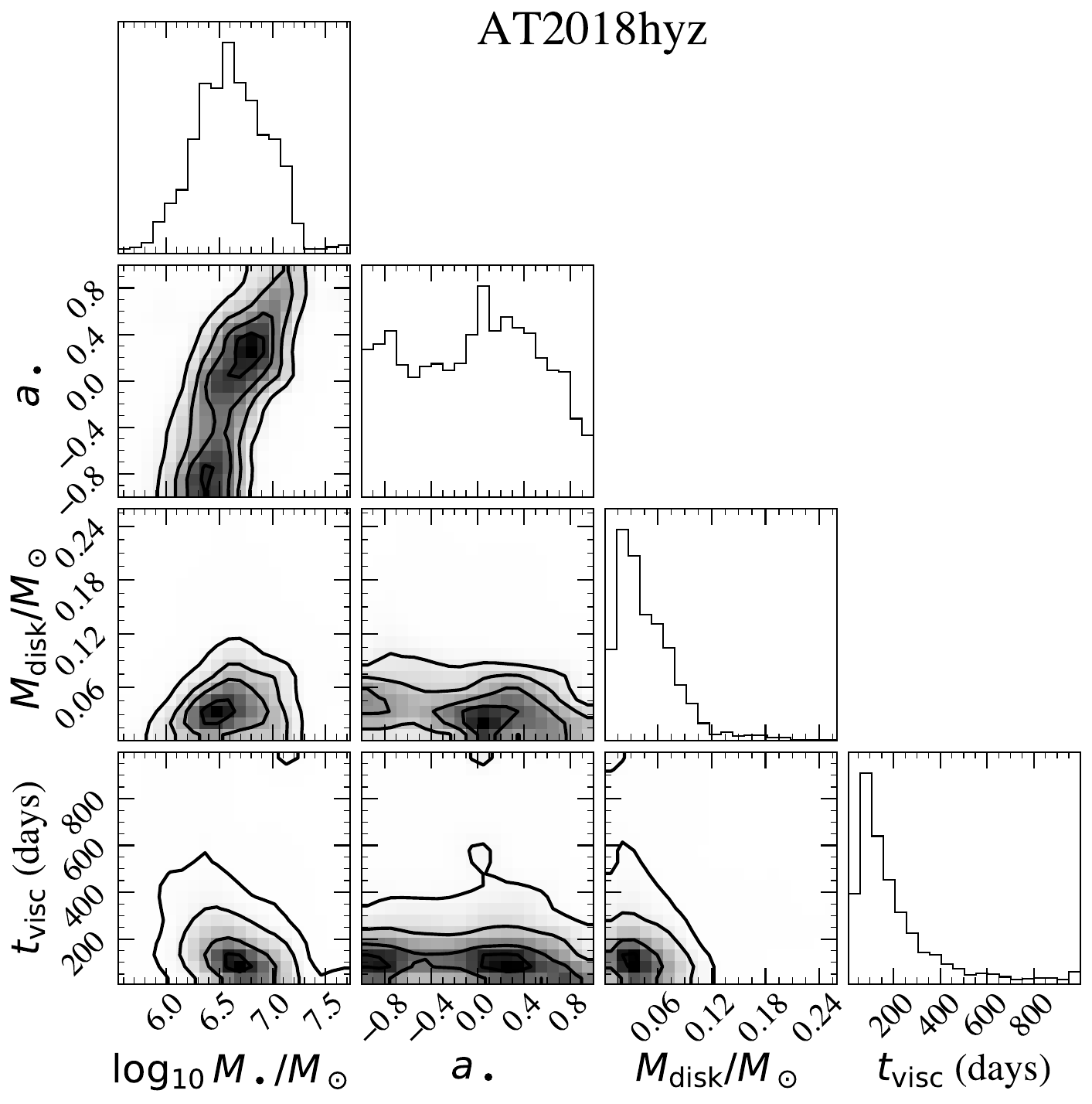}
    \includegraphics[width=0.45\linewidth]{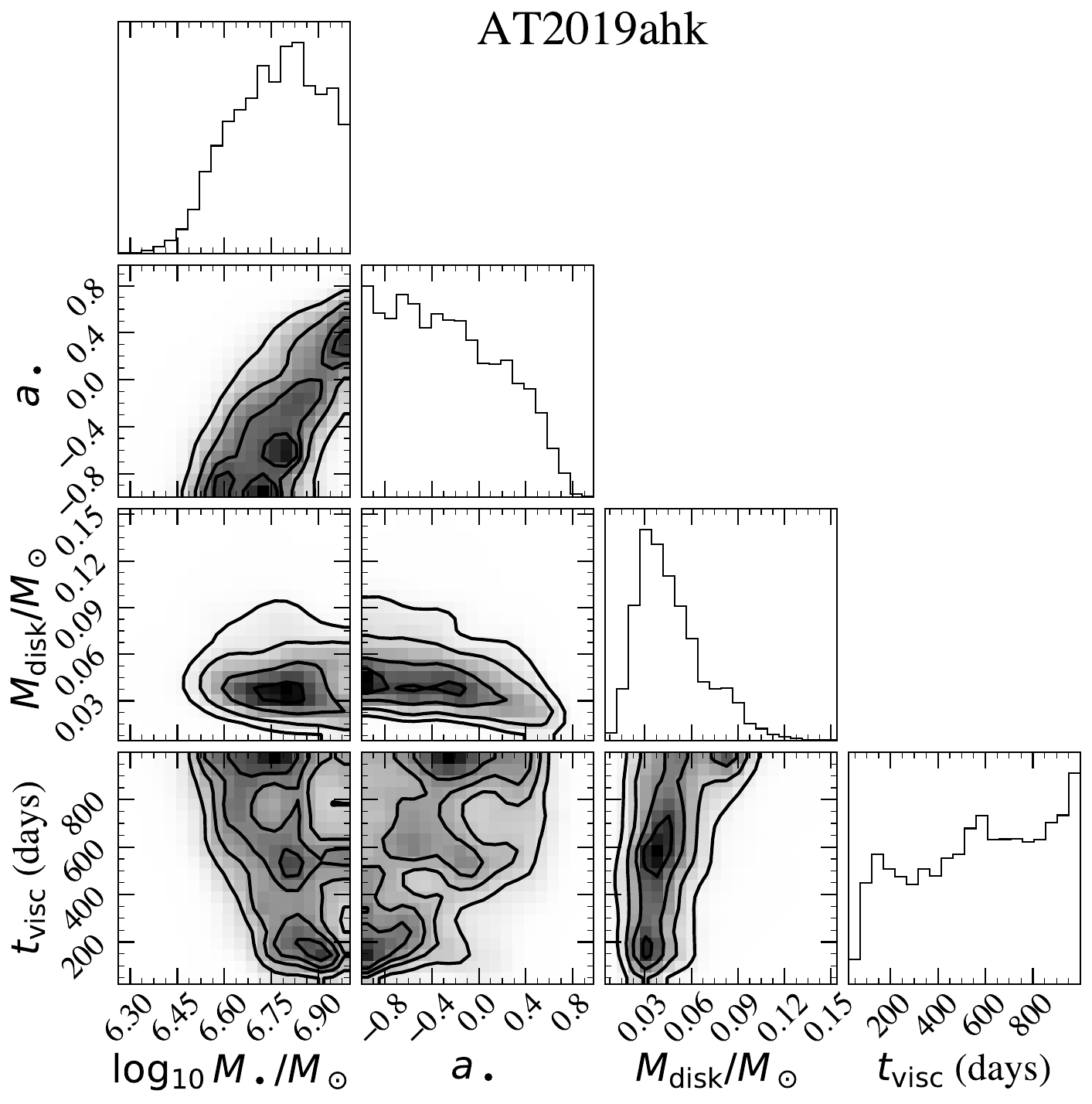}
    \includegraphics[width=0.45\linewidth]{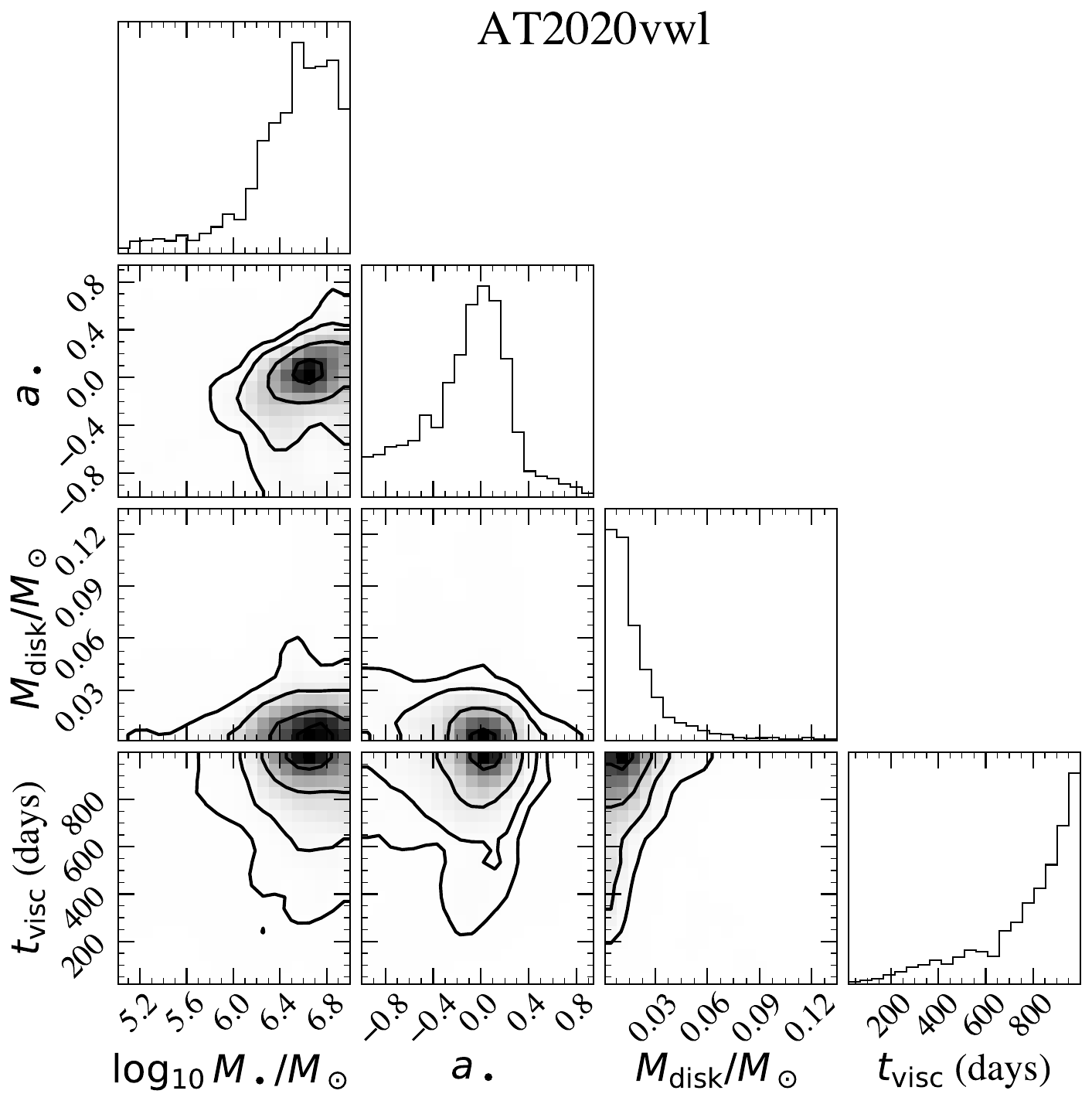}
    \includegraphics[width=0.45\linewidth]{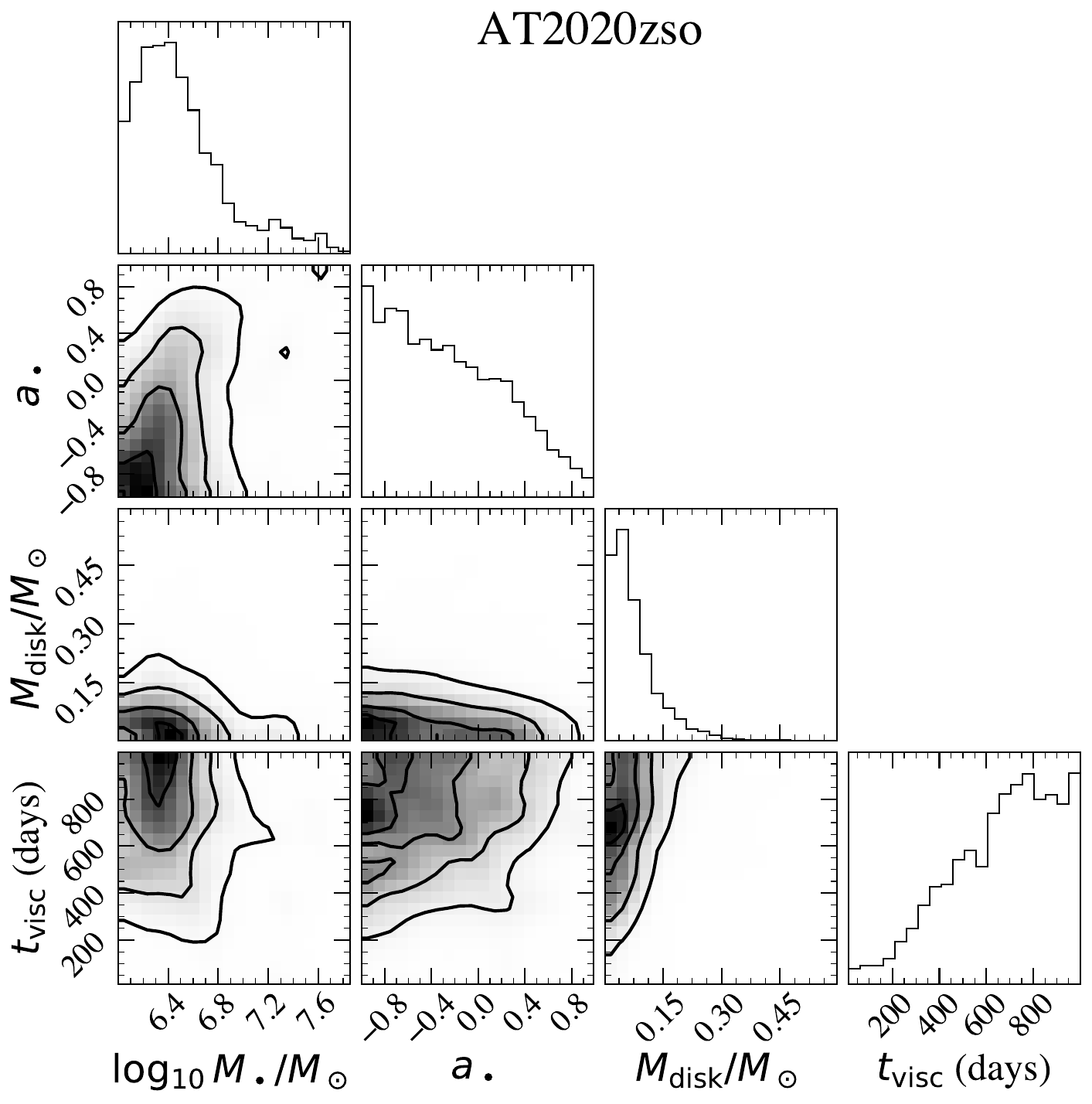}
    \caption{The posterior distributions of the parameter values which set the accretion rate of the disk at late times, for the five TDE systems from our gold sample which cannot be constrained at early epochs owing to a lack of X-ray emission. The name of each TDE system is displayed on each corner plot. }
    \label{fig:corners2}
\end{figure}

The individual disk model fits for each of the ``gold" TDEs are shown in Figures \ref{fig:models_1}--\ref{fig:models_9}. Corner plots of the posterior distributions for the ten TDEs which are in our ``gold'' sample are displayed in Figures \ref{fig:corners1} and \ref{fig:corners2}. We only display the posterior distributions of the {\tt FitTeD} parameters which affect the accretion rate/bolometric luminosity at the epochs of interest, for ASASSN-14li, ASASSN-15oi, AT2019dsg, AT2019azh and eJ2344 this includes the disk formation radius (as early radio flares probe the disk structure at times when the early evolution is important), while for the remaining five (ASASSN-14ae, AT2018hyz, AT2019ahk, AT2020vwl and AT2020zso) we do not display the poorly-constrained $r_0$ parameter, as it has no impact on the late-time inner-disk accretion rate which has lost memory of the initial condition. 

\subsubsection{Population synthesis}
In the paper we demonstrate that outflows are launched from TDE systems at two characteristic Eddington ratio scales, $\dot m \simeq 1$ and $\dot m \simeq 0.02$. An important question regards the fraction of sources which reach these two Eddington ratio scales, and the typical time-scale at which they are reached. To answer this question we perform TDE population synthesis, following the procedure developed in \cite{Mummery_et_al_2024,MummeryVV25}. 

We must sample different stellar and black hole parameters, before marginalising over them to produce distributions of quantities of interest. For the black holes we choose an agnostic black hole spin distribution, assuming a flat distribution covering the entire range of possible values:
\begin{equation}\label{pbha}
p_{a_\bullet} \propto 1 , \quad -1 < a_\bullet < 1.
\end{equation} 
A more important distribution is the black hole mass function. In this work we use the parametrised TDE black hole mass distribution constrained by \cite{MummeryVV25}, with the following functional form \citep{Schechter76}
\begin{equation}
    p_{M_\bullet}(m_\bullet) \propto  {m_\bullet^{\alpha_2} \over  1 + (M_c/m_\bullet)^{\alpha_1-\alpha_2} } \exp\left[- \left({m_\bullet\over M_g}\right)^\gamma\,\right] .
\end{equation}
This represents a broken power-law with a high mass exponential cut off. At low black hole masses $(M_\bullet \ll M_c)$ this function is well approximated by a single power law with index $p_{M_\bullet} \sim m_\bullet^{\alpha_2}$, while for higher masses $(M_c \ll M_\bullet \ll M_g)$ the power-law slope becomes $p_{M_\bullet} \sim m_\bullet^{\alpha_1}$, with the transition scale occurring at black hole masses $M_\bullet \sim M_c$. The very high mass behaviour is that of an exponential decay owing physically to the scarcity of high mass galaxies. We take $M_g = 6.4\times 10^7 M_\odot$ and $\gamma = +0.49$, which are the values presented in \cite{Shankar04}. Our results are only weakly sensitive to the values of $\gamma$ and $M_g$, as the high mass tidal disruption event rate is principally controlled by Hills mass \cite{Hills75} suppression (to be discussed shortly). We take the posterior median values inferred from the optical/UV TDE population by \cite{MummeryVV25}, namely $\alpha_1 = -0.85$, $\alpha_2 = 2.6$ and $\log_{10}M_c/M_\odot = 5.8$. This parametrised black hole mass function reproduces the late-time UV, early time optical and peak X-ray luminosity distributions of tidal disruption events \cite{MummeryVV25}, and is therefore robustly constrained.

For the stellar parameters we use the Kroupa initial mass function \citep{Kroupa01} to determine the intrinsic rate at which stars of different masses are formed. We assume that this equals the probability of a given star existing in the galactic centre (though not the probability of it being disrupted). The Kroupa IMF takes the form of a multiply broken power-law, with each power-law section taking the form
\begin{equation}
p_{\rm IMF}(M_\star) \propto M_\star^{k_i}.
\end{equation}
The values of $k_i$ are the following: $k_1 = -1.8$ for $M_\star < 0.5 M_\odot$; $k_2 = -2.7$ for $0.5 M_\odot < M_\star < M_\odot$; and $k_3 = -2.3$ for $M_\star > M_\odot$. We do not include stars with masses $M_\star < 0.08 M_\odot$ in our sample. The intrinsic rate at which TDEs occur for different stellar parameters, for a given black hole mass and spin, is a quantity which may be calculated theoretically. We use the rate calculation of \citet{Wang04} (see also \citet{Magorrian99} and \citet{Rees88})  whereby the intrinsic rate of tidal disruptions scales as 
\begin{equation}
p_{\rm rate}(M_\star, R_\star) \propto M_\star^{-1/3} R_\star^{1/4} .
\end{equation}
In other words this result encapsulates the intuitive result that more massive stars are harder to disrupt, while stars with larger radii are easier to disrupt. The masses and radii of stars on the main sequence are related. We use the mass-radius relationship of \citet{Kippenhahn90} 
\begin{equation}\label{star_mass}
R_\star(M_\star) \propto 
\begin{cases}
R_\odot \left({M_\star / M_\odot}\right)^{0.56}, \quad M_\star \leq M_\odot, \\
\\
R_\odot \left({M_\star / M_\odot}\right)^{0.79}, \quad M_\star > M_\odot. 
\end{cases} 
\end{equation}
We do not include giant (evolved) stars in our analysis, which are significantly more rare than main sequence stars.  This then allows us to determine the rate at which stars of different masses enter our TDE distribution, namely:
\begin{equation}\label{pstar}
p_\star(M_\star) \propto p_{\rm IMF}(M_\star) \times p_{\rm rate}(M_\star, R_\star(M_\star)) . 
\end{equation}
We sample stellar masses for our simulation from $p_\star$, and then use equation (\ref{star_mass}) to compute stellar radii. 

We sample stellar and black hole parameters from the above distributions. For each star and black hole we sample an incoming direction of the stars velocity vector (with respect to the black hole spin axis) assuming that the galactic-centre stellar population are isotropically distributed. We then use the analytical Hills mass solution of \cite{Mummery24} to determine whether the tidal radius lies inside or outside of the event horizon of this black hole. If the tidal disruption event is observable (i.e., the tidal radius is outside of the event horizon), we assume the disk circularises at twice the tidal radius (see above). 

The final parameter we need determine is the so-called ``viscous'' timescale of an evolving accretion flow. Which is given by the following simple expression:
\begin{equation}
t_{\rm visc} = \alpha^{-1} \left({H\over R}\right)^{-2} t_{\rm orbital} = \alpha^{-1} \left({H\over R}\right)^{-2} \sqrt{r^3 \over GM_\bullet} .
\end{equation}
where $r$ is the radius at which the flow begins, $M_\bullet$ is the black hole's mass, $H/R$ is the disc aspect ratio, $\alpha$ the \citet{SS73} alpha parameter, and $t_{\rm orbital}$ is the time it takes for the disc material to complete one orbit of the black hole at radius $r$.  A simple substitution of the circularisation radius (equation \ref{disk_form}) into the expression for the viscous timescale demonstrates that 
\begin{equation}
t_{\rm visc, c} = \alpha^{-1}\left({H\over R}\right)^{-2} \sqrt{8R_\star^3 \over \beta^3GM_\star} ,
\end{equation}
and we see that any explicit dependence on the properties of the black hole has dropped out of this expression.  The ratio of the viscous and orbital timescales is a dimensionless number, expected to be large $(\gg 1)$
\begin{equation}
{t_{\rm visc, c}  \over t_{\rm orbital}} = \alpha^{-1} \left({H\over R}\right)^{-2}  \equiv {\cal V} .
\end{equation}
The value of ${\cal V}$ will change over a population of TDEs, and we anchor our values of ${\cal V}$ in the range of values observed in disk-fits to the TDE population. To be explicit, we uniformly sample values of $\log_{10} {\cal V}$ in the range 
\begin{equation}\label{pv}
p_{\log_{10}\cal V} \propto 1, \quad 2 < \log_{10}{\cal V} < 4. 
\end{equation}
See \cite{Mummery_et_al_2024} for a discussion of the TDE disk fitting which results in this choice for ${\cal V}$. We note that this choice is entirely consistent with the fitting results of this paper.  With these black hole, stellar and viscous parameter choices determined, the {\tt FitTeD} code can be used to generate posterior distributions of various Eddington ratio parameters, as discussed in the text.

\subsection{Observational evidence for jet-like properties of the delayed flares}\label{sec:discussion}

Analyses of the few TDEs that have shown both prompt and delayed outflow behaviour to date have found different energetics in the radio properties of the two flares \citep{Hajela2025,Goodwin25,Christy2025}. Modelling the radio lightcurve evolution of the TDEs in our sample, overall we find the rise indices of the prompt and delayed outflow populations are bi-modal, with prompt outflows rising with indices 1--2.5, whereas delayed outflows rise with indices $>4$ (see Section \ref{sec:radio_lcs}). In the case of a ballistically expanding outflow without constant energy injection, the radio lightcurve evolution is driven primarily by the density of the ambient medium the shock propagates through. Assuming the ambient density is stratified, with $\rho (r) \sim r^{-k}$, the optically thick radio luminosity evolves as $\nu L_{\nu}\sim t^{\frac{k+8}{4}}$ \citep{Matsumoto2024}. For typical values of $k$ observed for TDEs of 2--2.5 \citep[e.g.][]{Alexander16,Cendes2021_dsg,Goodwin2023_opy}, this would imply a rise index of 2.5--2.7. Such a rise index is consistent with the observed rise indices of the prompt outflows in our sample. Conversely, lightcurve rise indices significantly greater than $3$, as observed for the delayed flares in our sample, cannot be explained by changes in the ambient medium, and instead require sustained energy injection into the outflow \citep[e.g.][]{Matsumoto2024,Goodwin25}. Such energy injection is consistent with the presentation of a compact jet formed at the base of the accretion disk in the low-hard accretion disk state.

In XRBs, compact jets that switch on during the soft to hard state transition generally show inverted or flat radio spectra \citep{Corbel2001,Fender01} due to the presence of multiple synchrotron-emitting components within the jet structure. In our sample of TDEs, all delayed flares show optically thin radio spectra down to $\approx$2\,GHz, in contrast to the typical radio spectrum of a compact jet. 
Nevertheless, complex cases have been observed in XRBs, especially at early times after the compact jet switches on, including changes in the location of the spectral break for the hard-state jet \citep{Russell2014} and optically thin spectra at early times that evolves to an optically thick spectrum as the jet structure grows and evolves \citep{Corbel2013}. 
As derived by \citep{Blandford1979}, the spectrum of a radio jet is flat up to a cut-off frequency that depends on the brightness temperature of the radio source. The timescale for an SMBH jet to evolve would be significantly larger than for an XRB (simply reflecting the factor $\sim 10^5-10^6$ increase in the natural length scale of the system), which likely explains why all late-flaring TDEs detected to date have shown optically thin radio spectra down to $\approx$2\,GHz, as the jet likely has not evolved to a flat spectrum within the observed frequency bands.

A natural multi-wavelength expectation of this late-flaring radio TDE paradigm is the presence of a hard X-ray emitting corona which is anticipated to be observed coincident with the late time radio re-flaring. Indeed, ASASSN-15oi showed a significant increase in hard X-ray emission approximately coincident with the second radio flare observed \citep{Guolo24,Hajela2025}. ASASSN-14ae also showed evidence of hard X-ray emission 4.8\,yr after discovery \citep{Jonker20}, consistent with a corona being present after our inferred delayed radio outflow launch date of $\approx$4.1\,yr. Similarly, hard X-ray emission was reported from AT2019teq $\approx952$\,d post optical flare, consistent with a corona present after our inferred (although uncertain) radio launch date of $\approx700$\,d. Other TDEs not included in our sample due to a lack of public radio spectral data have shown late-time X-ray hardening and hints of corona-like emission such as AT2018fyk \citep{Wevers2021} and AT2021ehb \citep{Yao2022ApJ_22ehb}. A late-time radio detection of AT2018fyk was reported \citep{Cendes2024_fykATel}, strengthening the jet-corona connection in this source. 
We note that some TDEs, such as AT2020vwl \citep{Goodwin25} remain undetected in X-rays during the delayed radio flaring episode. It is important to note however that the X-ray emission from hard-state TDE disks can be rather under-luminous, and the sources are typically at large distances. If AT2020vwl formed a corona with luminosity equal to that seen in ASASSN-15oi upon state transition, it would still be  below detectable thresholds with current instruments at the distance of AT2020vwl. 

In our analysis, we only included TDEs for which both the disk parameters and radio outflow launch date could be robustly constrained by detailed modelling of the radio spectral peak evolution over time. We note that due to the dependence on environment and timing of observations, a single frequency radio lightcurve alone is not a good tracer of the outflow launch time, which requires robust constraints on the emitting region size over time via radio spectral observations.  However, for some TDEs in our sample there are single frequency radio observations that indicate a re-brightening at late times, but do not allow for detailed outflow radius modelling. For example, AT2019dsg is known to have re-brightened in the radio at late times \cite{Cendes2024}, but with only poorly-constrained radio spectral information. Nevertheless, the estimated launch time of this second radio flare \cite{Cendes2024} is consistent with the time at which AT2019dsg is inferred by our disk modelling to have passed through $\dot m = 0.02$. This means that every source in our sample with a disk which is well-constrained at early times and is consistent with having reached super-Eddington accretion rates is observed to have launched a prompt radio outflow, and every source which is predicted by our disk-modelling to have transitioned through $\dot m = 0.02$ shows observational evidence of a delayed radio flare.

\begin{appendices}

\section{Extracting Eddington ratios post-fit}
We are interested in the bulk accretion state of these TDE disks at the time at which outflows are inferred to be launched from the system. Global states of accretion flows are normally characterised by their Eddington ratio. There are in principle two ``Eddington ratios'' one can compute in a disk, the Eddington ratio of the bolometric luminosity, or of the accretion rate (one could also choose e.g., the ``Eddington ratio of the X-ray luminosity'', or indeed any observed luminosity, but this is effectively meaningless as it does not correct for the fraction of the total luminosity caught in a given band, a particularly pertinent distinction for TDEs which emit most of their luminosity in the unobservable extreme UV). 

In a steady-state accretion disk the mass accretion rate and the bolometric luminosity of the disk are linearly proportional, $L_{\rm bol} = \eta(a_\bullet) \dot M c^2$, with a spin dependent efficiency factor. This is only true when the disk is in inflow equilibrium, which cannot be true in a TDE with a limited mass budget deposited on small scales. Indeed, particularly at early times, TDE disks are as far from inflow equilibrium as is possible, and so we must be much more careful in how we compute disk Eddington ratios in this work. 

Once a posterior distribution of disk and black hole parameters (which we denote $p_\Theta$) is constrained from the multi-band light curves, we can, for each element of the MCMC chain, compute both the bolometric disk luminosity $L_{\rm bol}$, and disk mass accretion rate $\dot M$, at any time. Marginalising over the {\tt FitTeD} parameter distributions at a given time then produces a time-dependent posterior distribution of Eddington ratios (both mass flux and bolometric luminosity). 

The bolometric luminosity of the disk, as emitted from its surface, can be simply computed from the disk temperature $T(r, t)$ as 
\begin{equation}
    L_{\rm bol}(t) = 2 \int_{\rm disk} 2\pi r \, \sigma T^4(r, t) \, {\rm d}r, 
\end{equation}
where the additional factor 2 here results from the two (upper and lower) surfaces of the disk which radiate, and the integral is over the entire disk radial extent. We define the Eddington luminosity in the usual way
\begin{equation}
    L_{\rm edd} = {4\pi G M_\bullet m_p c \over \sigma_T} \simeq 1.26 \times 10^{38} \, \left({M_\bullet \over M_\odot}\right) \, {\rm erg}/{\rm s} , 
\end{equation}
where $m_p$ is the proton mass and $\sigma_T$ is the Thomson scattering cross section. For a given posterior distribution we can then construct the time dependent posterior distribution of the luminosity Eddington ratio 
\begin{equation}
    \ell \equiv {L_{\rm bol}(t) \over L_{\rm edd}} , 
\end{equation}
by repeated sampling of the disk temperature implied by the posterior distributions of the disk fit.

The disk's local mass accretion rate (a function of both radius and time), is formally given by 
\begin{equation}
    \dot M(r, t) \equiv 2 \pi r U^r \Sigma, 
\end{equation}
where $U^r$ is the inflow velocity of the disk fluid, which can be written explicitly in terms of the disk surface density and turbulent stress as \citep[e.g.,][]{EardleyLightman75, Balbus17}
\begin{equation}
    \dot M(r, t) = - {2\pi  U^0 \over U_\phi'} {\partial \over \partial r} \left({r \Sigma W^r_\phi \over U^0}\right) , 
\end{equation}
where the prime denotes a radial gradient. The quantities $U^0$ and $U_\phi'$ are the disk fluids' Lorentz factor and angular momentum gradient respectively (see \cite{mummery2024fitted} for more details). Despite the angular momentum gradient of the flow vanishing at the ISCO this quantity converges to a finite limit as $r \to r_I$ (see \cite[][]{Mummery23a}). 

\begin{figure}
    \centering
\includegraphics[width=0.3\linewidth]{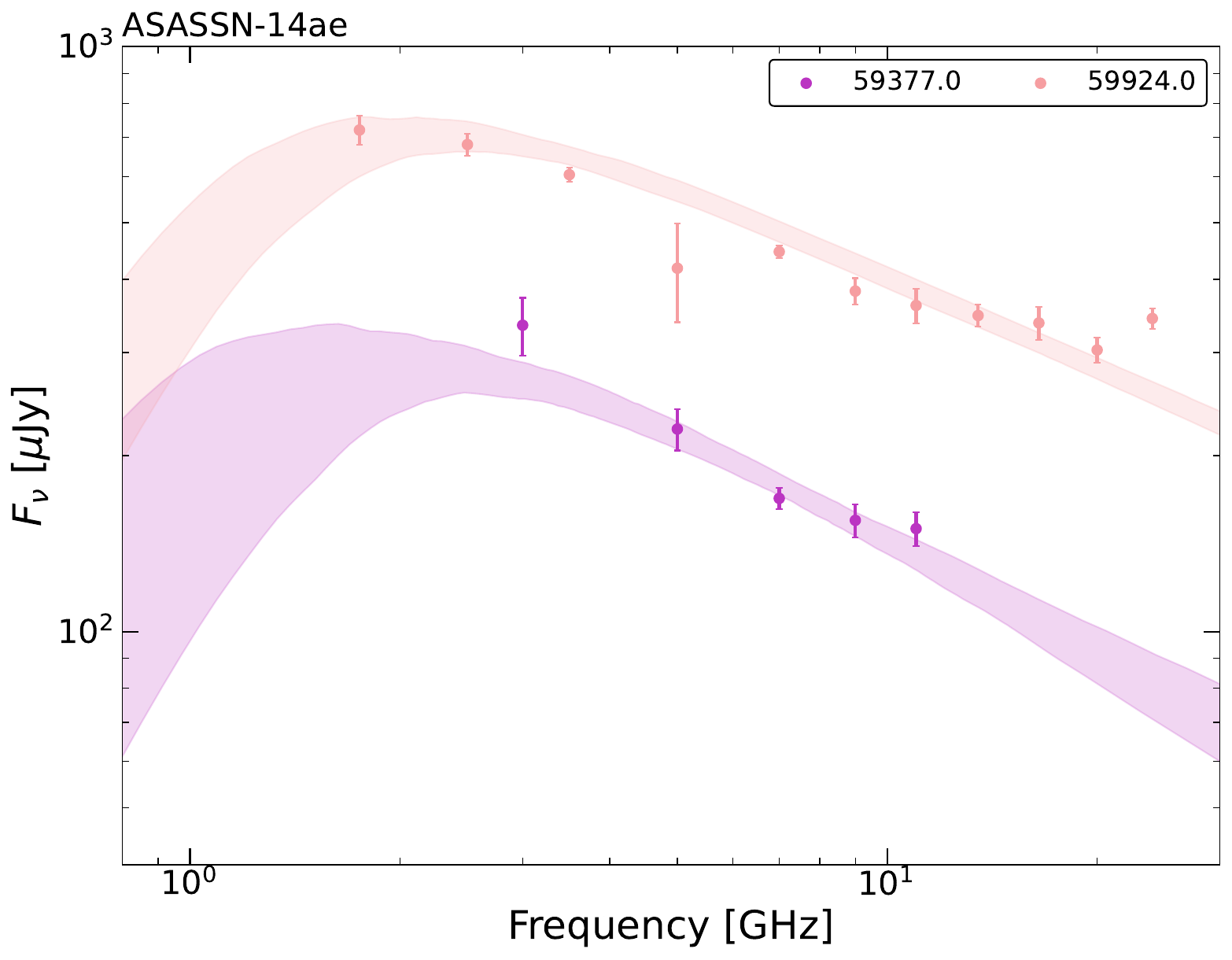}
    \includegraphics[width=0.3\linewidth]{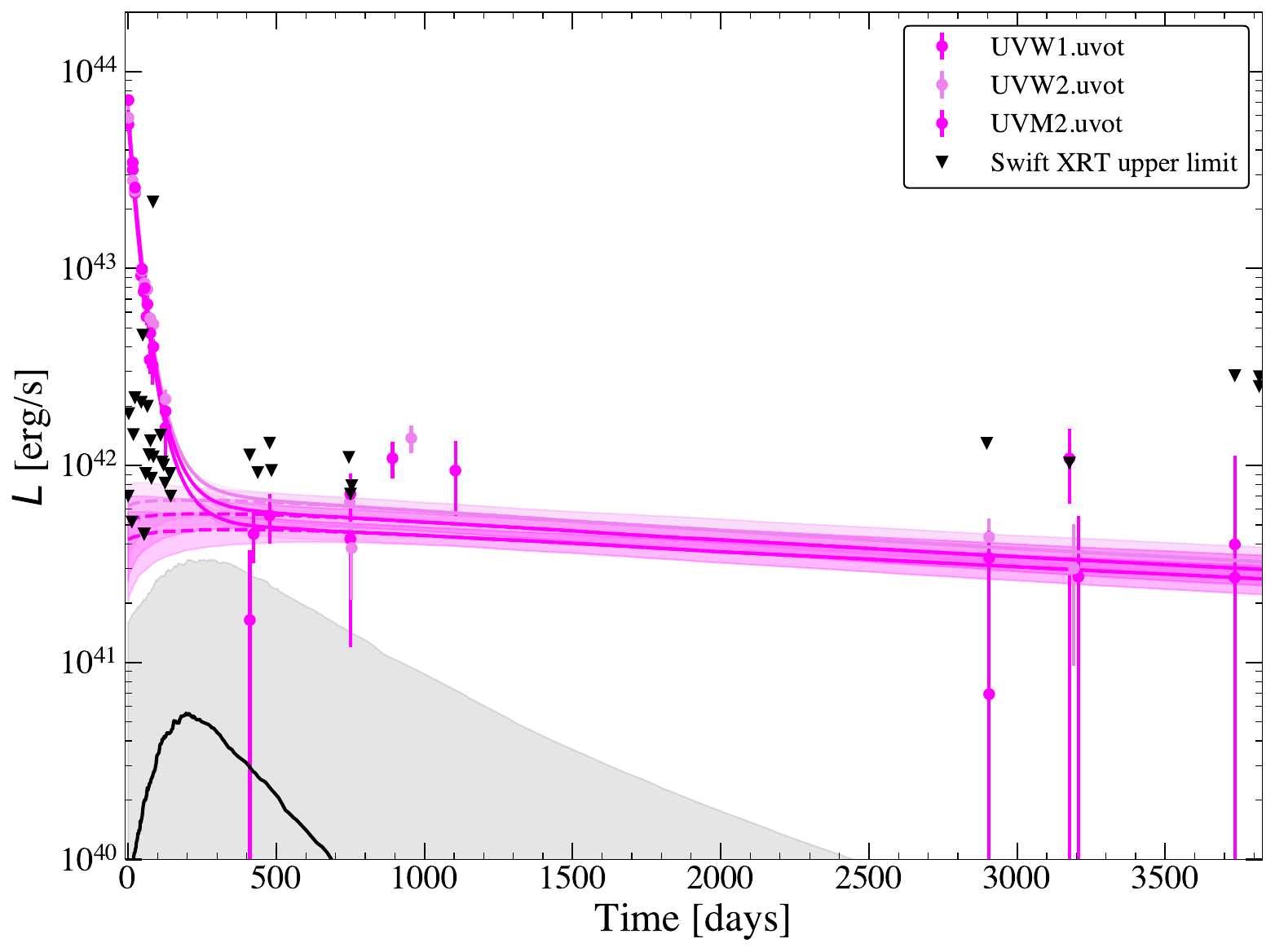}
    
    \includegraphics[width=0.3\linewidth]{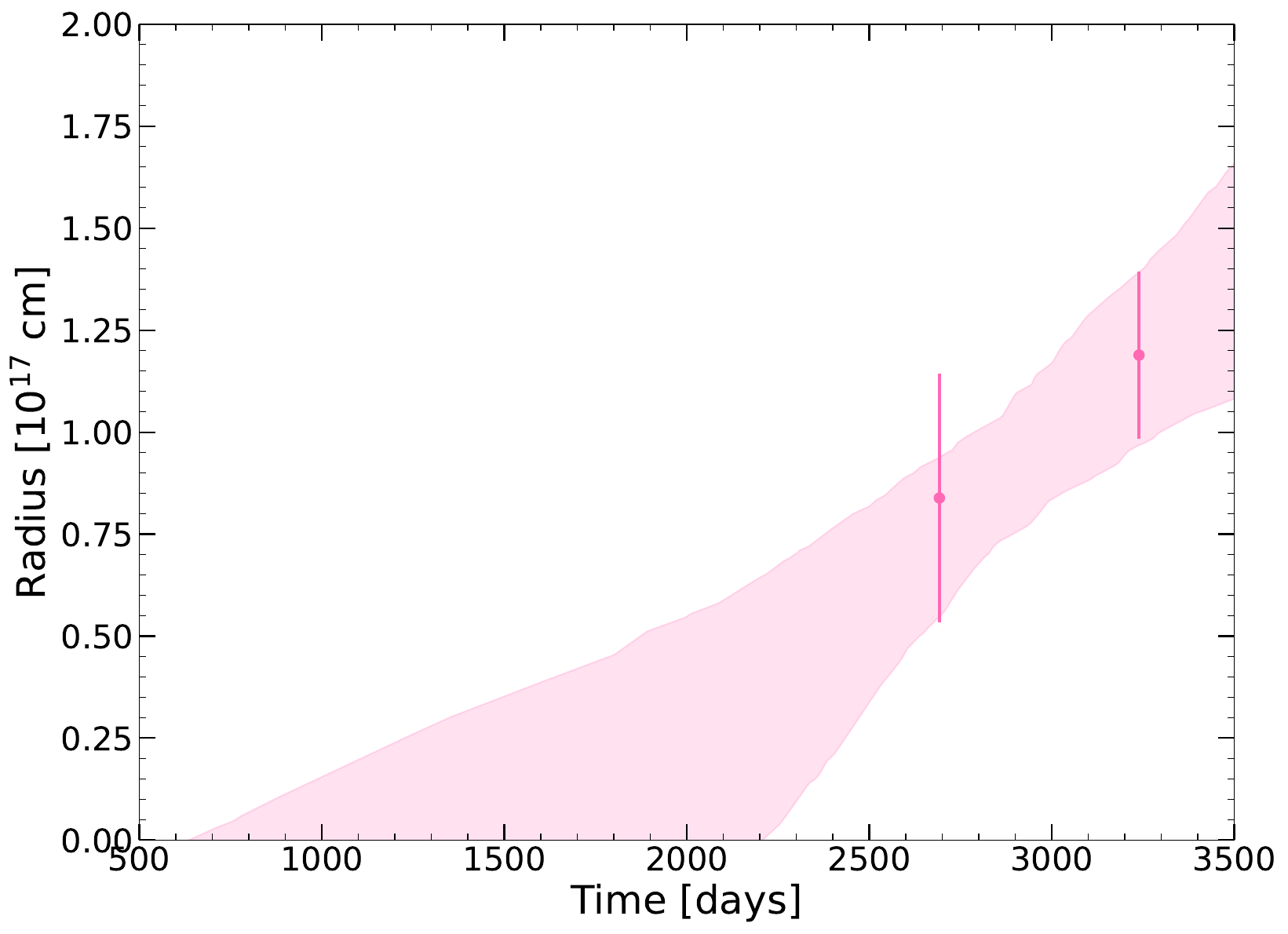}
    \includegraphics[width=0.3\linewidth]{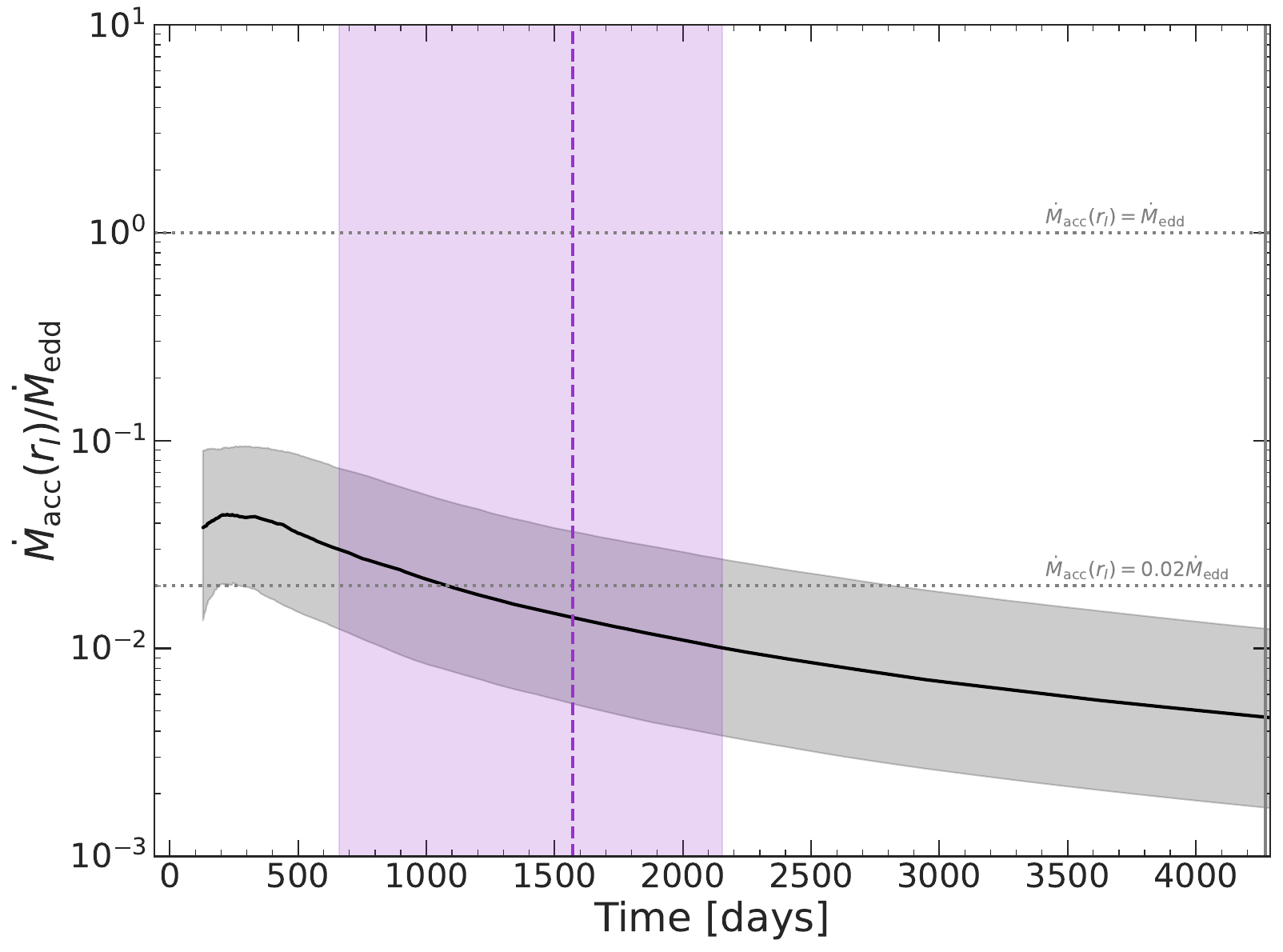}
    \includegraphics[width=0.4\linewidth]{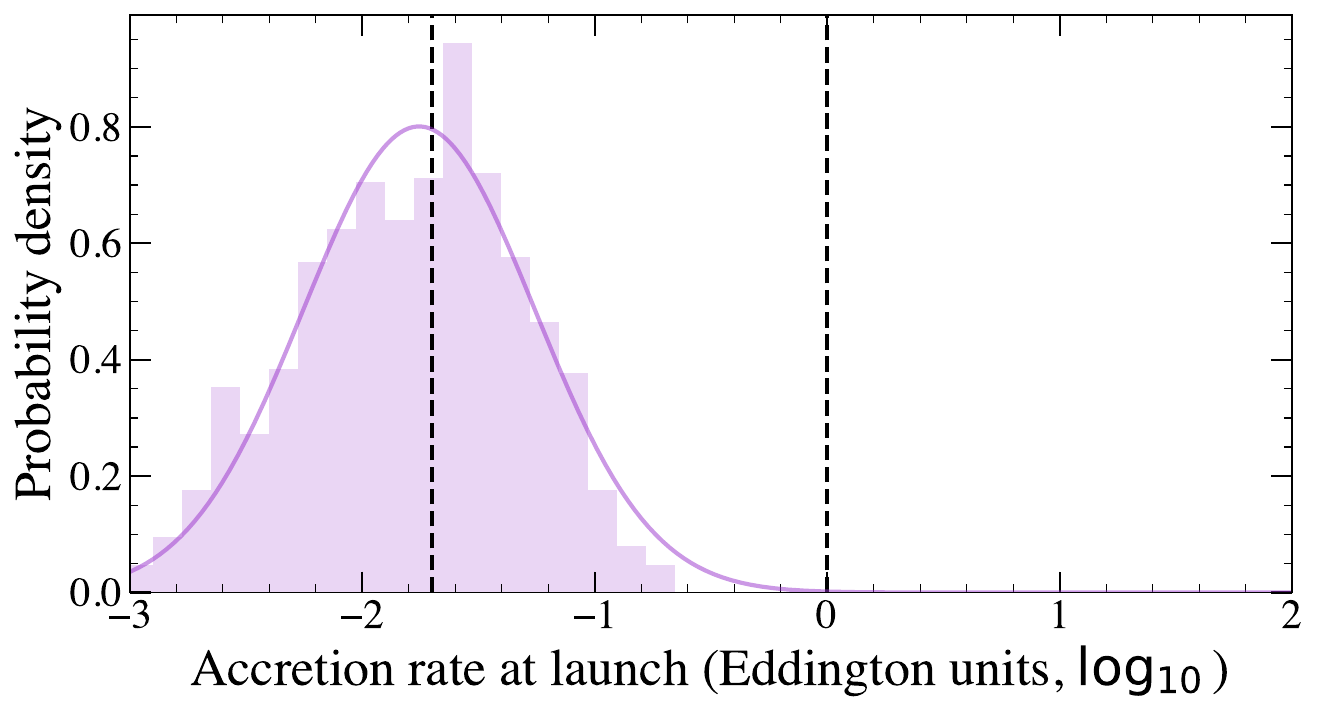}
    \caption{The same as Figure \ref{fig:models_14li} but for ASASSN-14ae. Accretion rate only plotted after times at which the disk model dominates the flux in optical/UV bands.  }
    \label{fig:models_1}
\end{figure}

\begin{figure}
    \centering
\includegraphics[width=0.3\linewidth]{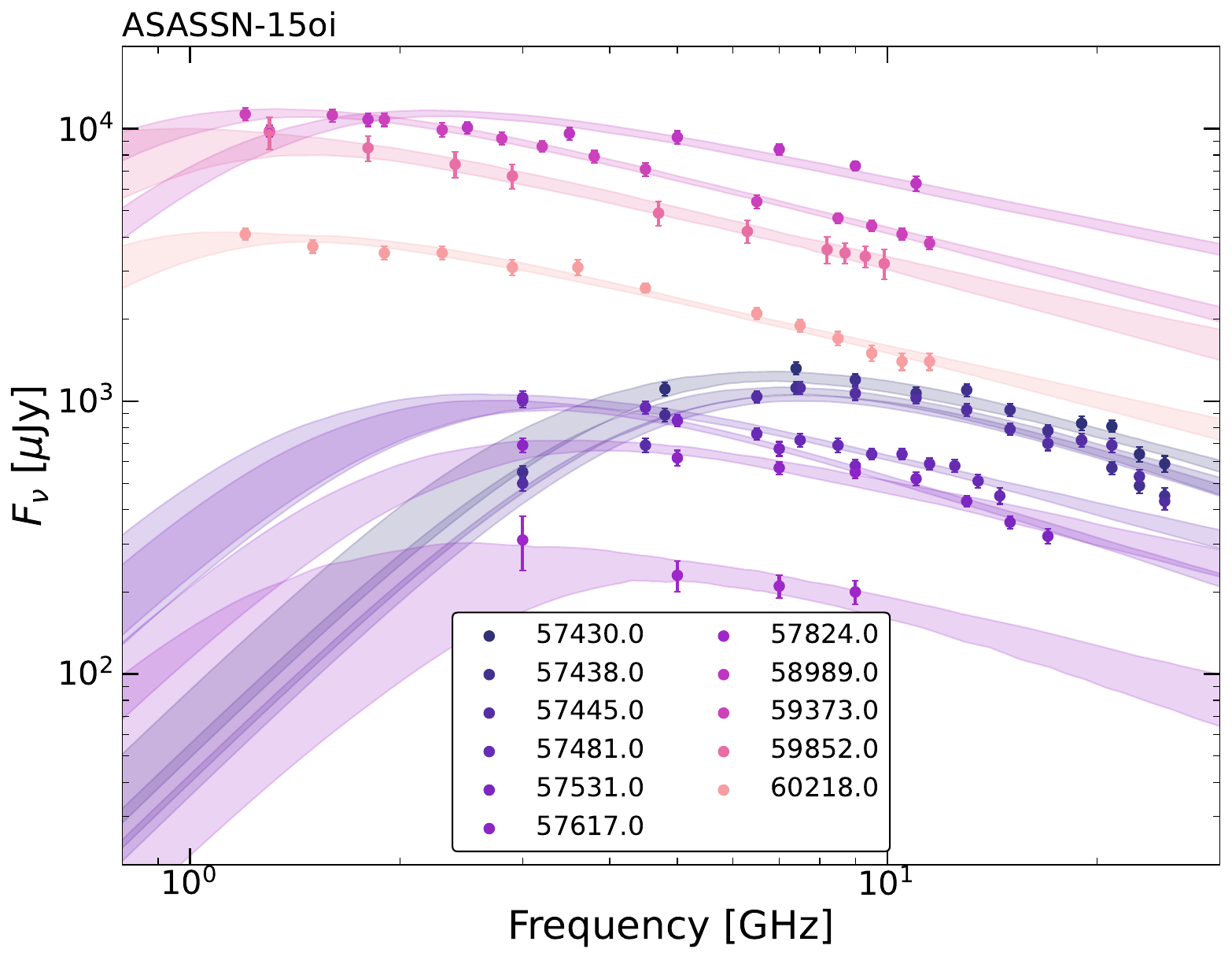}
    \includegraphics[width=0.3\linewidth]{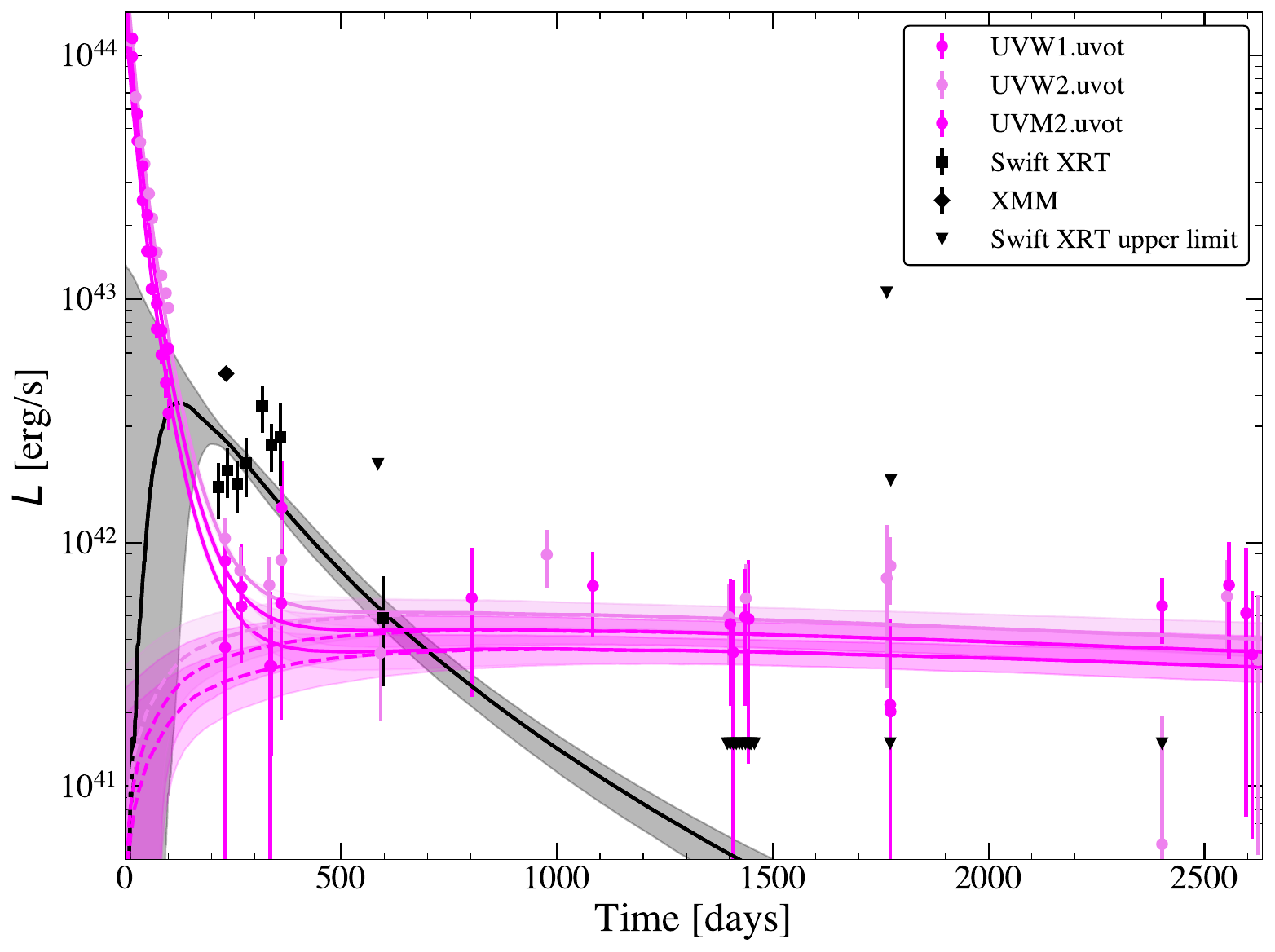}
    
    \includegraphics[width=0.3\linewidth]{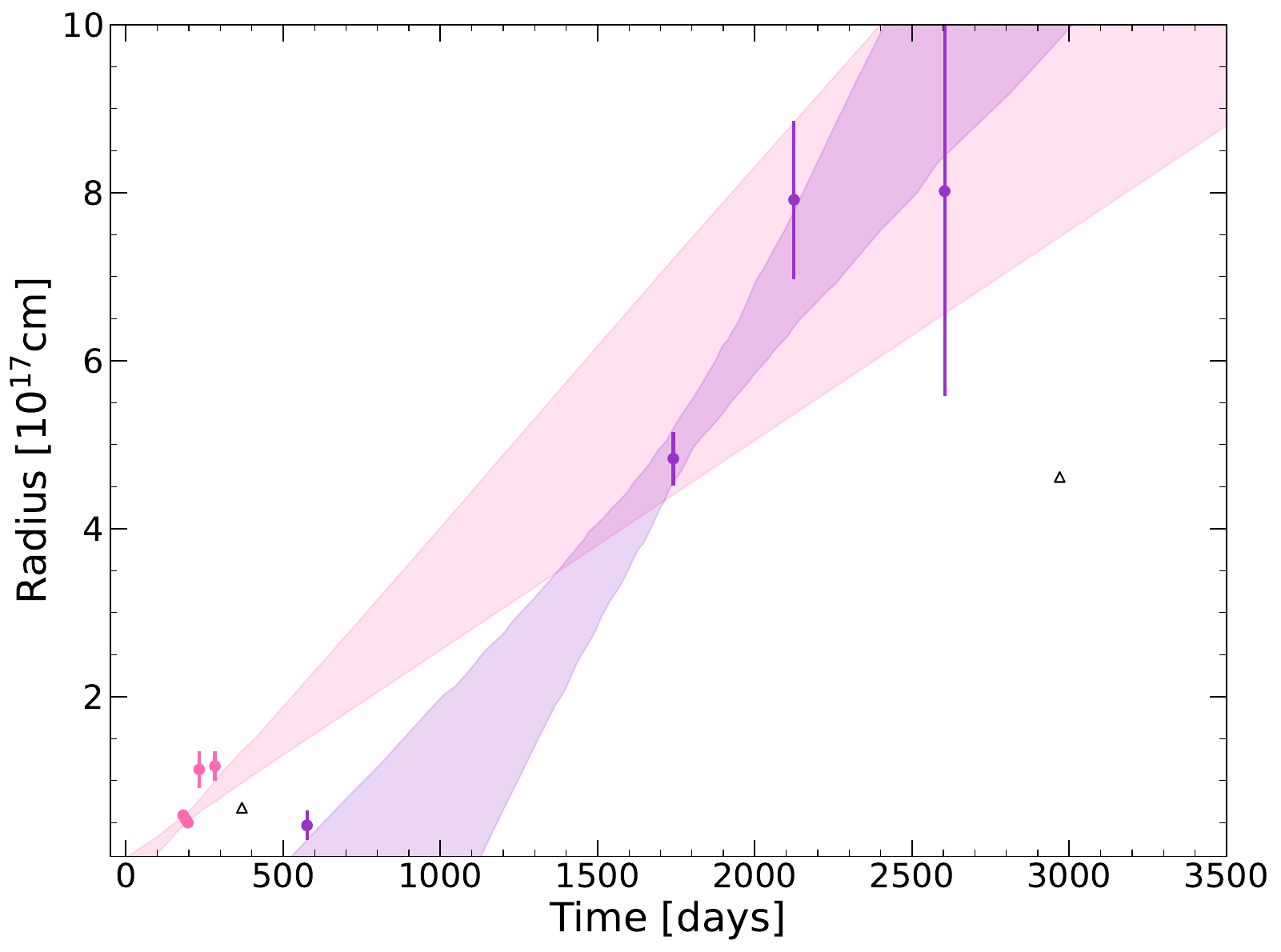}
    \includegraphics[width=0.3\linewidth]{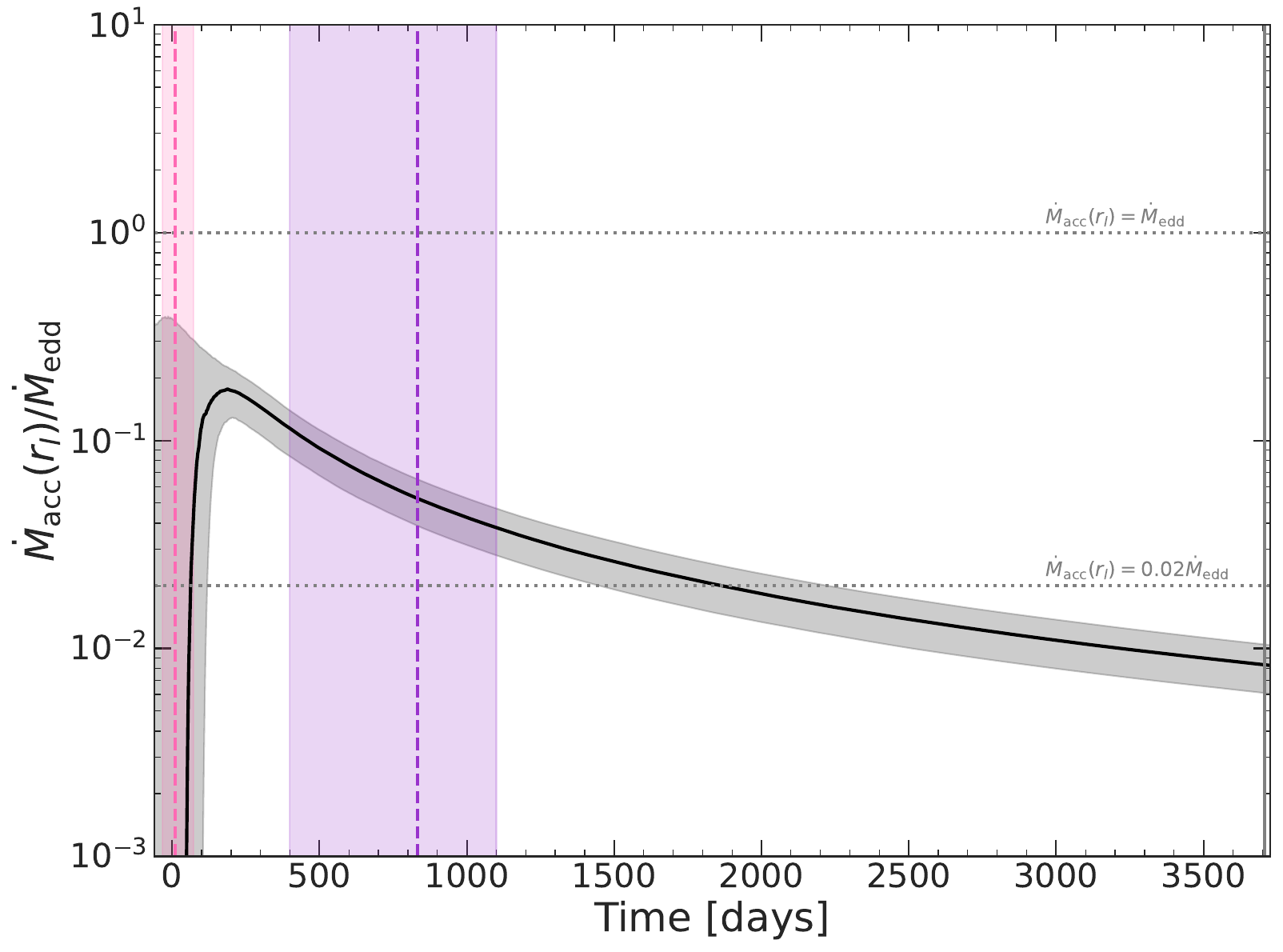}
    \includegraphics[width=0.4\linewidth]{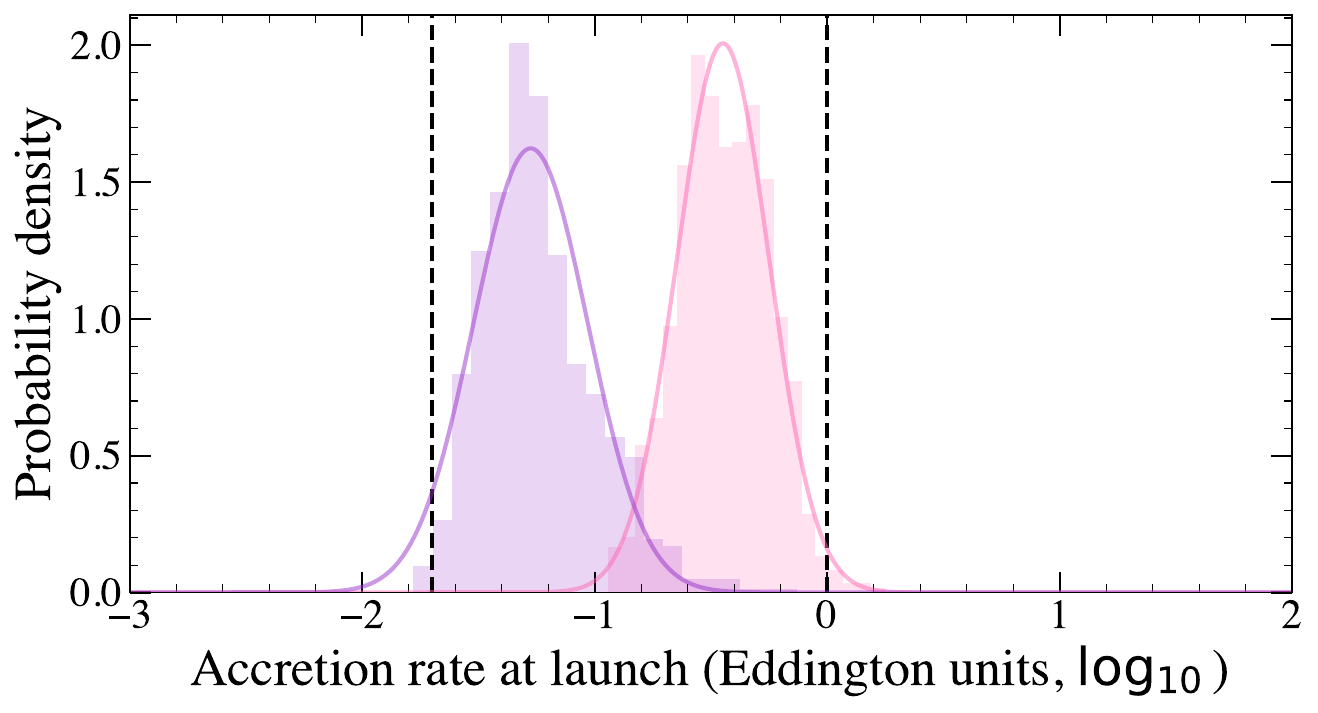}
    \caption{The same as Figure \ref{fig:models_14li} but for ASASSN-15oi. }
    \label{fig:models_2}
\end{figure}

\begin{figure}
    \centering
\includegraphics[width=0.3\linewidth]{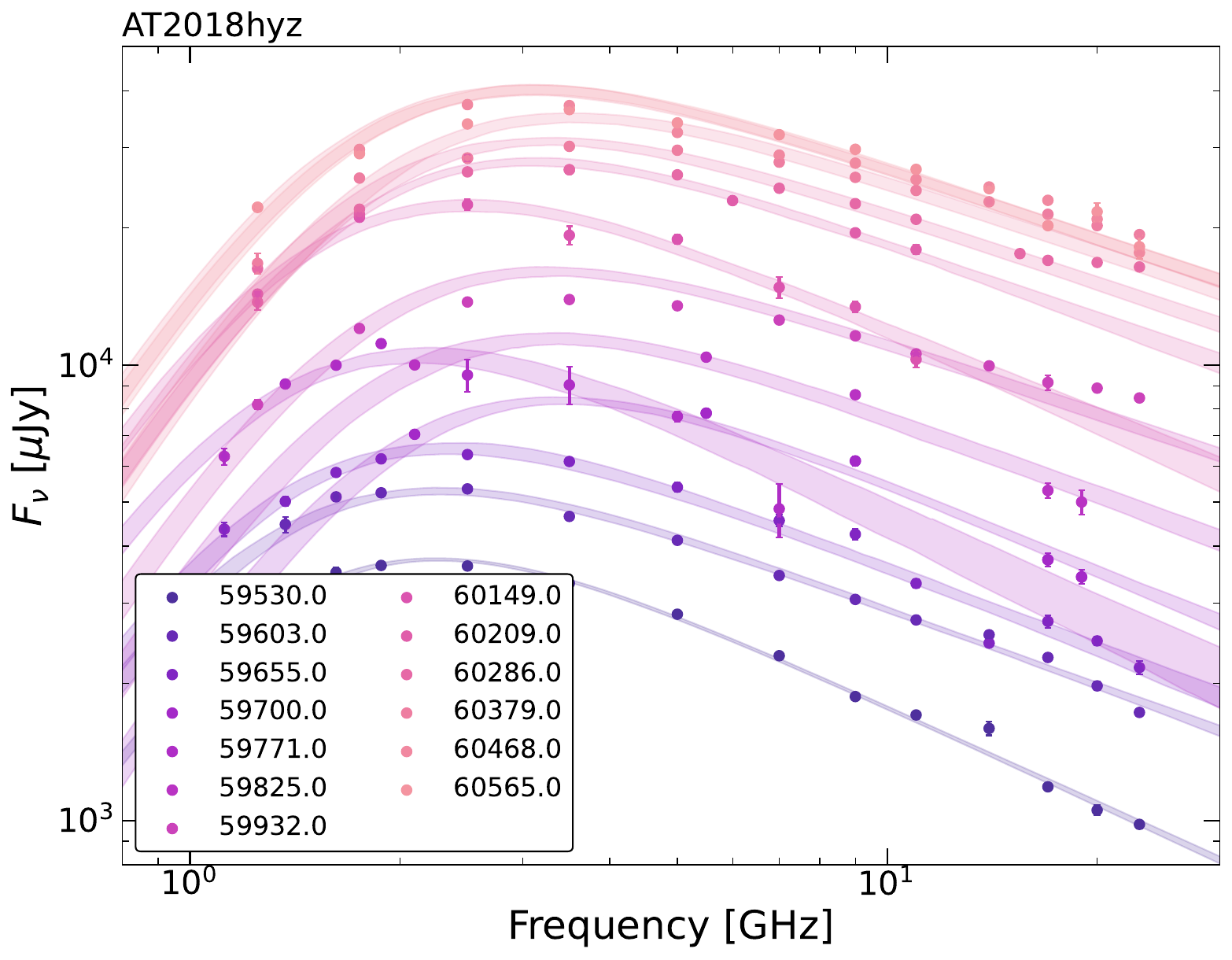}
    \includegraphics[width=0.3\linewidth]{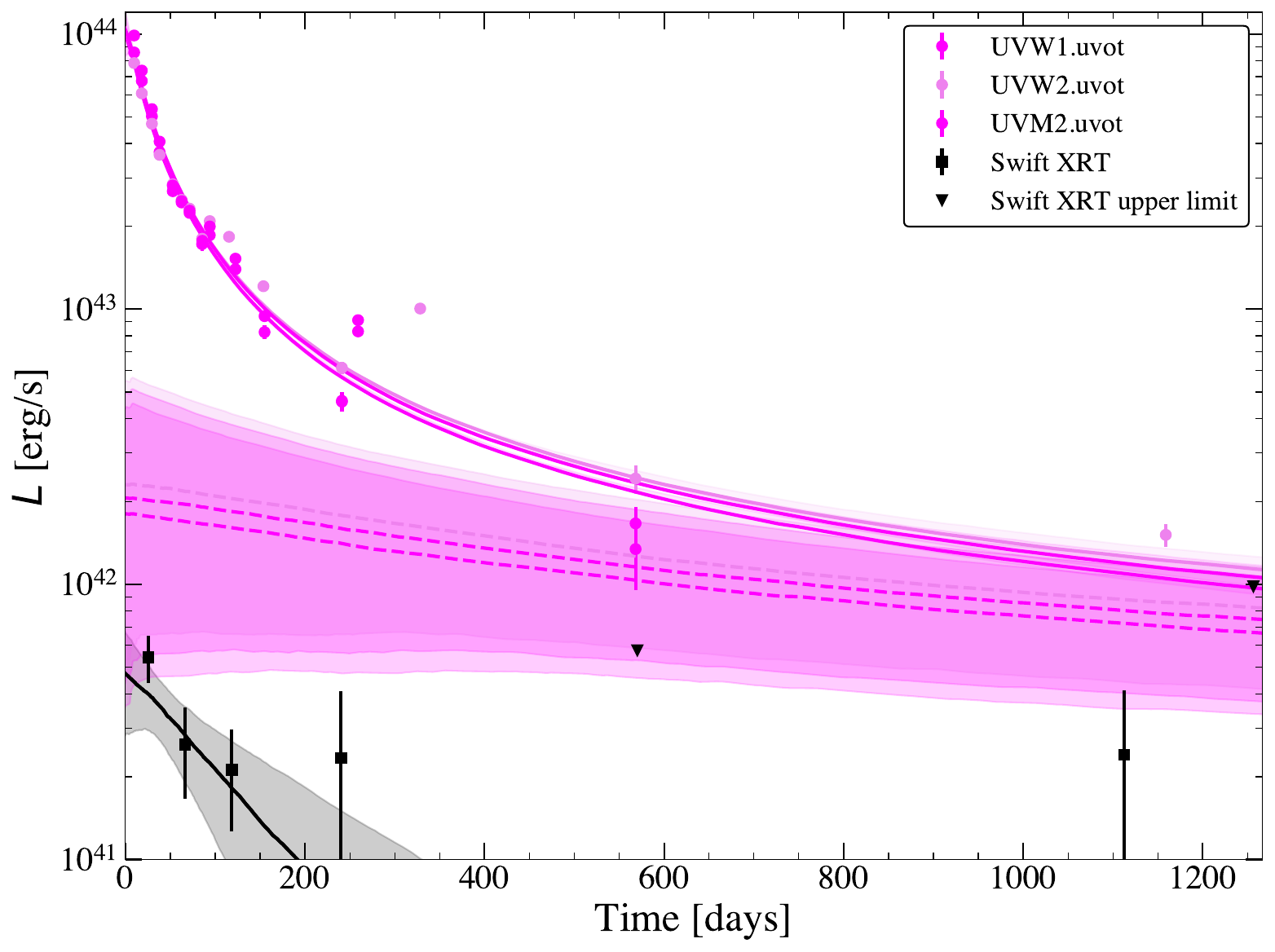}
    
    \includegraphics[width=0.3\linewidth]{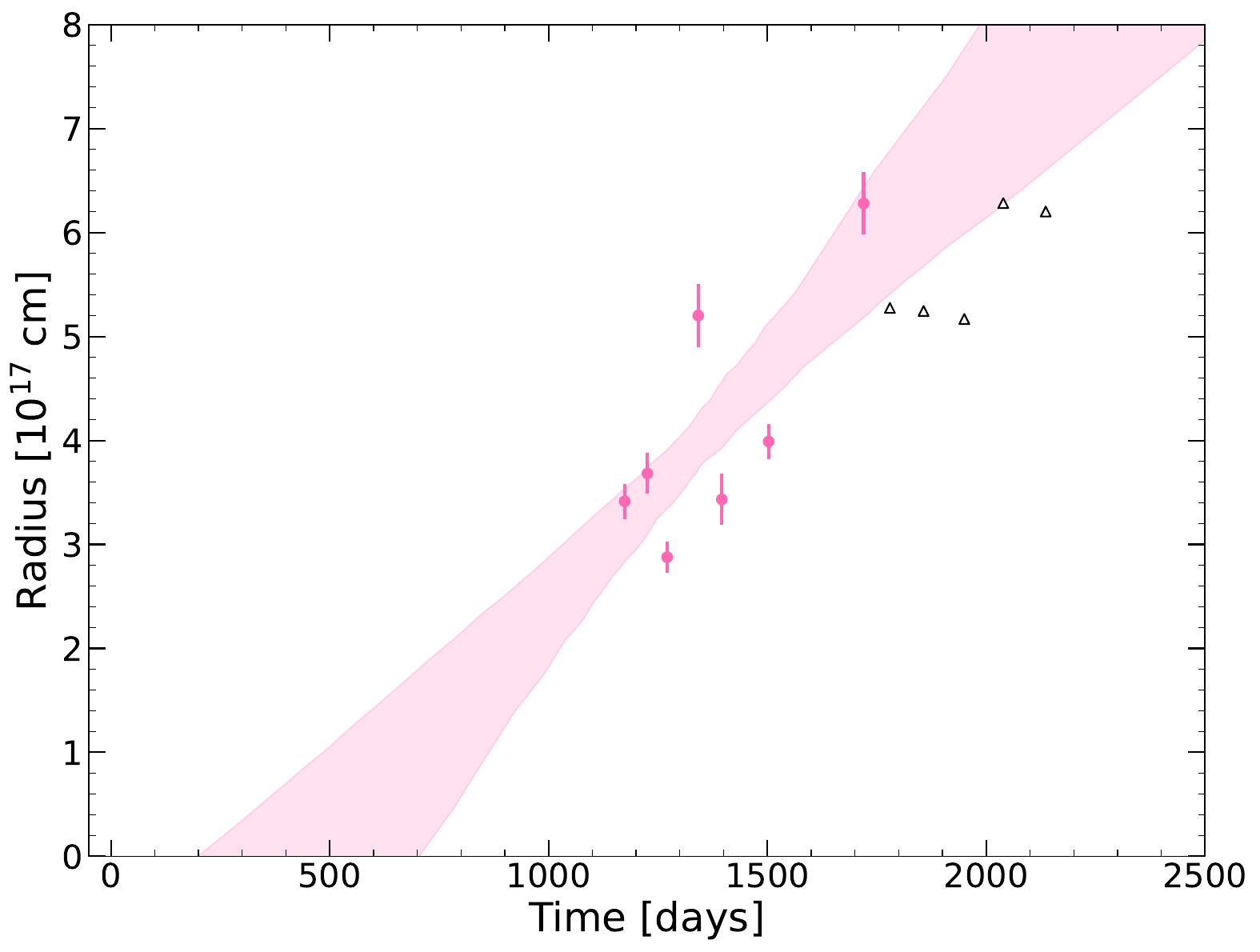}
    \includegraphics[width=0.3\linewidth]{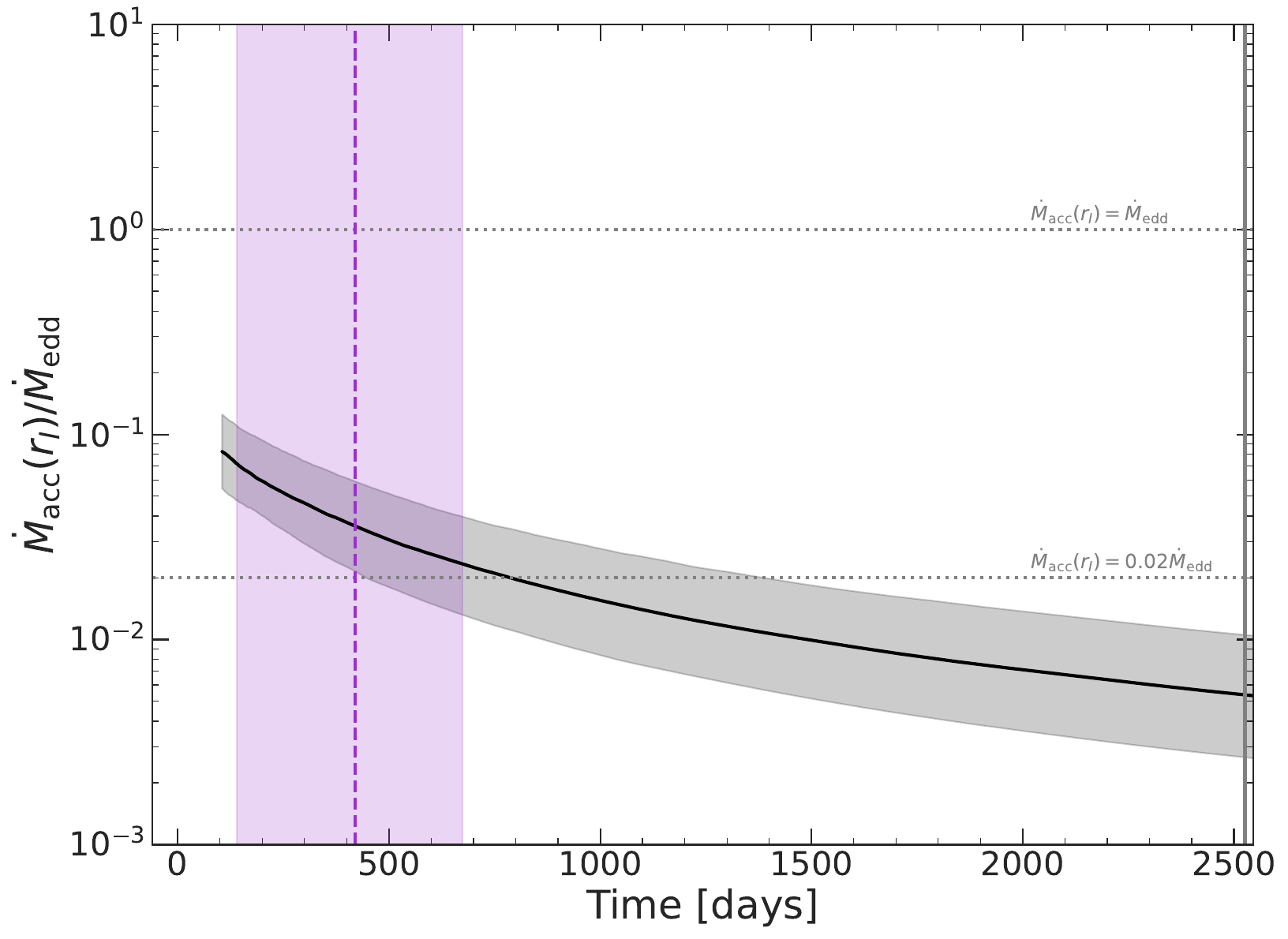}
    \includegraphics[width=0.4\linewidth]{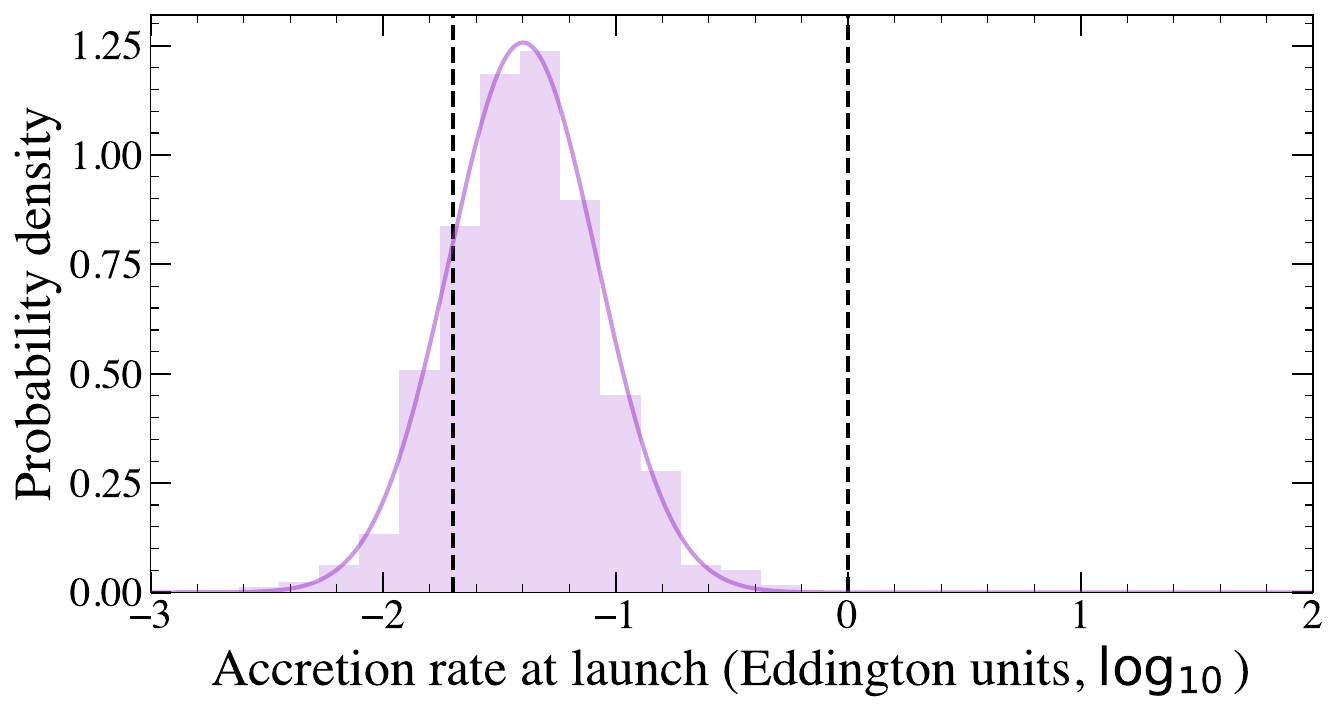}
    \caption{The same as Figure \ref{fig:models_14li} but for AT2018hyz. Accretion rate only plotted after times at which the disk model dominates the flux in optical/UV bands.  }
    \label{fig:models_3}
\end{figure}

\begin{figure}
    \centering
\includegraphics[width=0.3\linewidth]{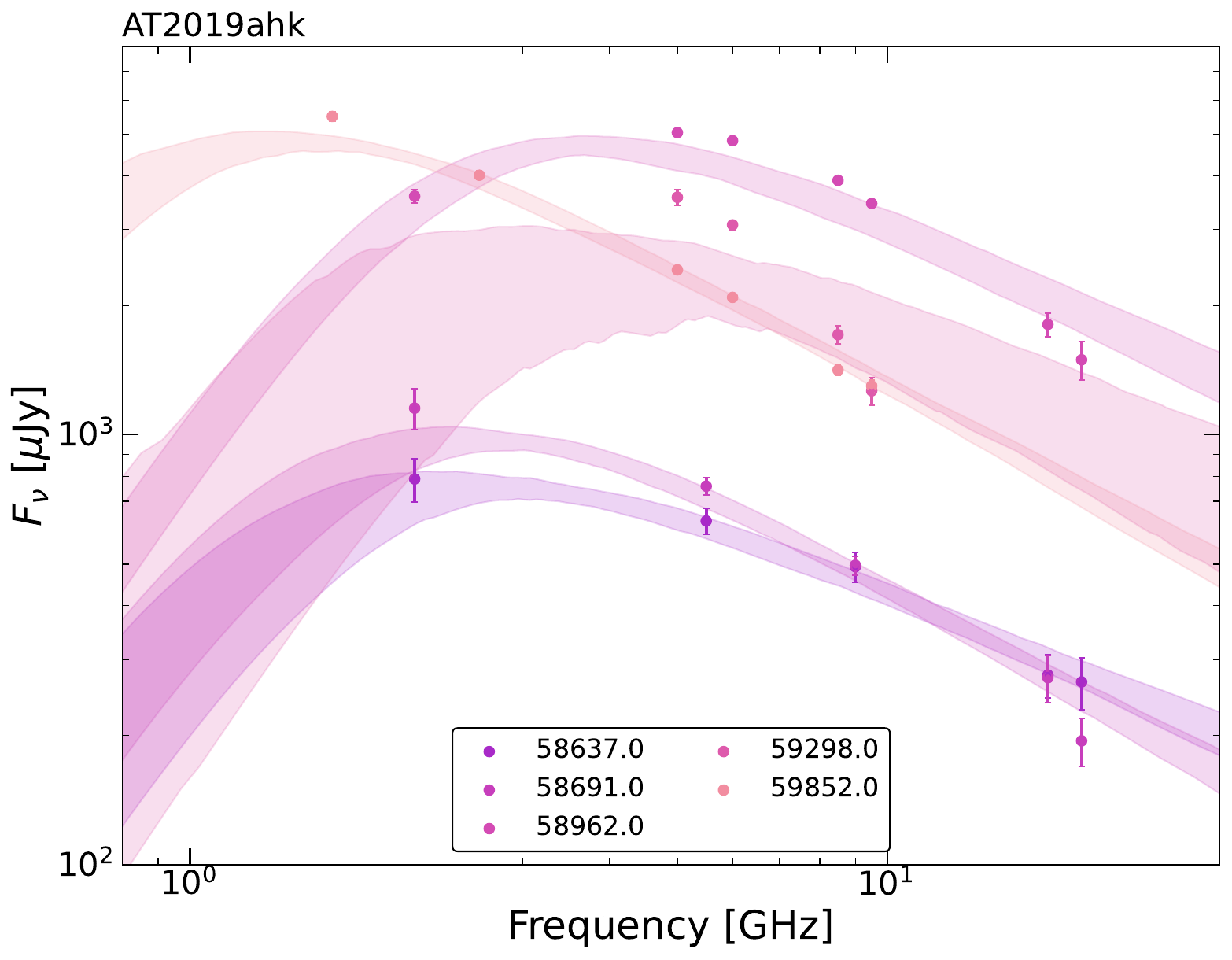}
    \includegraphics[width=0.3\linewidth]{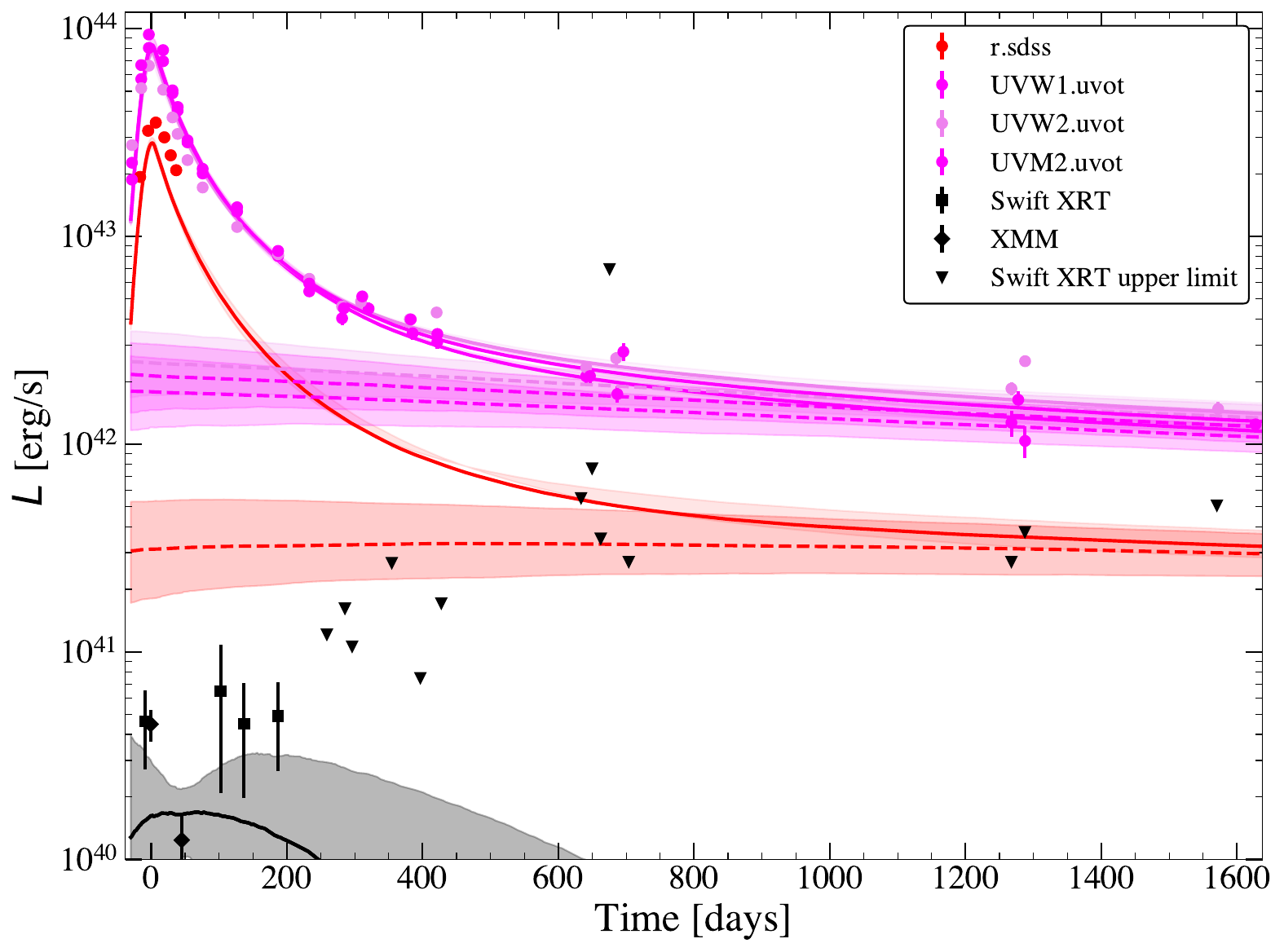}
    
    \includegraphics[width=0.3\linewidth]{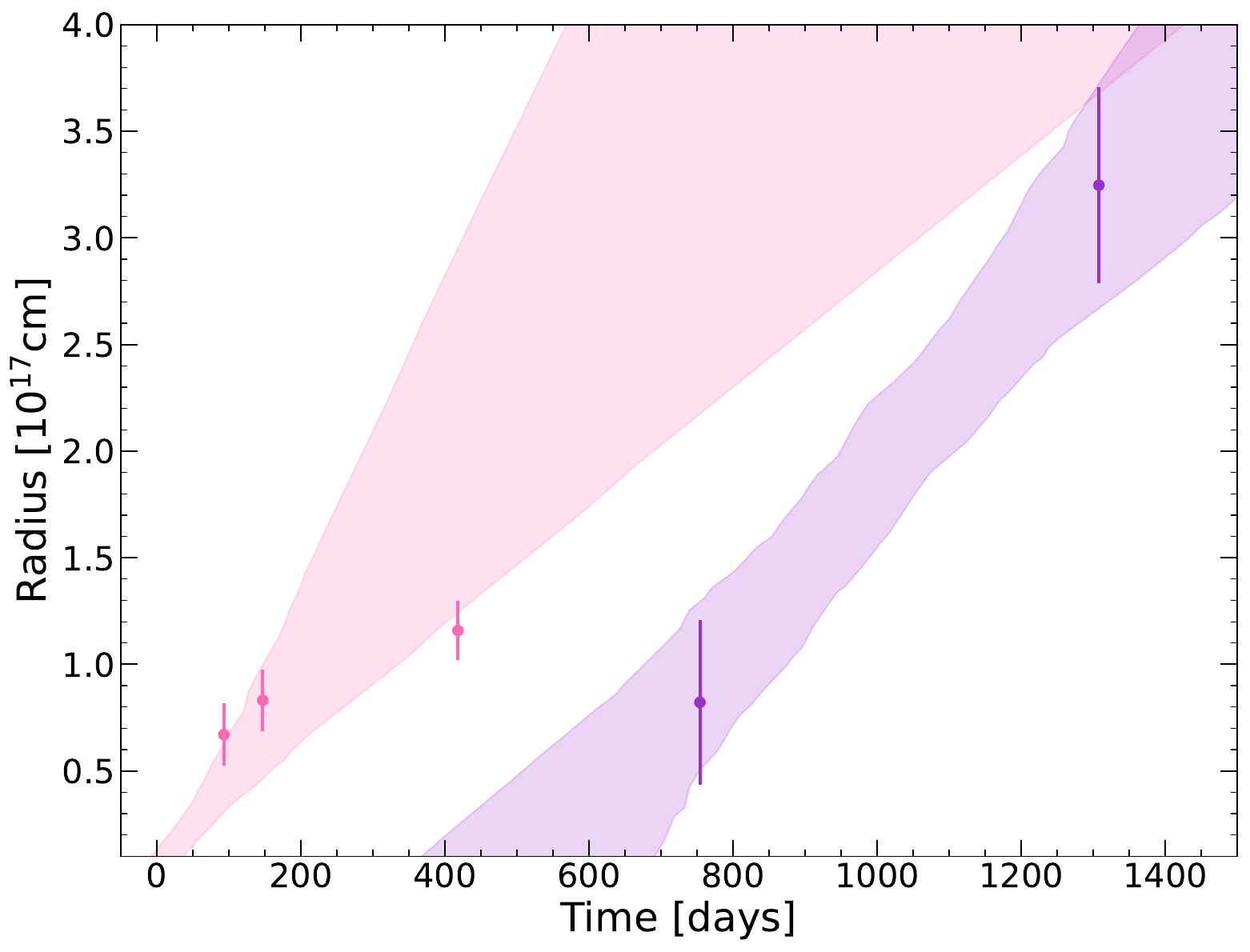}
    \includegraphics[width=0.3\linewidth]{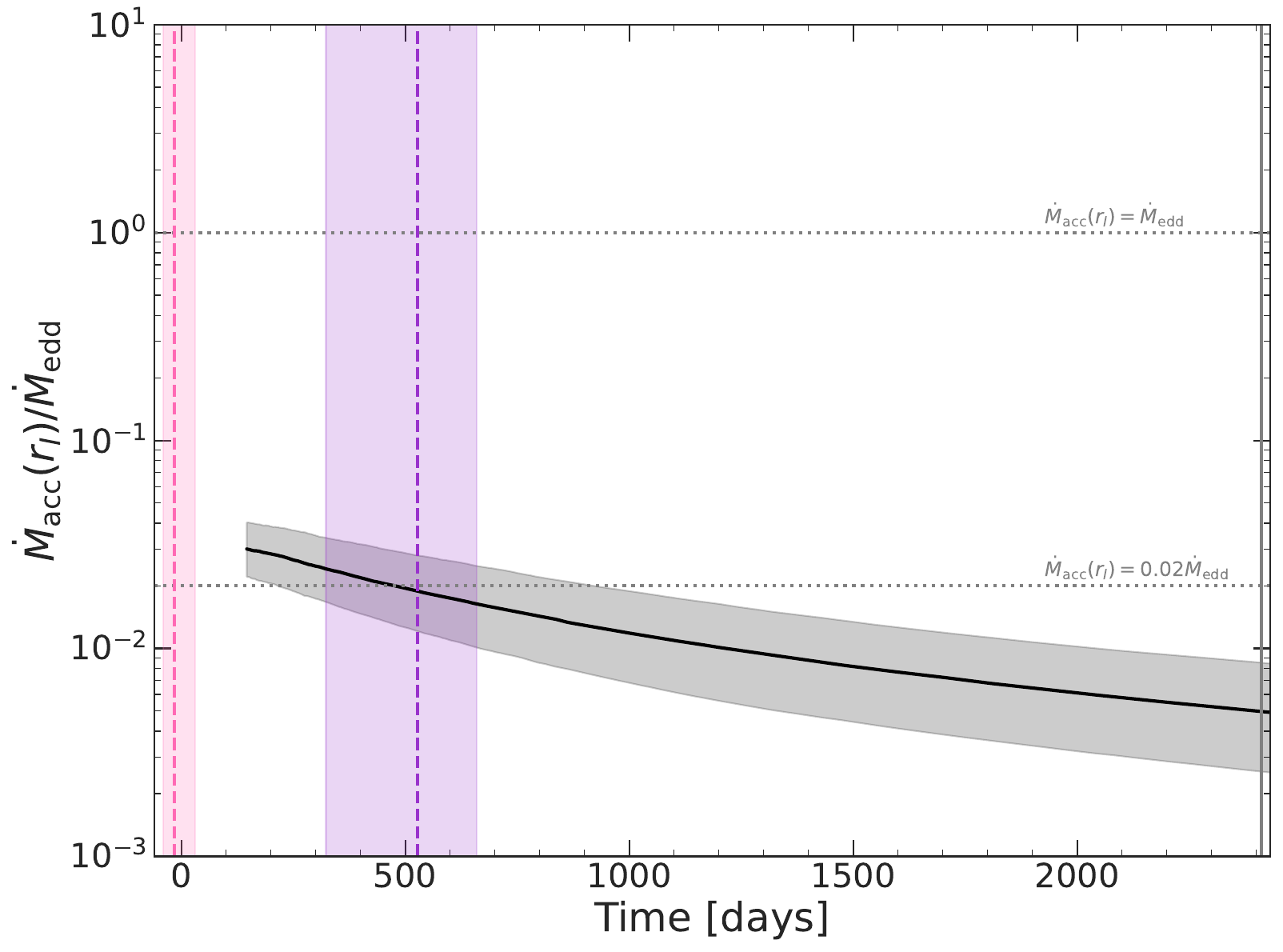}
    \includegraphics[width=0.4\linewidth]{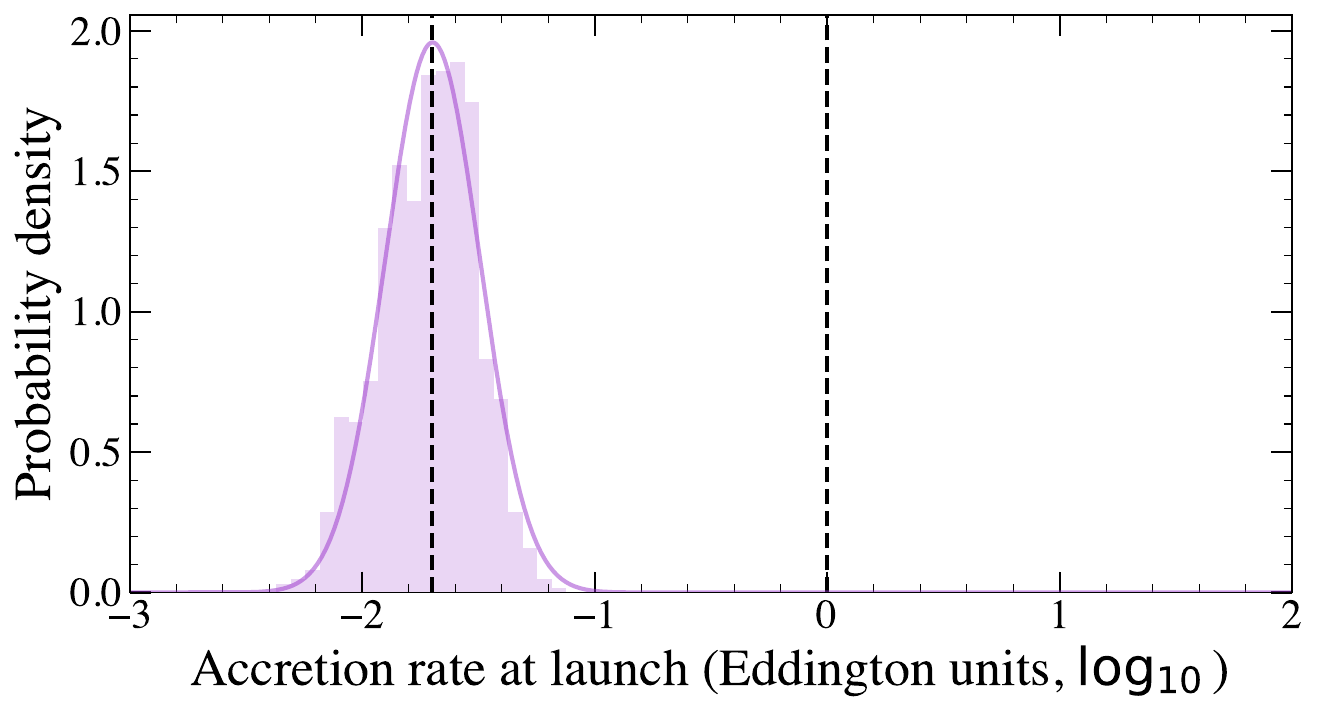}
    \caption{The same as Figure \ref{fig:models_14li} but for AT2019ahk. Accretion rate only plotted after times at which the disk model dominates the flux in optical/UV bands.  }
    \label{fig:models_4}
\end{figure}

\begin{figure}
    \centering
\includegraphics[width=0.3\linewidth]{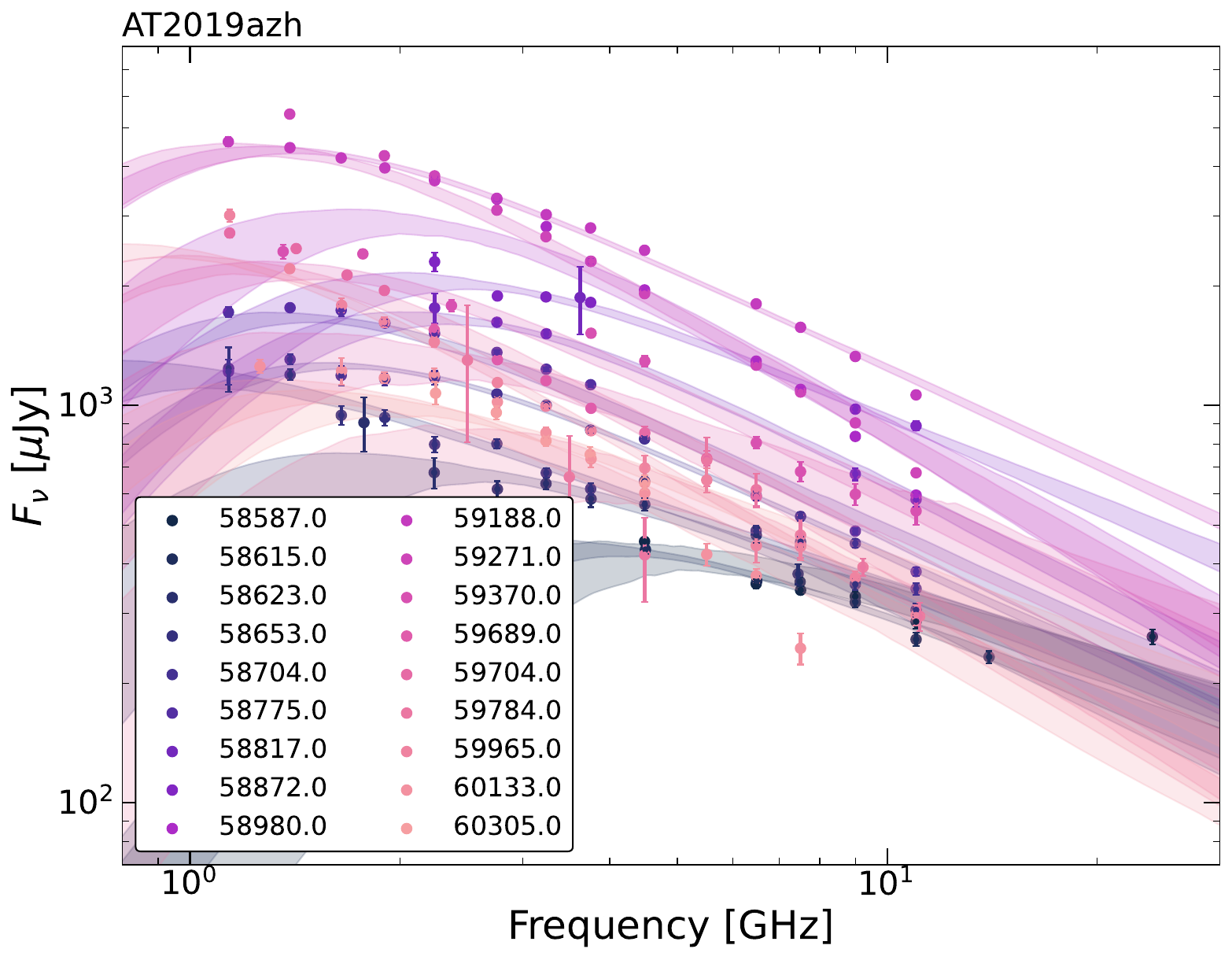}
    \includegraphics[width=0.3\linewidth]{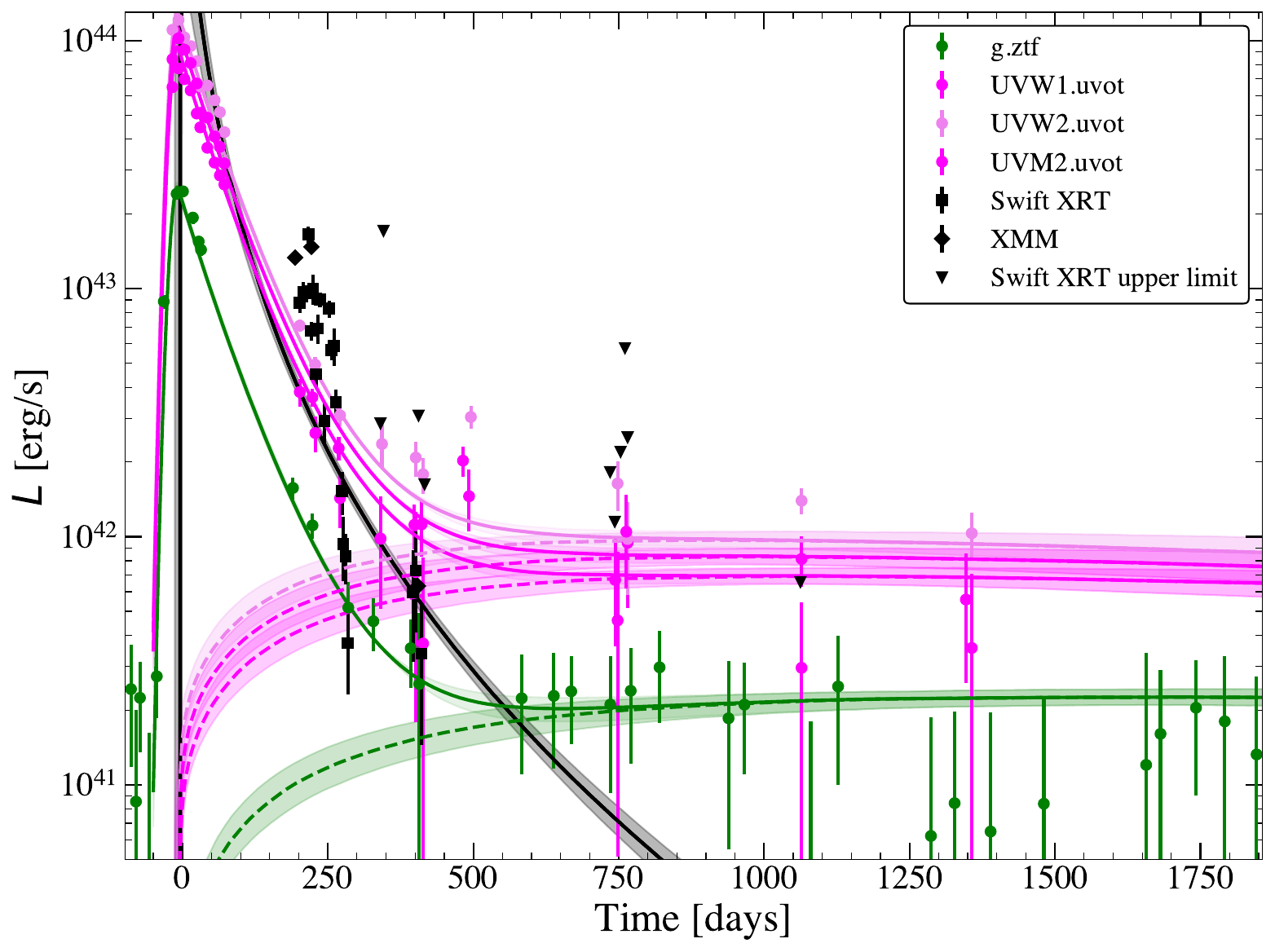}
    
    \includegraphics[width=0.3\linewidth]{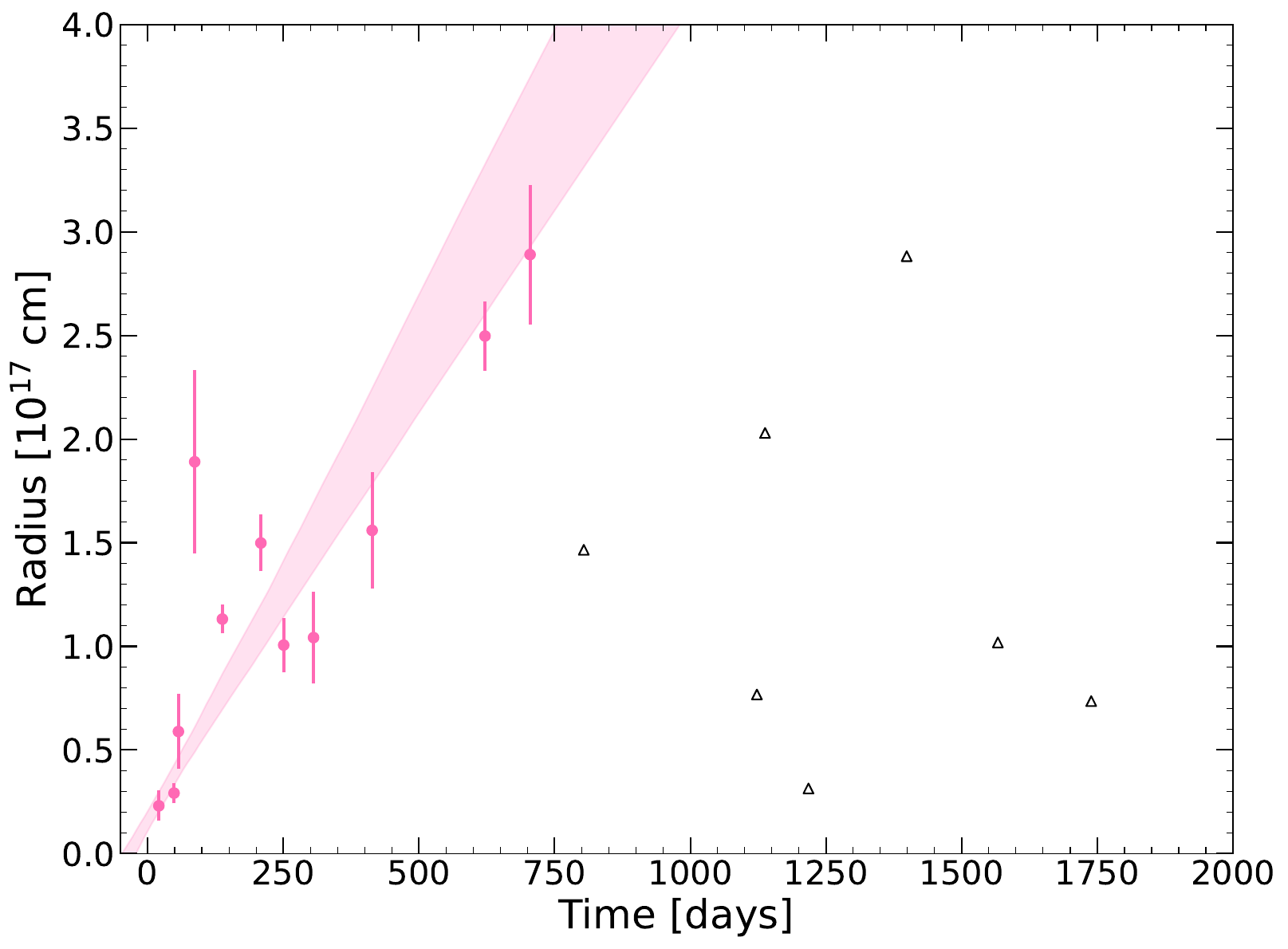}
    \includegraphics[width=0.3\linewidth]{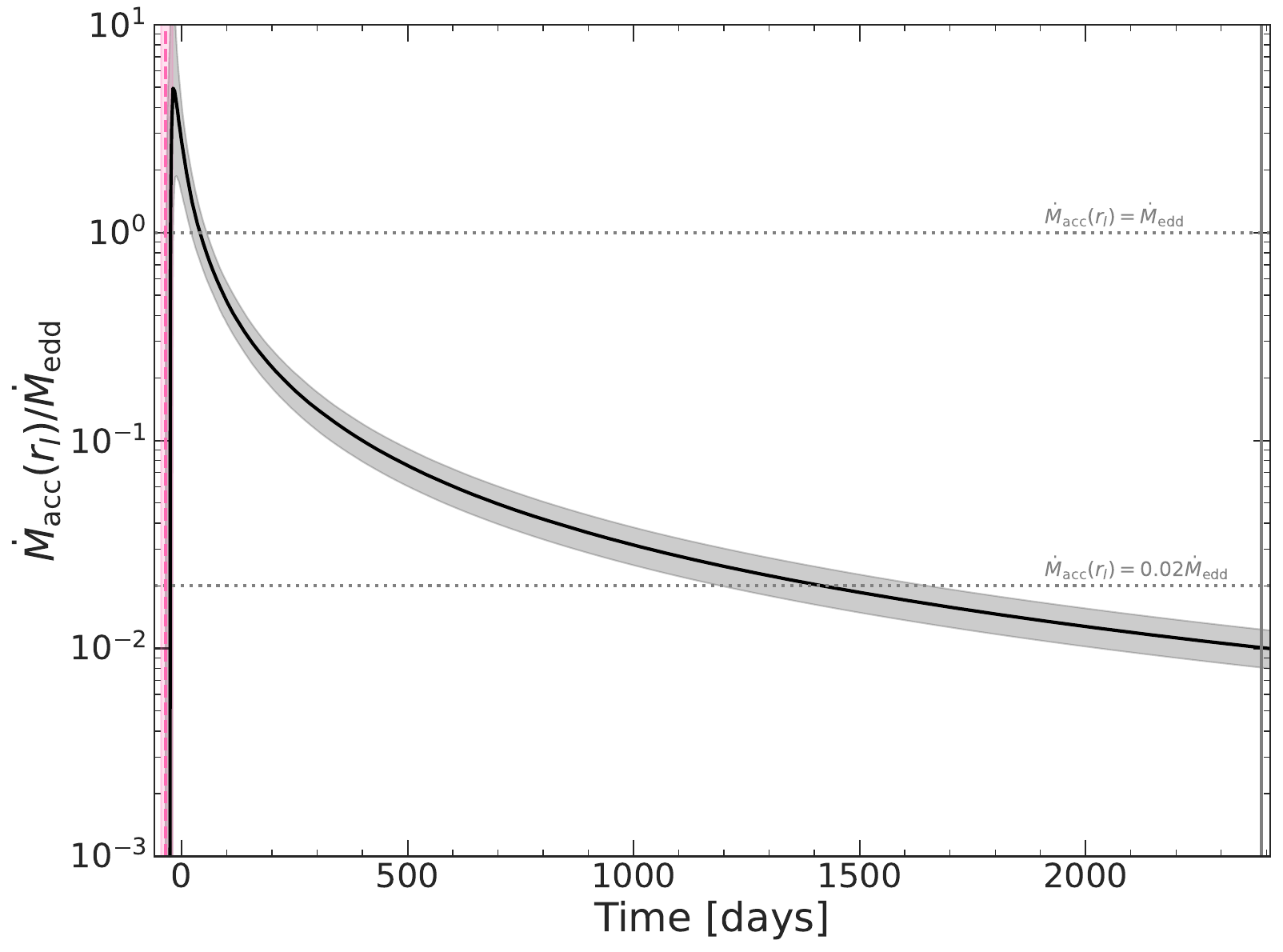}
    \includegraphics[width=0.4\linewidth]{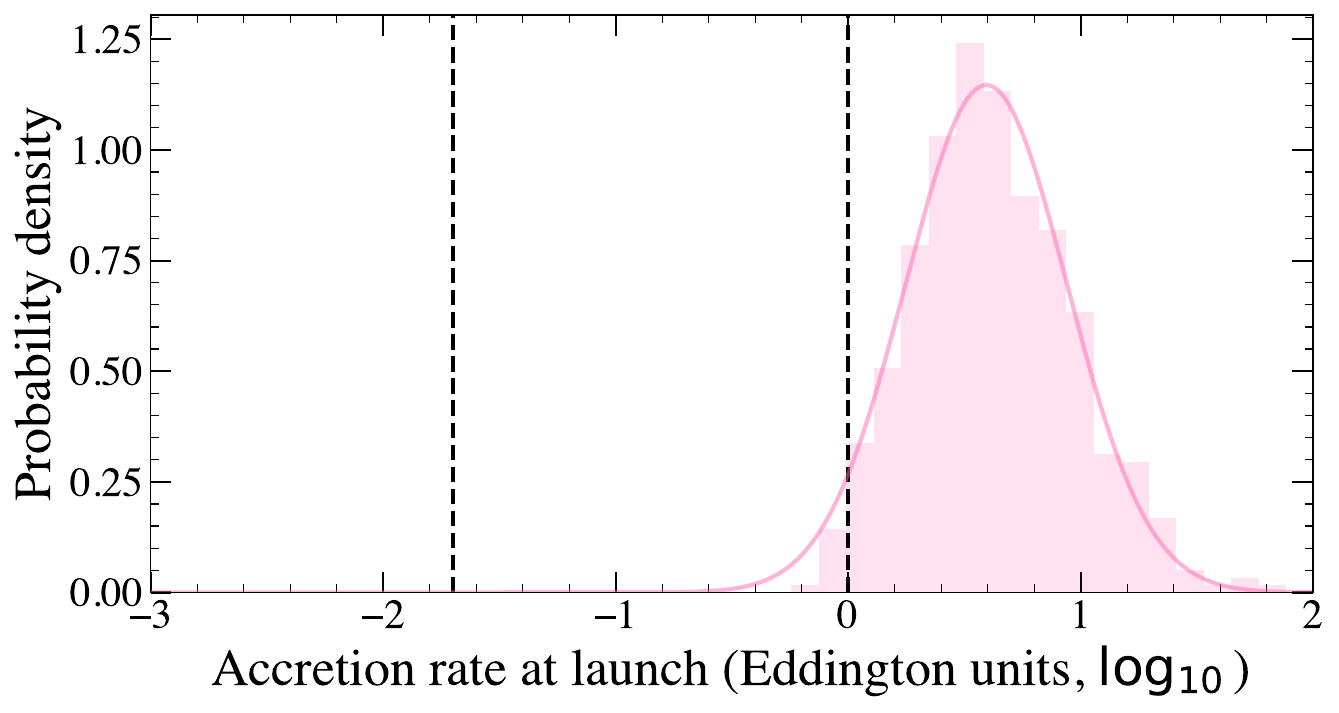}
    \caption{The same as Figure \ref{fig:models_14li} but for AT2019azh. }
    \label{fig:models_5}
\end{figure}

\begin{figure}
    \centering
\includegraphics[width=0.3\linewidth]{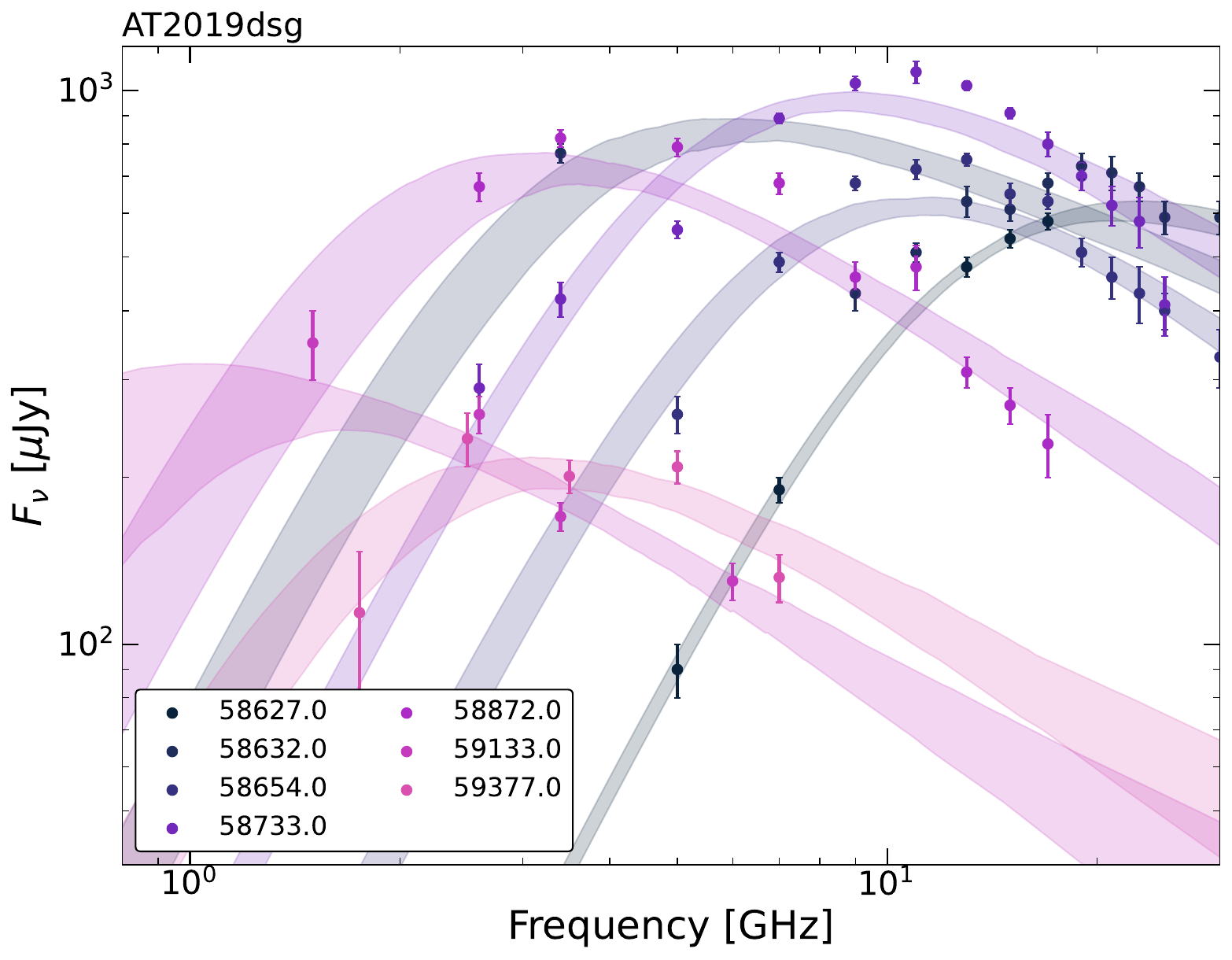}
    \includegraphics[width=0.3\linewidth]{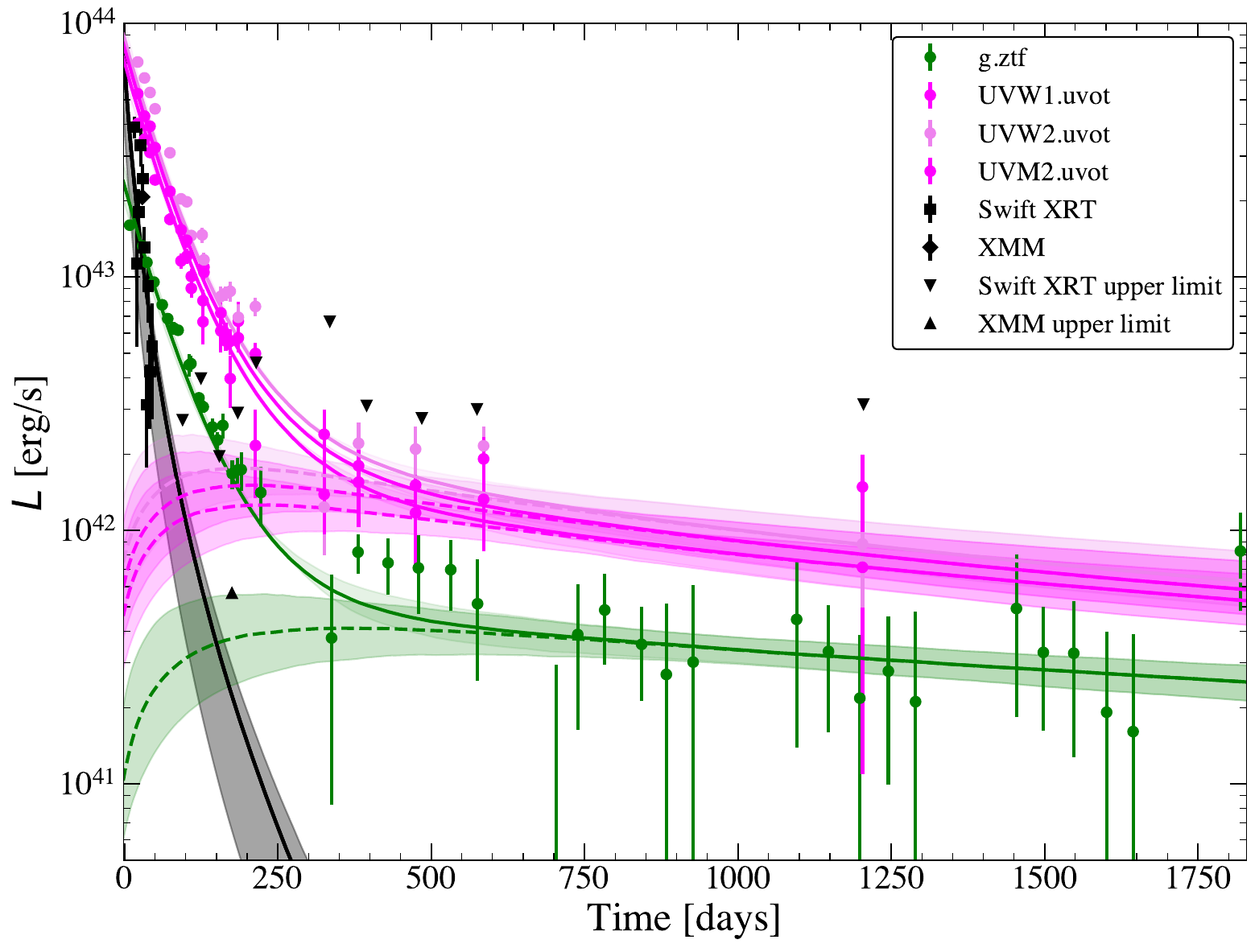}
    
    \includegraphics[width=0.3\linewidth]{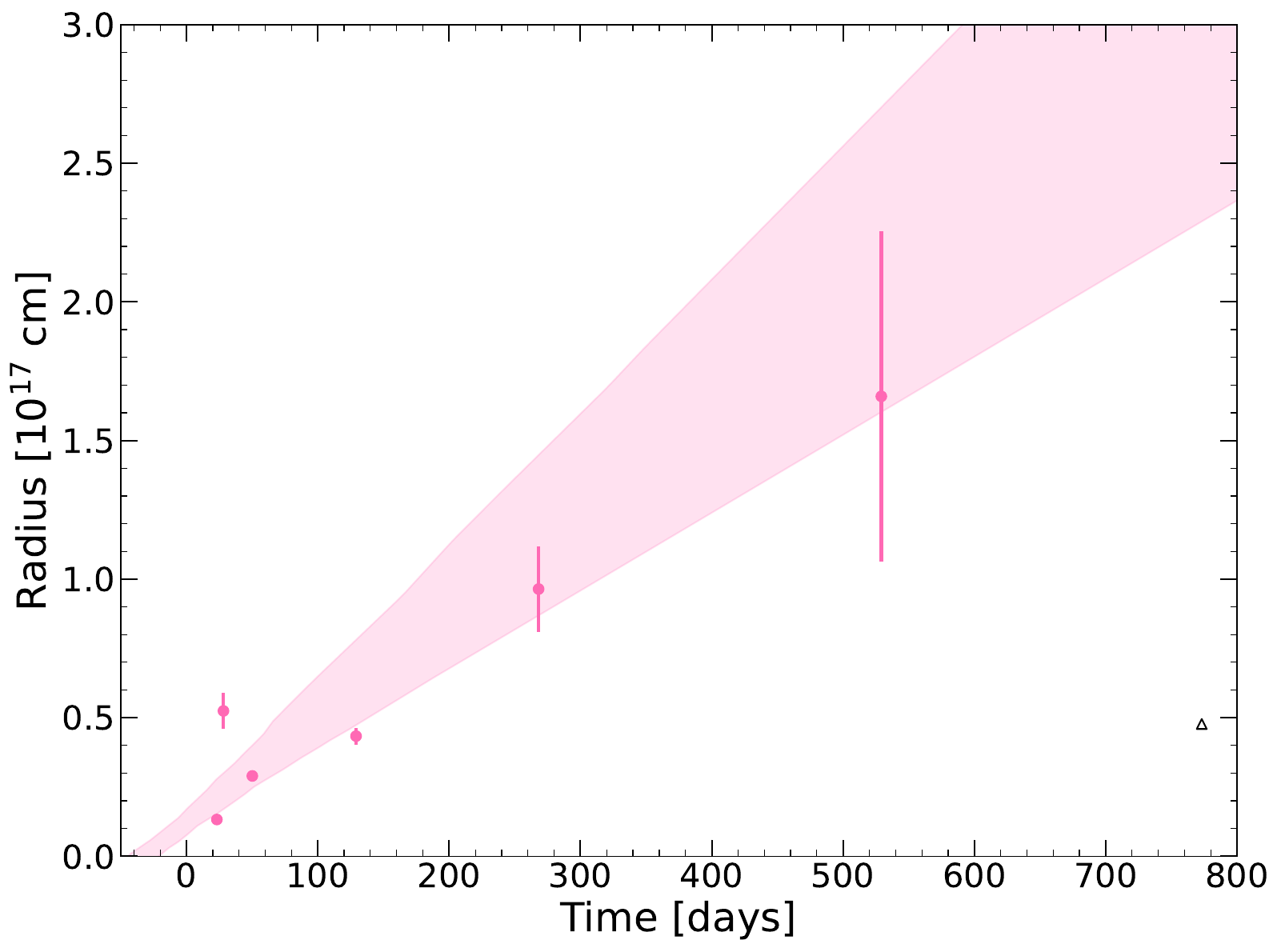}
    \includegraphics[width=0.3\linewidth]{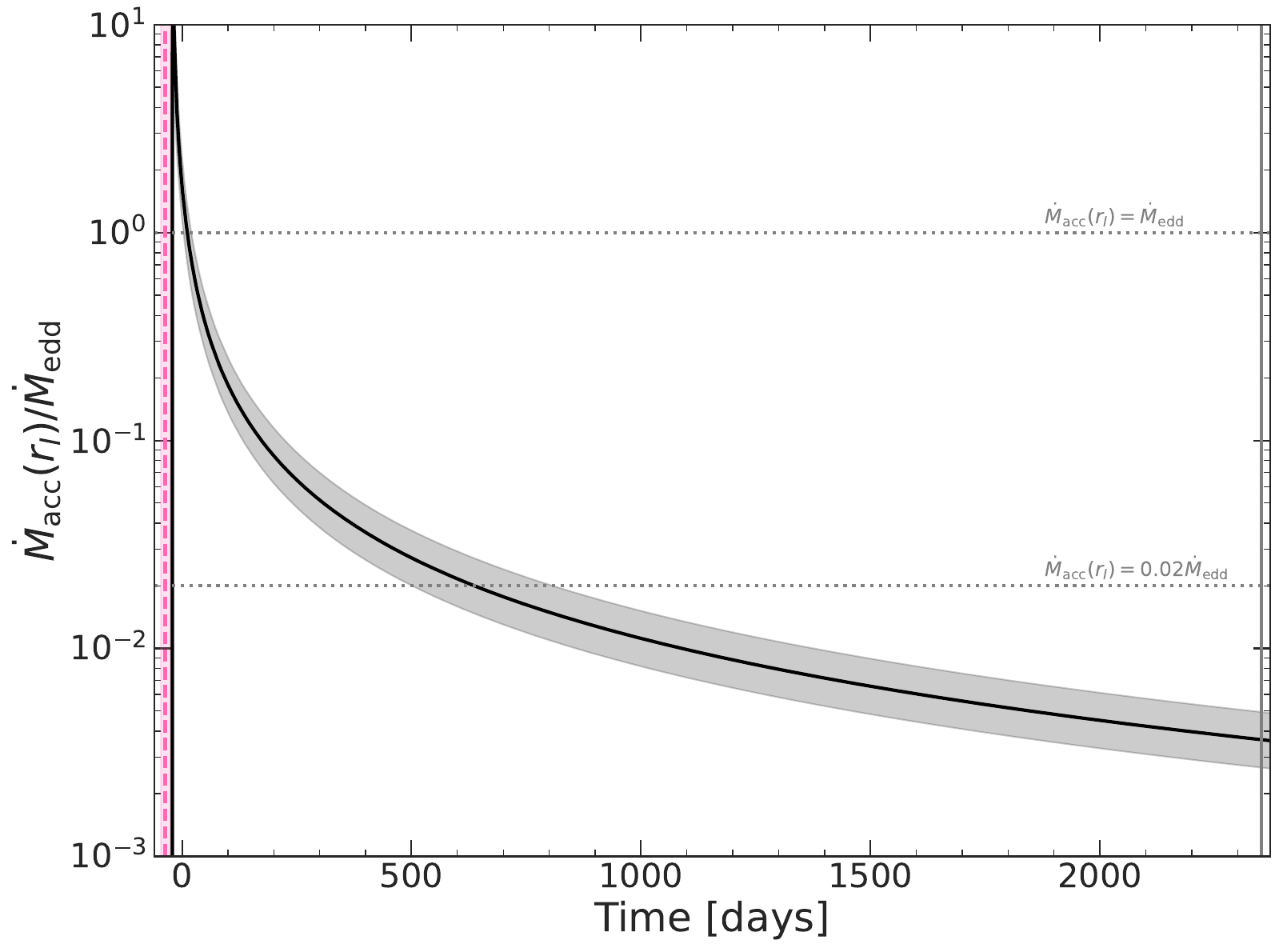}
    \includegraphics[width=0.4\linewidth]{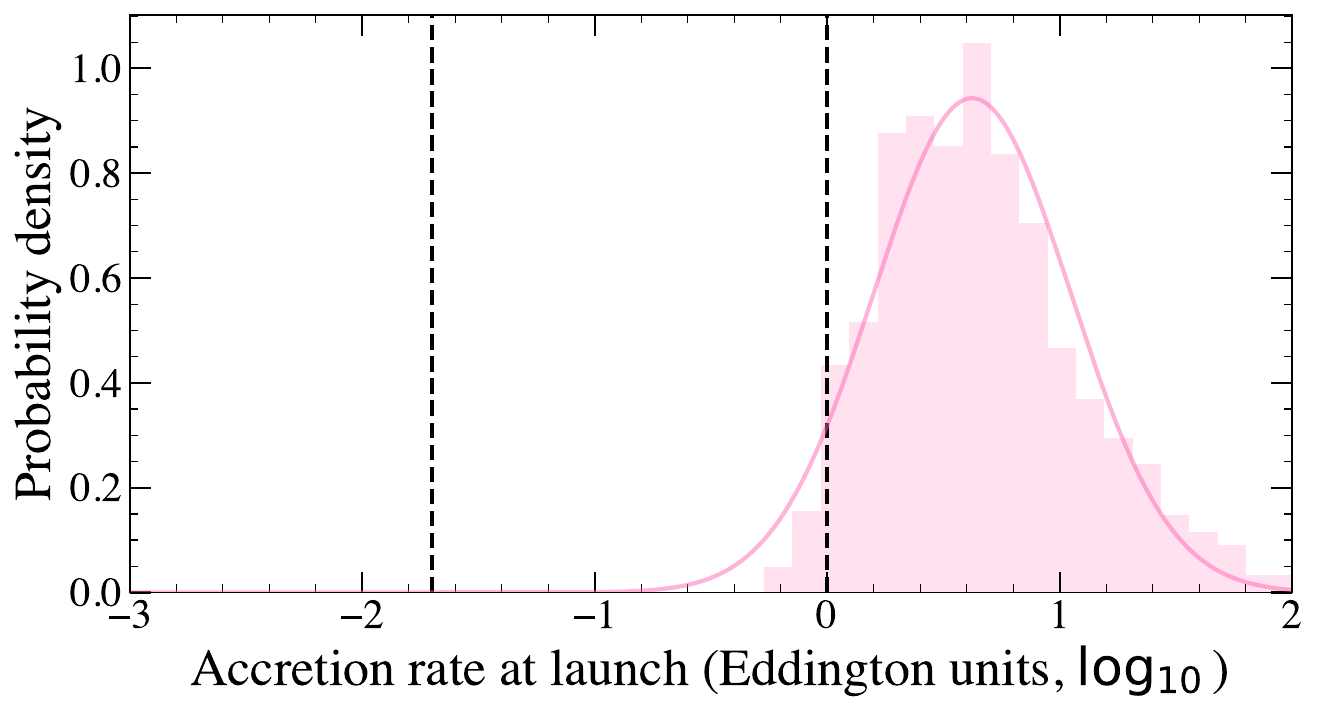}
    \caption{The same as Figure \ref{fig:models_14li} but for AT2019dsg.}
    \label{fig:models_6}
\end{figure}

\begin{figure}
    \centering
\includegraphics[width=0.3\linewidth]{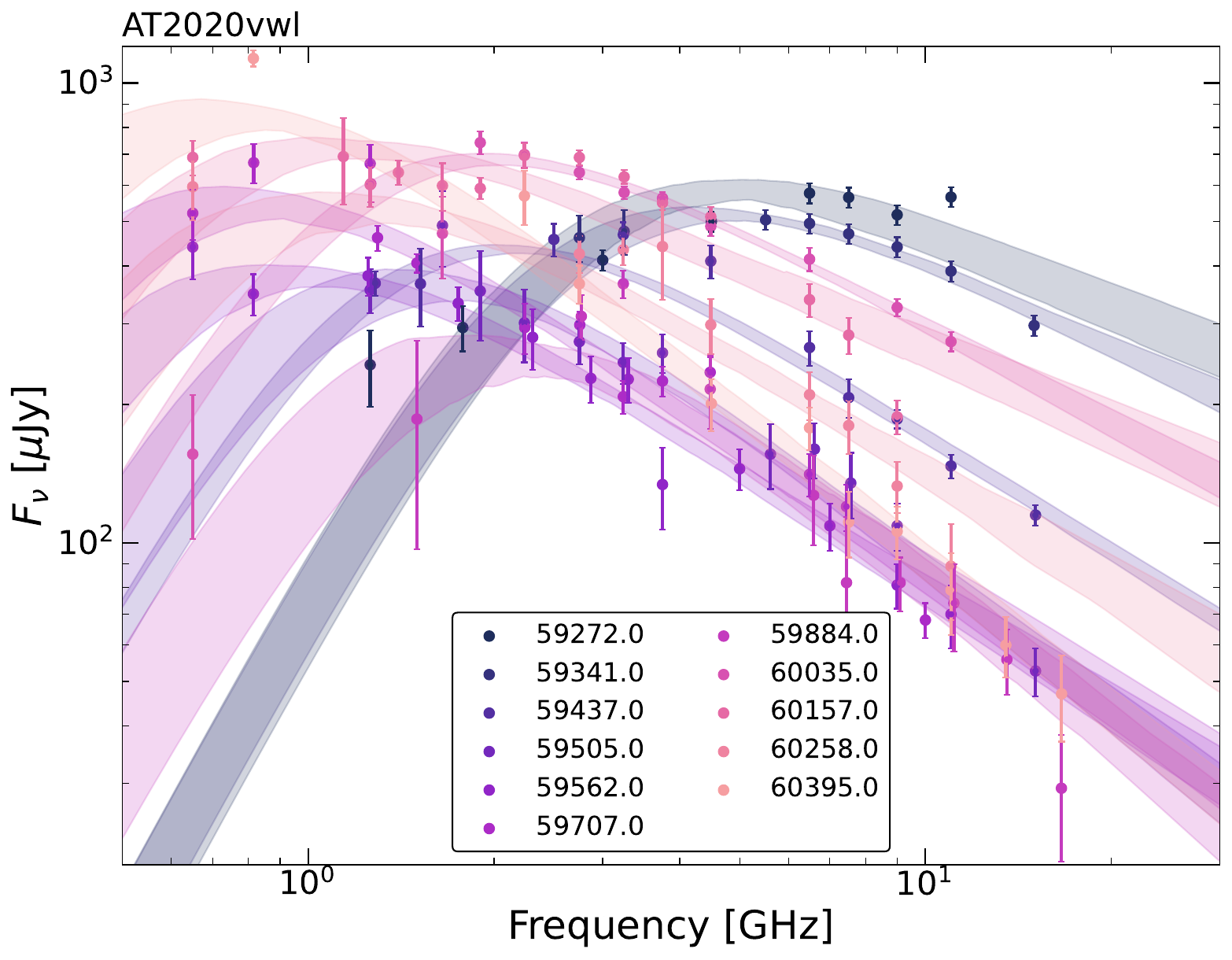}
    \includegraphics[width=0.3\linewidth]{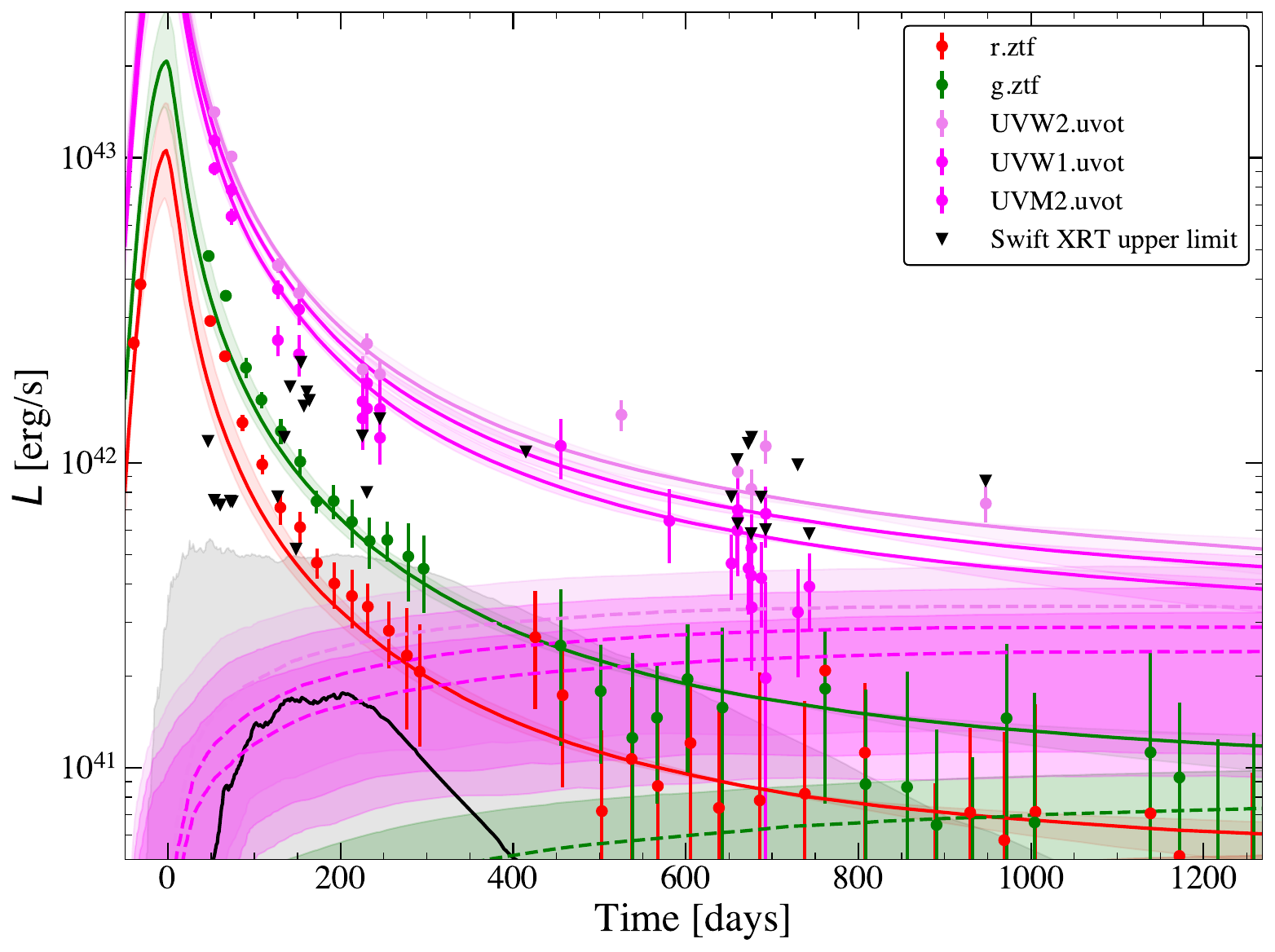}
    
    \includegraphics[width=0.3\linewidth]{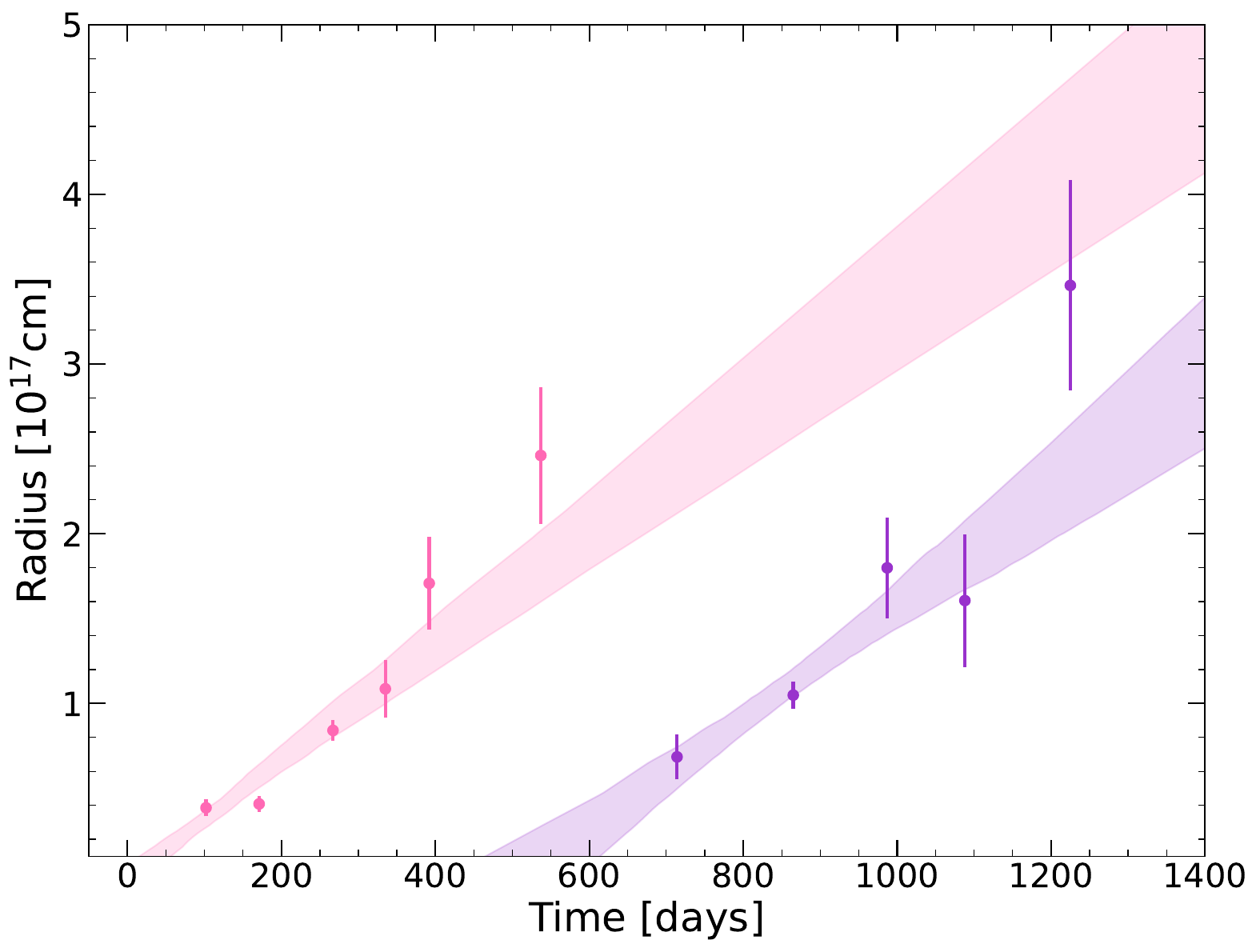}
    \includegraphics[width=0.3\linewidth]{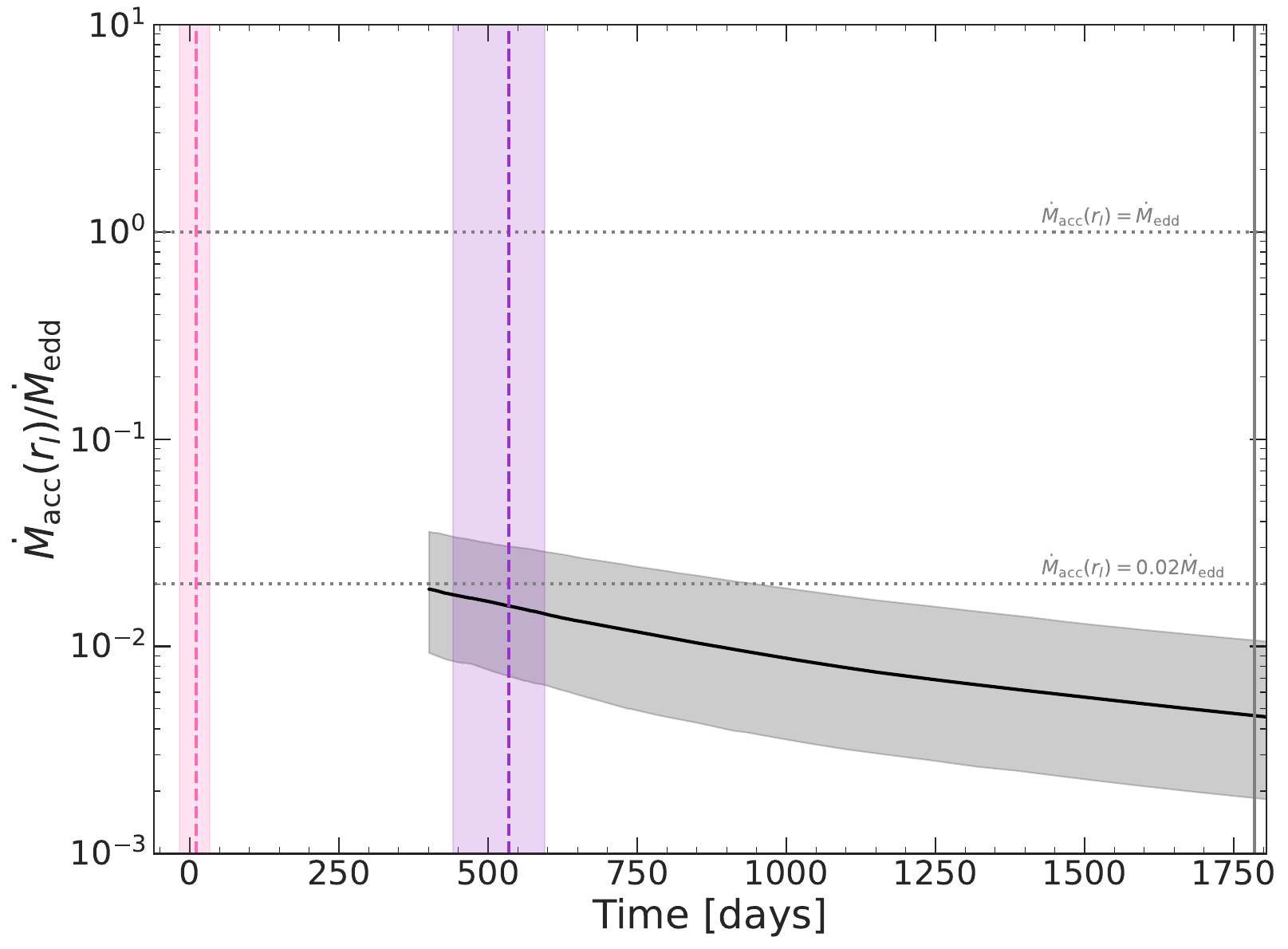}
    \includegraphics[width=0.4\linewidth]{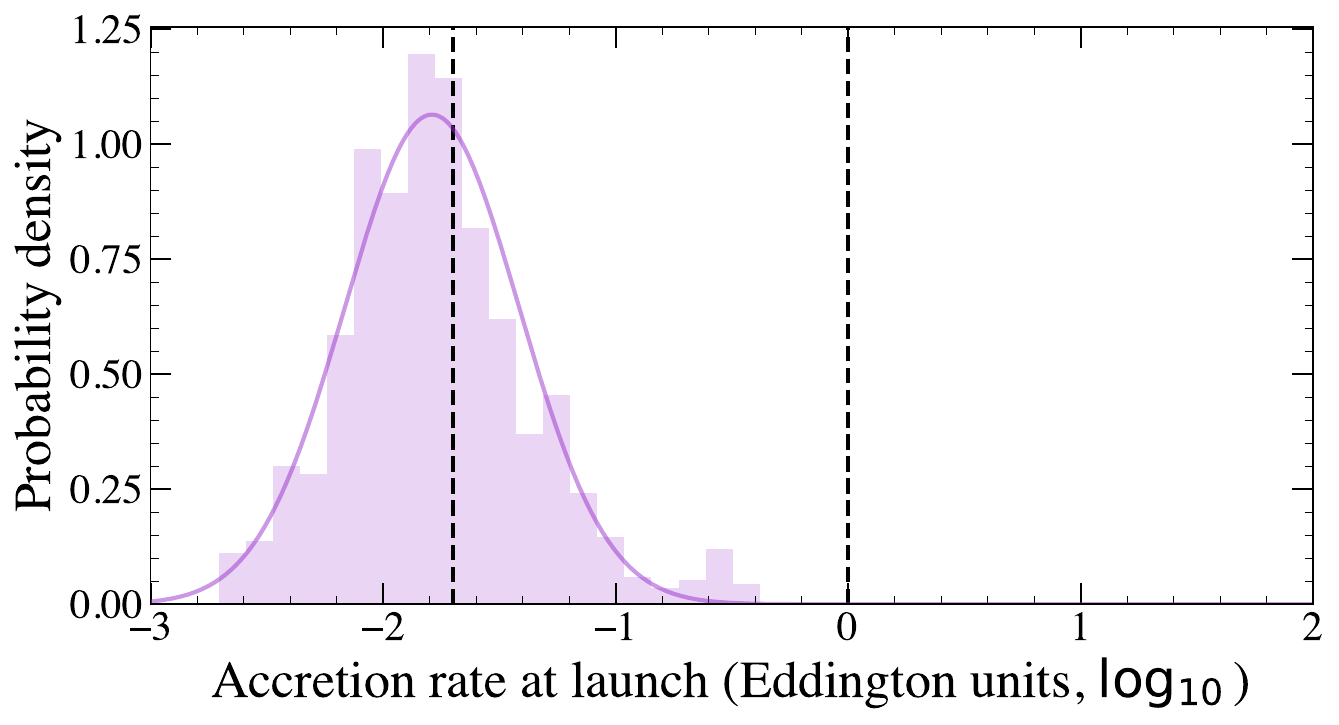}
    \caption{The same as Figure \ref{fig:models_14li} but for AT2020vwl. Accretion rate only plotted after times at which the disk model dominates the flux in optical/UV bands. }
    \label{fig:models_7}
\end{figure}

\begin{figure}
    \centering
\includegraphics[width=0.3\linewidth]{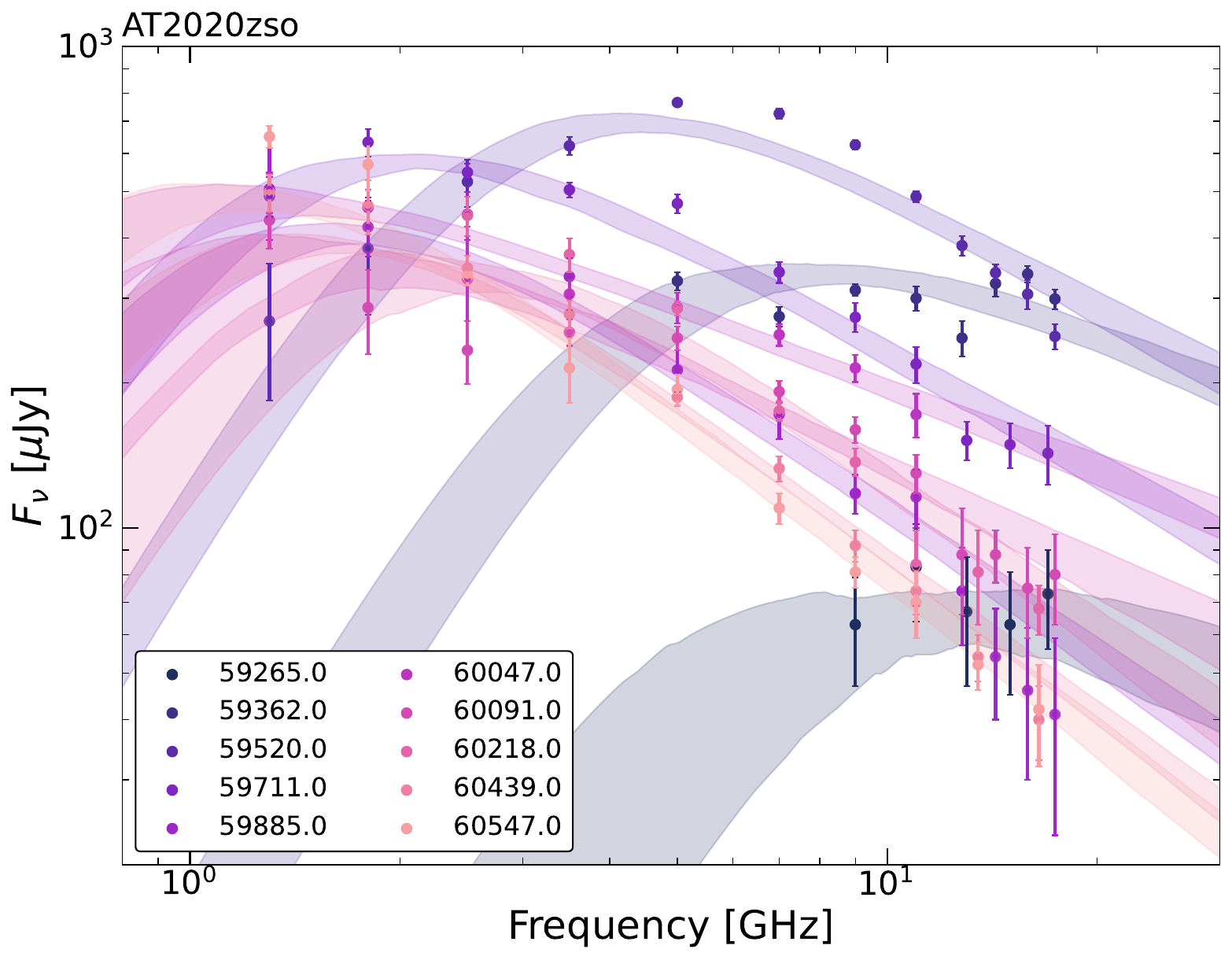}
    \includegraphics[width=0.3\linewidth]{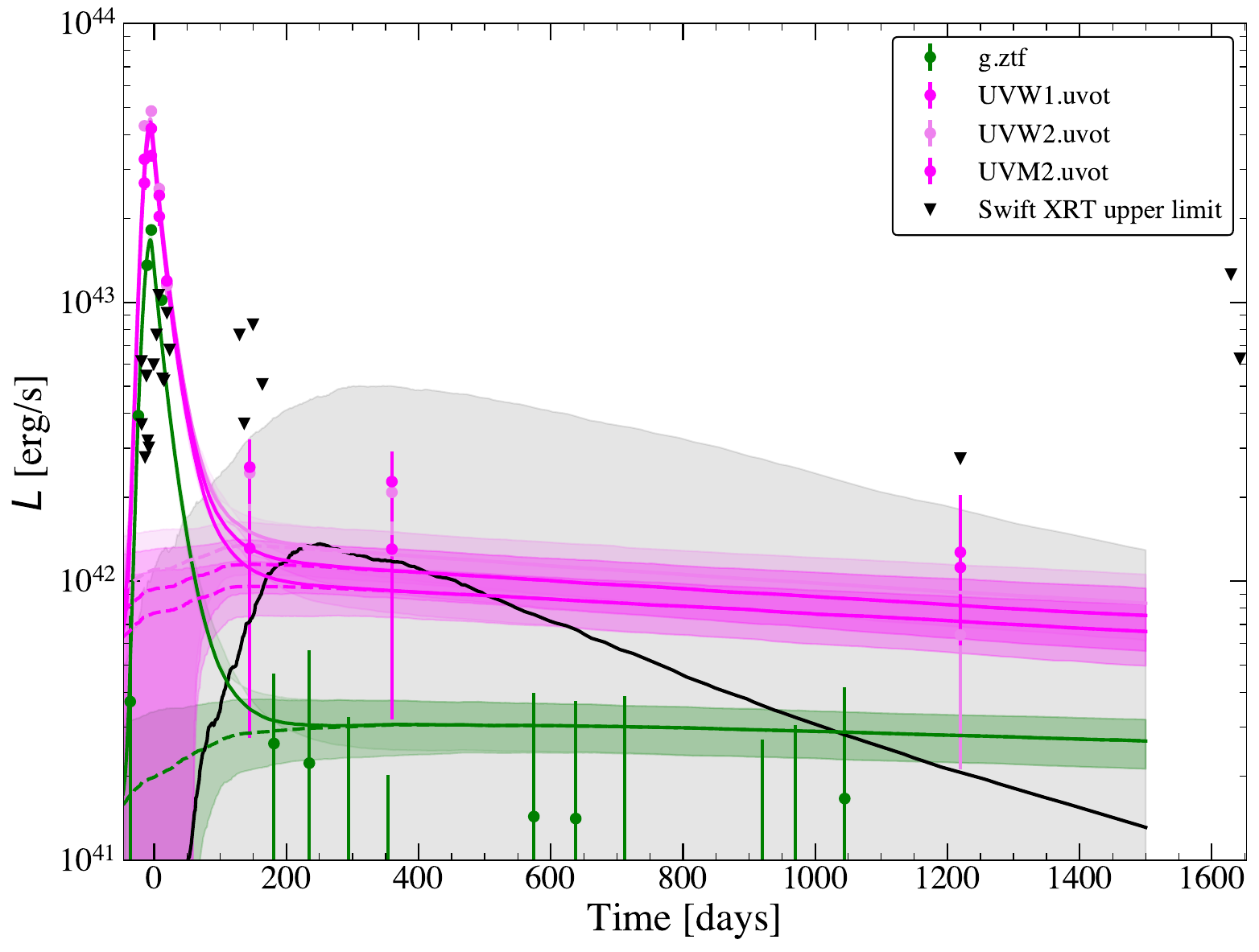}
    
    \includegraphics[width=0.3\linewidth]{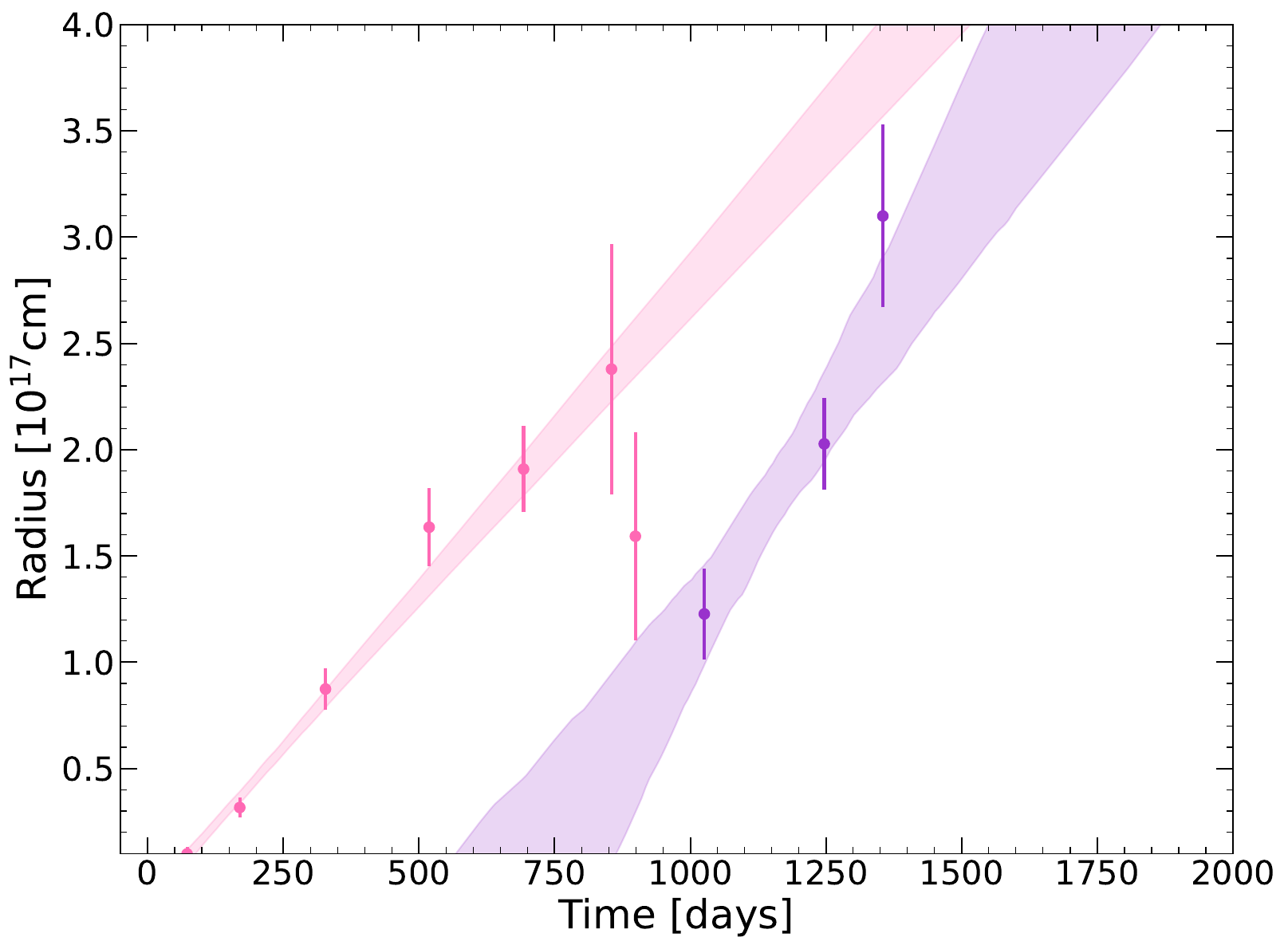}
    \includegraphics[width=0.3\linewidth]{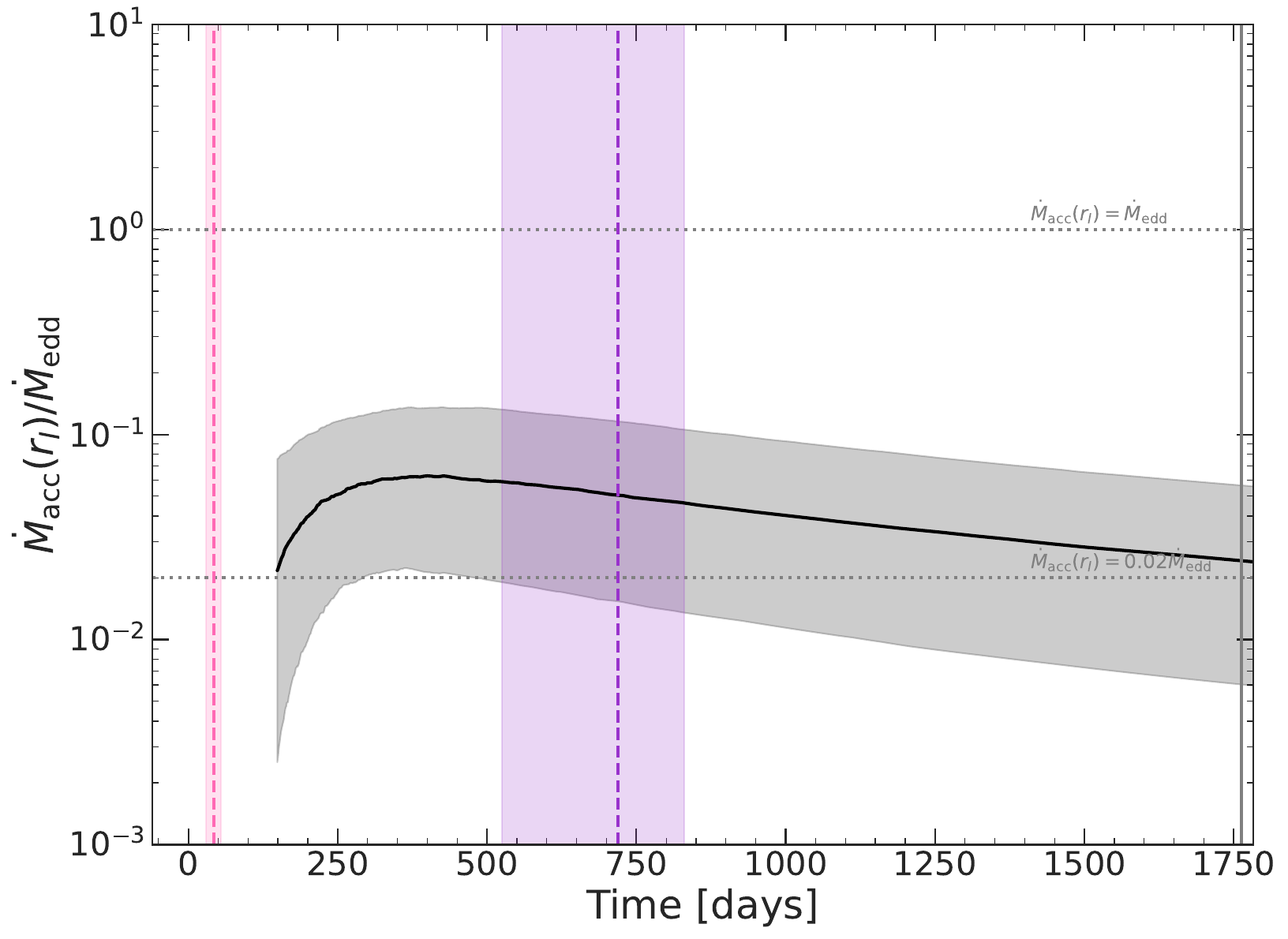}
    \includegraphics[width=0.4\linewidth]{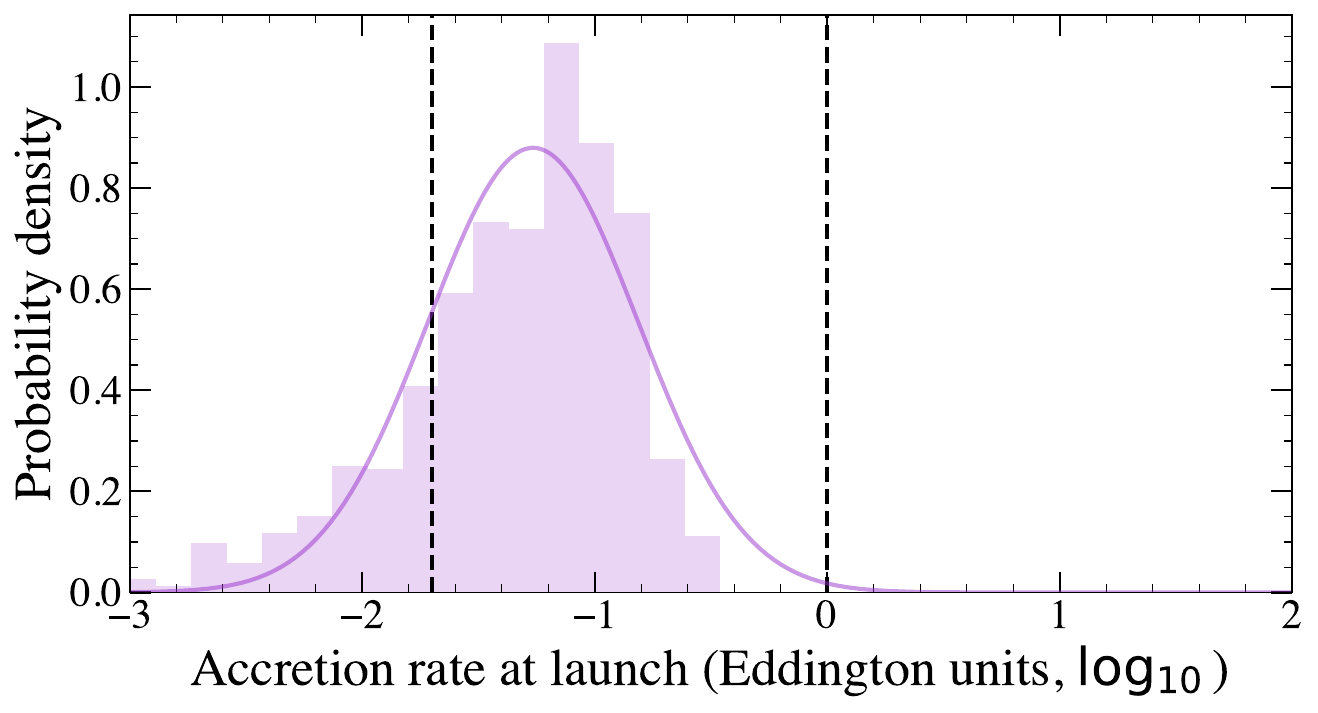}
    \caption{The same as Figure \ref{fig:models_14li} but for AT2020zso. Accretion rate only plotted after times at which the disk model dominates the flux in optical/UV bands.  }
    \label{fig:models_8}
\end{figure}

\begin{figure}
    \centering
\includegraphics[width=0.3\linewidth]{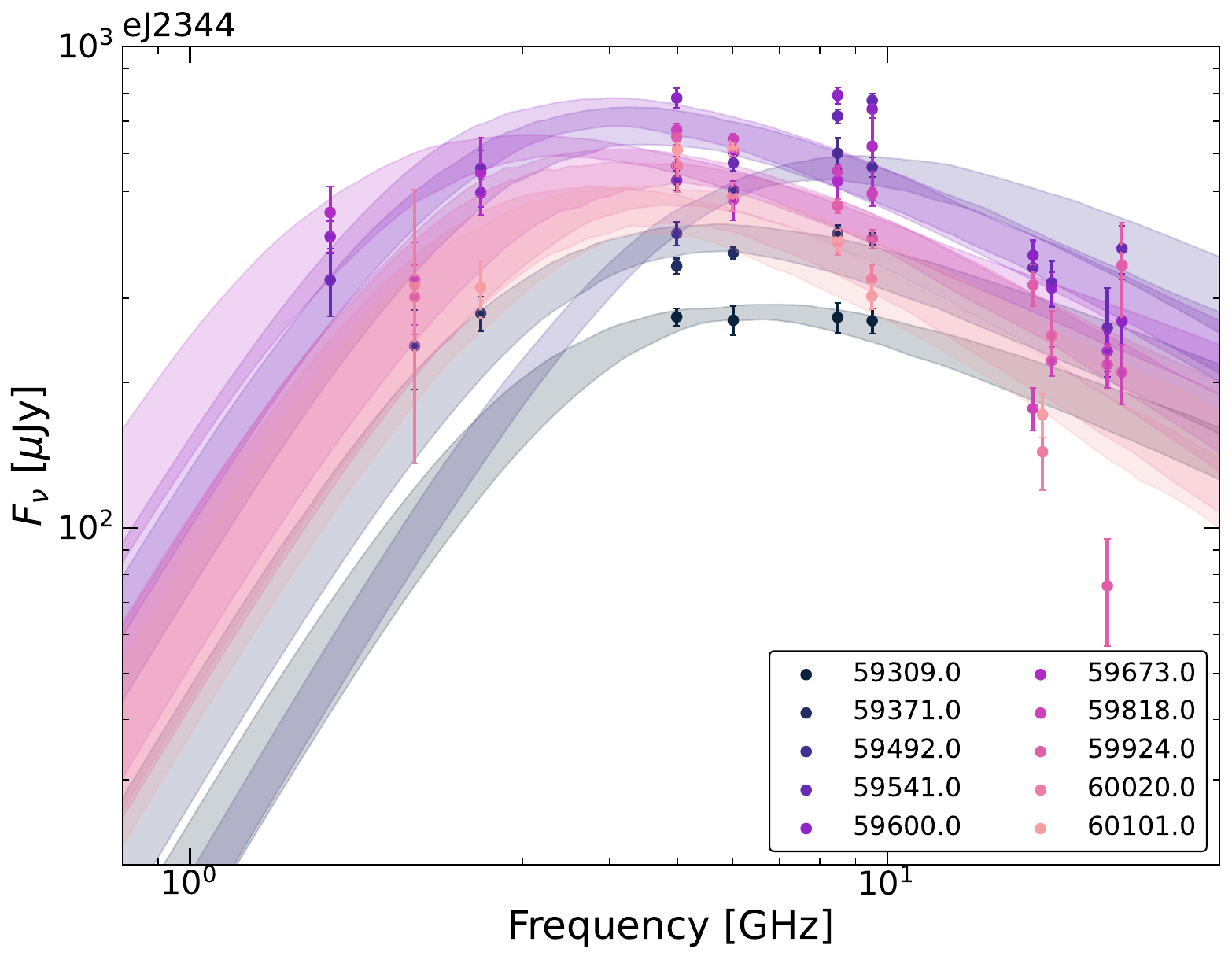}
    \includegraphics[width=0.3\linewidth]{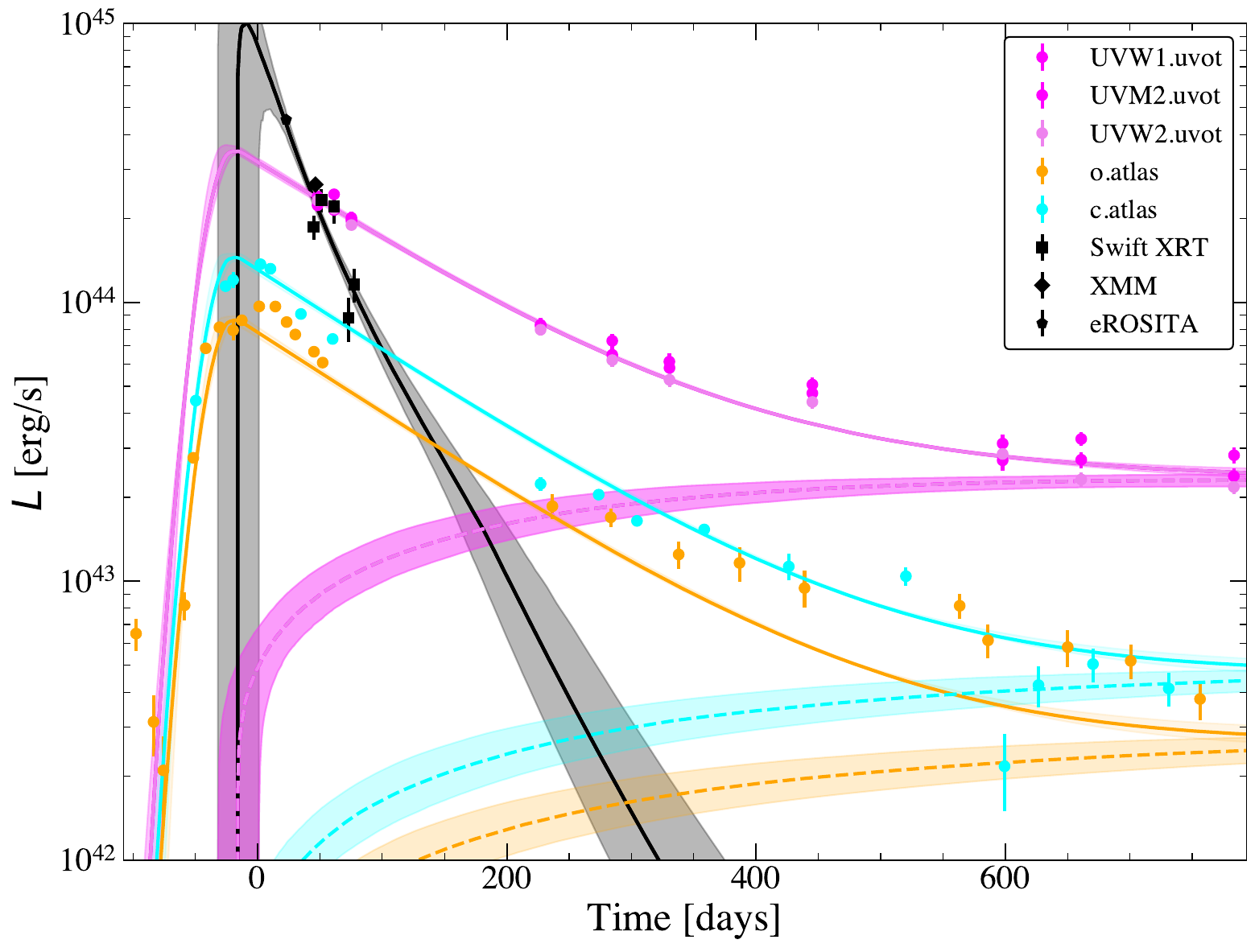}
    
    \includegraphics[width=0.3\linewidth]{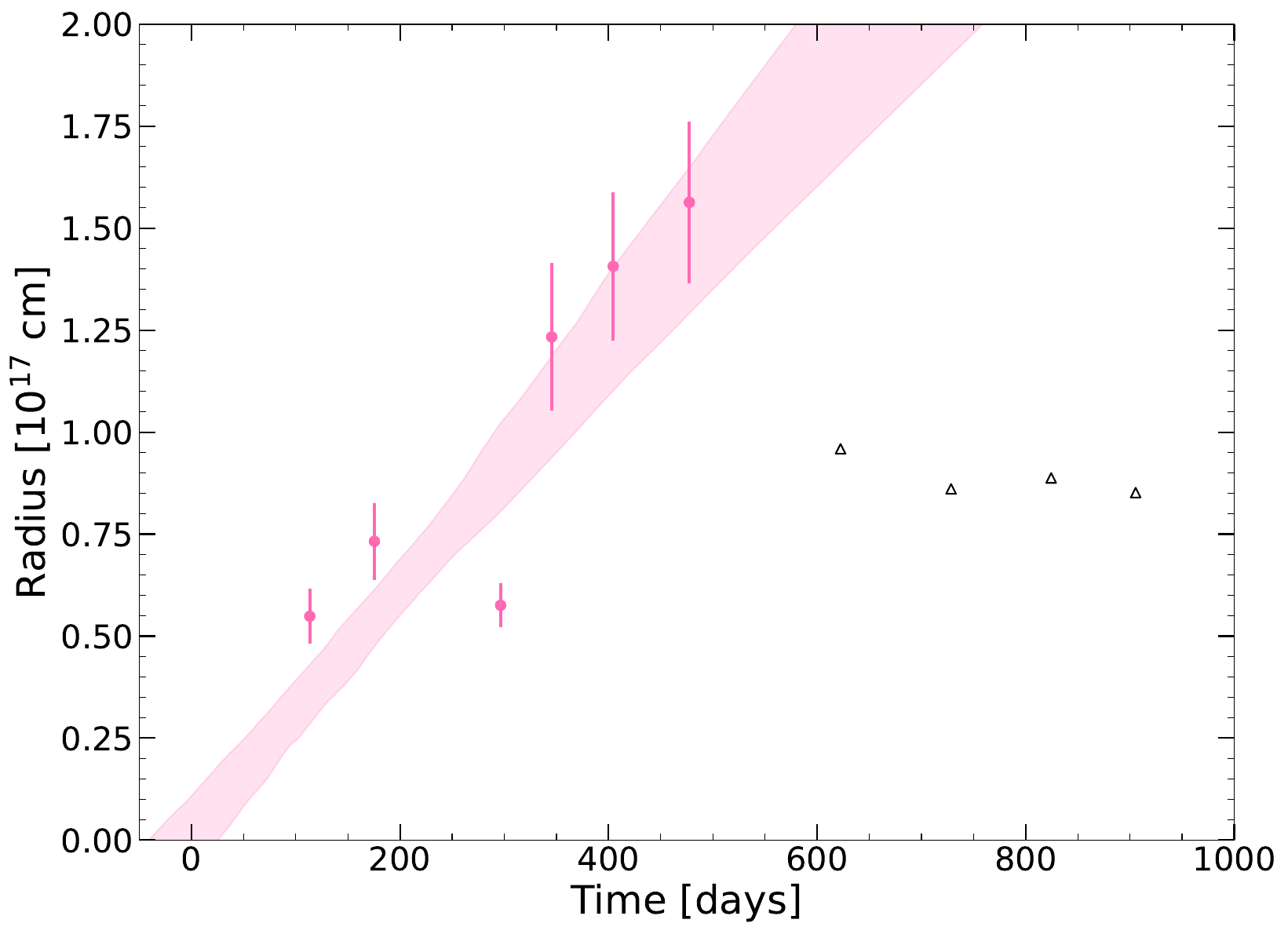}
    \includegraphics[width=0.3\linewidth]{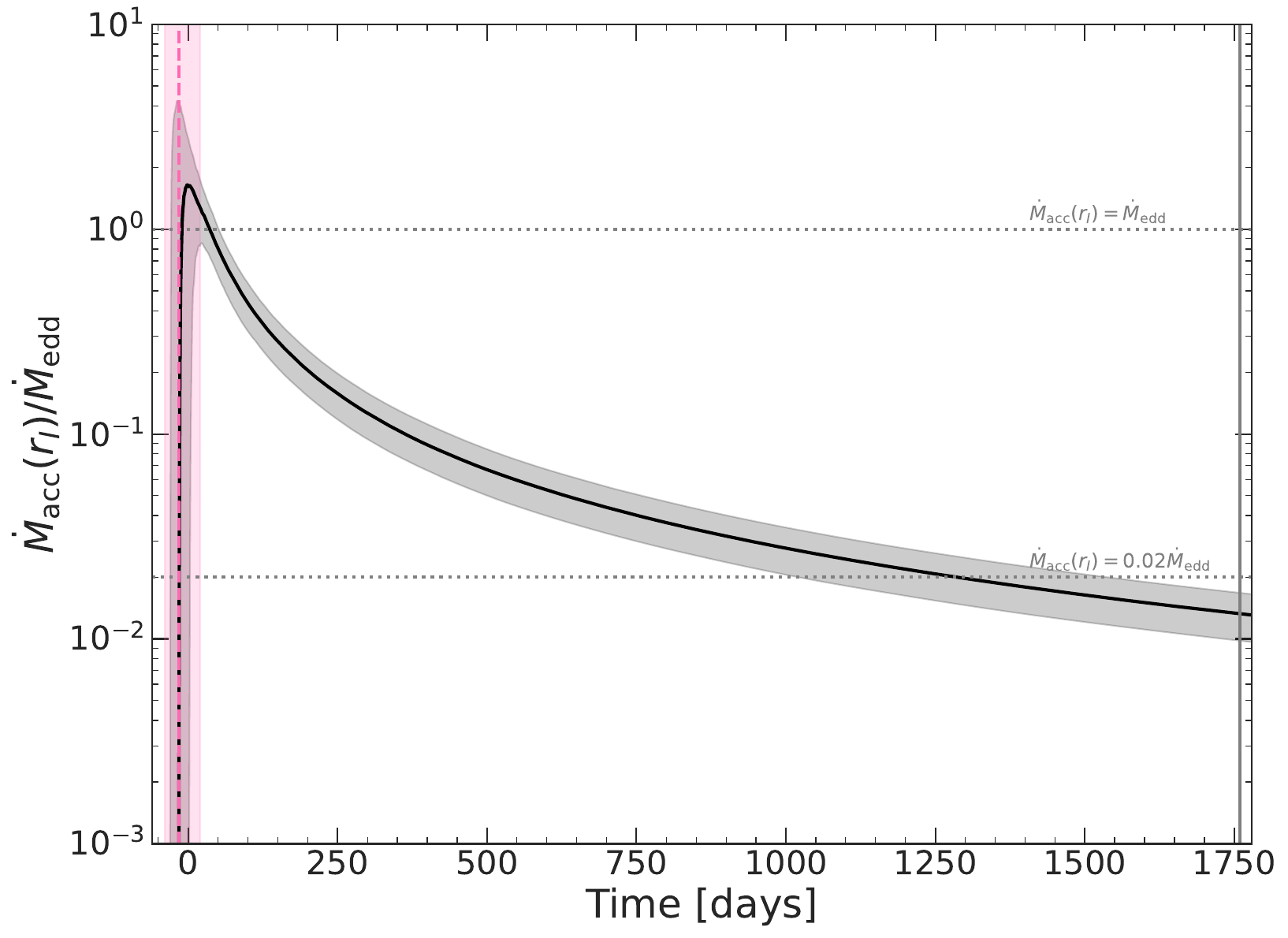}
    \includegraphics[width=0.4\linewidth]{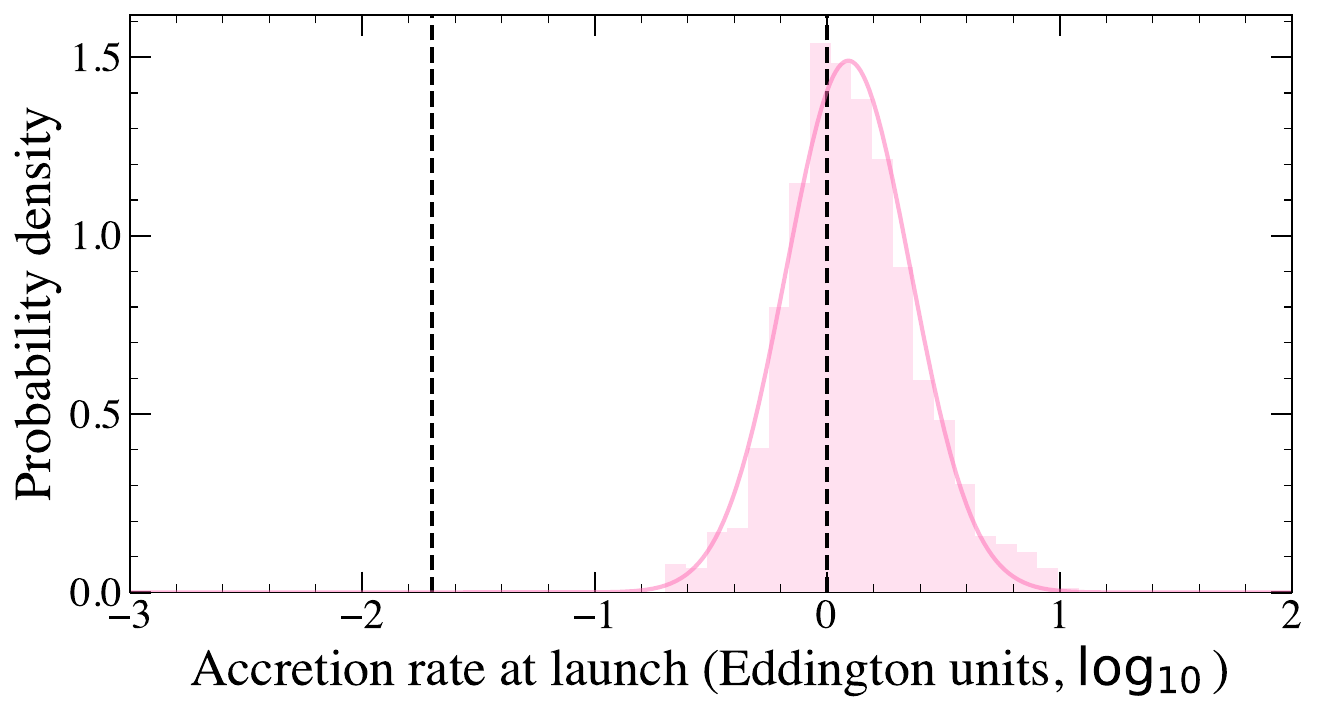}
    \caption{The same as Figure \ref{fig:models_14li} but for eJ2344. }
    \label{fig:models_9}
\end{figure}

We wish to compare the mass accretion rate (at the disk's inner edge) with the Eddington mass accretion rate. We define the Eddington mass accretion rate as 
\begin{equation}
    \dot M_{\rm edd} \equiv {L_{\rm edd} \over \eta c^2 }, 
\end{equation}
where 
\begin{equation}
    \eta(a_\bullet) = 1 - \left(1 - {2r_g\over 3r_I}\right)^{1/2}, 
\end{equation}
is the usual radiative efficiency of a steady-state black hole accretion flow as a function of black hole spin. The ISCO radius is computed in the standard manner following \cite{Bardeen72}. We define the mass accretion rate Eddington ratio as 
\begin{equation}
    \dot m \equiv {\dot M(r_I, t) \over \dot M_{\rm edd}}. 
\end{equation}
Again, for a given posterior distribution of disk and black hole parameters, we can construct a time-dependent posterior distribution of $\dot m$ by repeated sampling. 
The individual accretion rate profiles for each of the ``gold" TDEs are shown in Figures \ref{fig:models_1}--\ref{fig:models_9}. 

\section{Population-level sanity-checks of fit}
While for all 10 of our ``gold'' sources the disk models are converged (at the very least at late times), it is important to test that the parameters recovered from these fits reproduce population-level correlations which would be expected for a sample of TDE systems. This provides confidence that the fitted parameters are reliable.

\begin{figure}
    \centering
    \includegraphics[width=0.8\linewidth]{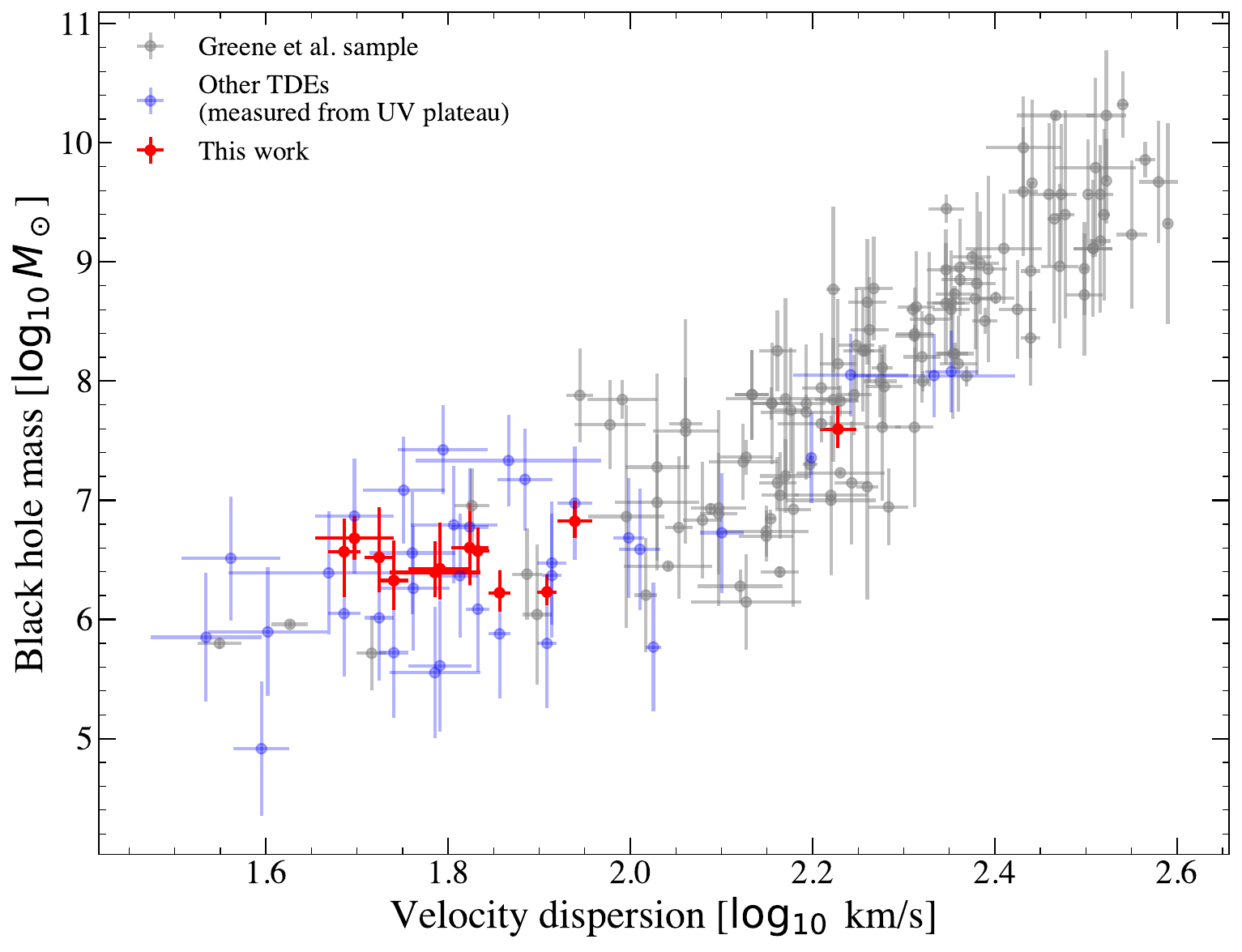}
    \includegraphics[width=0.8\linewidth]{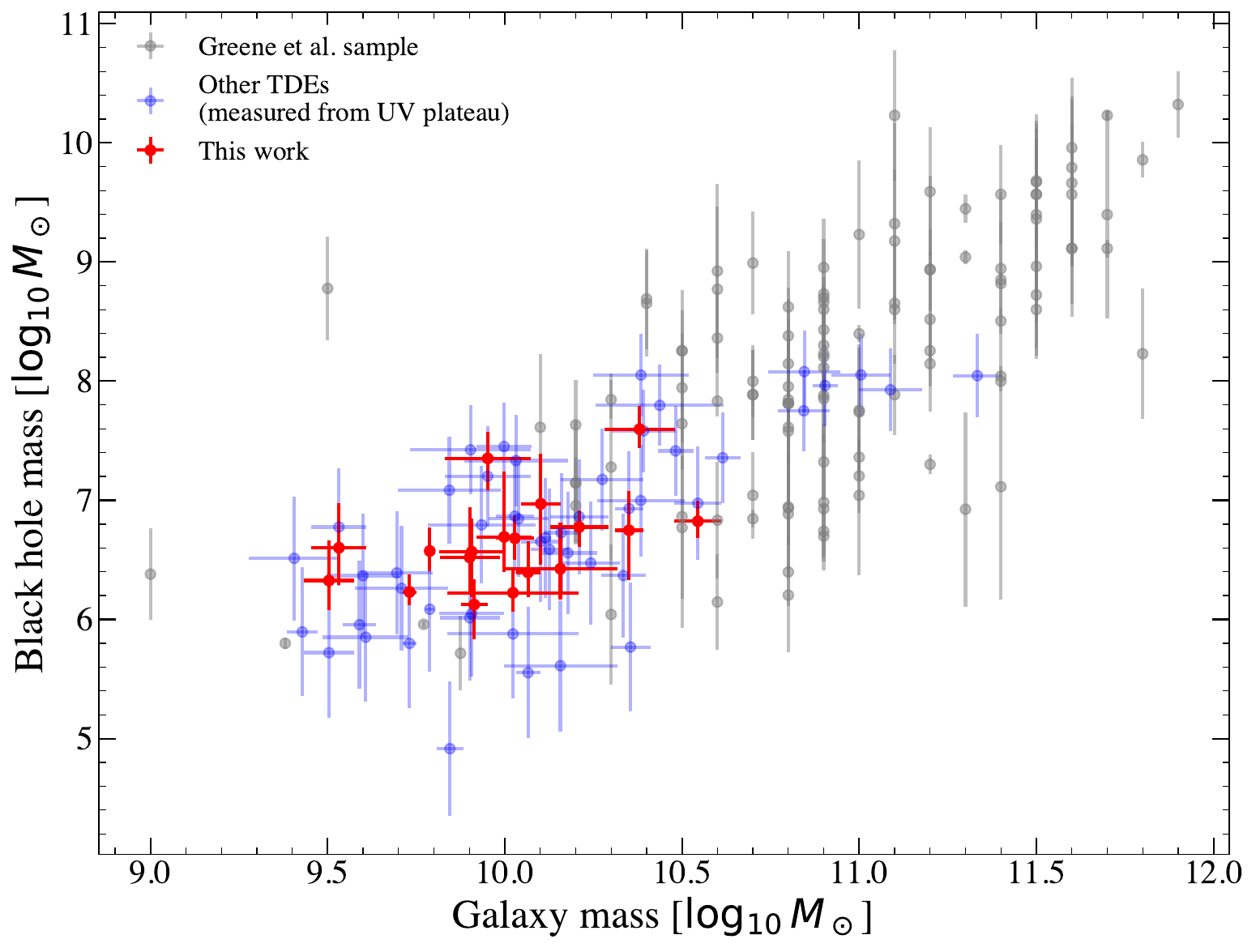}
    \caption{The correlations of the black hole masses we infer in this work (red points) with host velocity dispersion (upper panel) and galaxy mass (lower panel). Also shown for reference is the large sample of Greene et al. \cite{Greene20} (grey points), and the larger TDE sample of Mummery et al. \cite{Mummery_et_al_2024} (light blue points). We recover statistically significant correlations ($p$-values $< 0.05$) between black hole mass as inferred in this work and both galaxy properties, despite the relatively small sample size. Our results are consistent with, but more precise than, measurements made with only the late time UV emission from the TDE plateau. }
    \label{fig:galaxy}
\end{figure}

There are two natural population-level tests that can be performed. The first is that TDE disk modelling recovers known galactic scaling relationships between the black hole mass in the event and the mass of the host galaxy/the velocity dispersion of the host galaxy. We show the results of these tests in Figure \ref{fig:galaxy}, where we plot the black hole masses we infer in this work (red points) with host velocity dispersion (upper panel) and galaxy mass (lower panel). Also shown for reference is the large sample of Greene et al. \cite{Greene20} (grey points), and the larger TDE sample of Mummery et al. \cite{Mummery_et_al_2024} (light blue points). We recover statistically significant correlations ($p$-values $< 0.05$ for both the Kendall $\tau$ and Spearman rank correlation tests) between black hole mass as inferred in this work and both galaxy properties, despite the relatively small sample size. Our results are consistent with, but more precise than, measurements made with only the late time UV emission from the TDE plateau.  

A second natural test, relevant for those systems where we constrain the early-time evolution of the disk (i.e., ASASSN-14li, ASASSN-15oi, AT2019dsg, AT2019azh and eJ2344) is that we recover the expected anti-correlation between black hole mass and circularisation/disk formation radius in units of the gravitational radius. {\tt FitTeD} provides posterior measurements on both $M_\bullet$ and $r_0$, which is the (physical) disk formation radius $R_0$ in units of the gravitational radius of the black hole $r_g \equiv GM_\bullet/c^2$, $r_0\equiv R_0/r_g$. While we place no prior bounds, or coupling with $M_\bullet$, on $r_0$, we would expect to recover the disk-circularisation radius on the population level. 

This is because the tidal radius imprints a natural  formation radial scale on the disk 
\begin{equation}\label{disk_form}
R_{\rm circ} \;\approx\; \frac{2R_{T}}{\beta}
\;\approx\; \frac{92}{\beta } \left ( \frac{M_*}{M_{\odot}} \right)^{7/15} \left ( \frac{M_{\bullet}}{10^6 M_{\odot}} \right)^{-2/3} r_g,
\end{equation}
where $R_{T} \simeq R_{*}\,(M_{\bullet}/M_{*})^{1/3}$ is the tidal radius, $M_{*}$ and $R_{*}$ are the stellar mass and radius, $\beta$ is the impact parameter (the ratio of the pericentre of the incoming stars  orbit to the tidal radius), and we have assumed a main-sequence mass--radius relation in going to the final expression on the right.  If we assume that the {\tt FitTeD} parameter $r_0$ should be $r_0 \approx R_{\rm circ}/r_g$, then we see that we would expect $r_0 \sim M_\bullet^{-2/3}$, with possibly large scatter from the different stars $(M_\star, R_\star)$ and orbits $(\beta)$ in the population. 

\begin{figure}
    \centering
    \includegraphics[width=0.8\linewidth]{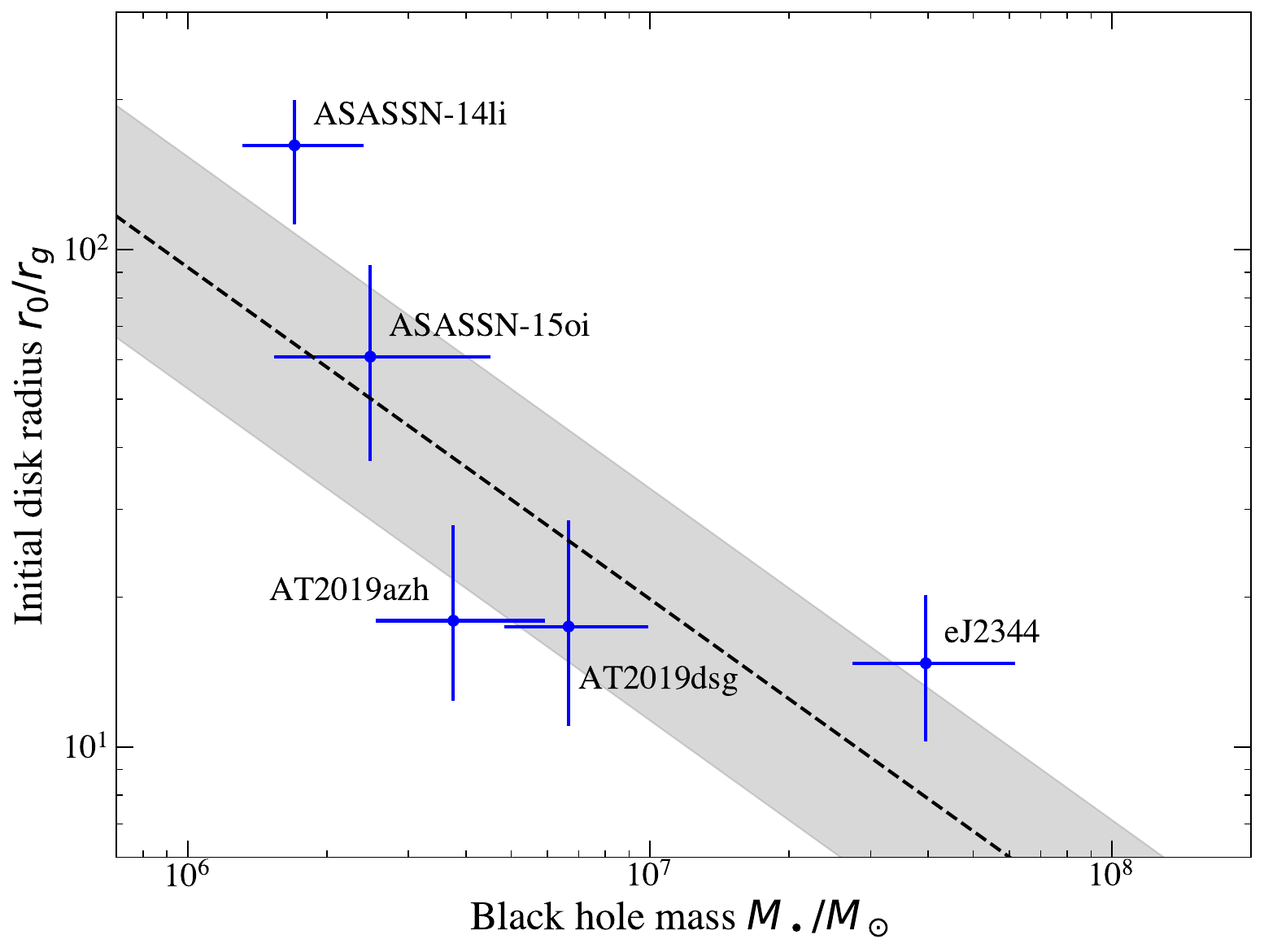}
    \caption{The initial radii for the {\tt FitTeD} model solutions plotted against the black hole mass for each model. The circularisation radius should imprint a characteristic $r_0/r_g \propto M_\bullet^{-2/3}$ profile, with scatter from
    the different stellar masses across the TDE population (grey shaded region). Our posterior distributions are consistent with this expected trend.}
    \label{fig:tidalradius}
\end{figure}

In Figure \ref{fig:tidalradius} we display the  $r_0$ constraints, plotted against $M_\bullet$ constraints, for the five TDEs with well-sampled early time X-ray light curves. The grey shaded region shows equation \ref{disk_form} with $\beta=1$ and solar masses ranging from $M_\star = 0.3M_\odot-3M_\odot$. The black dashed curve shows equation \ref{disk_form} evaluated for $M_\star = M_\odot, \beta=1$. All five sources in this work are consistent with the expected circularisation radius at worst at the $\lesssim 2\sigma$ level. Performing a Kendall's $\tau$ correlation test between $M_\bullet$ and $1/r_0$ (we invert $r_0$ so we can use standard tests for correlation) finds a statistically significant correlation, with $p$-value $< 0.05$. We conclude that the parameters inferred for this sample are reliable.

\section{Statistical tests}\label{sec:stats}
\subsection{The need for a bi-modal distribution}
We have constructed a distribution of Eddington-normalised accretion rates within TDE accretion disks, as inferred at the time at which an outflow was launched from the system. This distribution is in reality a list of posterior values from a numerical fit, which we denote $\{\dot m_i\}$. The claim in the paper is that this distribution is bi-modal, with outflows launched from TDE disks at the two accretion scales  $\dot m \simeq 1$ (for prompt radio ejecta) and $\dot m \simeq 0.02$ (for delayed radio ejecta). This can be simply tested, in a accretion-model agnostic fashion, by fitting a double Gaussian profile to this distribution of the logarithm of the Eddington-normalised accretion rates 
\begin{equation}
    p(\Theta) = A\,{\cal N}(\mu_{\rm high}, \sigma_{\rm high}) + (1-A) \,{\cal N}(\mu_{\rm low}, \sigma_{\rm low}) ,
\end{equation}
where ${\cal N}(\mu, \sigma)$ is the standard Normal distribution. The list of parameters which we denote $\Theta = \{A, \mu_{\rm high}, \sigma_{\rm high}, \mu_{\rm low}, \sigma_{\rm low}\}$ are all of intrinsic interest to this study. In particular, the parameter $A$, if statistically significantly different from $0$, represents a measurement of the confidence in which we can say that two distinct populations of accretion rates are present in our population. If the interpretation in the paper is correct, then indeed we expect the precise value $A = {\rm number\, of\, prompt\, flares\, in\, sample}/{\rm total\, sample\, size} = 5/11 \approx 0.45$. The two means of these Gaussians correspond to the (logarithms of) the critical accretion rates present in the data, namely $\dot m_{\rm high} = 10^{\mu_{\rm high}}$, and $\dot m_{\rm low} = 10^{\mu_{\rm low}}$. We therefore expect $\mu_{\rm high} \approx 0$ and $\mu_{\rm low} \approx \log_{10}(0.02) \approx -1.70$. 

To fit this distribution to the numerical distribution of accretion rates $\{\dot m_i\}$ one constructs the double-Gaussian likelihood ${\cal L}$ equal to 
\begin{multline}
    {\cal L} = \prod_i {A\over \sqrt{2\pi \sigma_{\rm high}^2}} \exp\left(-{(\log_{10}\dot m_i - \mu_{\rm high})^2\over 2\sigma_{\rm high}^2}\right)  \\ + {1-A\over \sqrt{2\pi \sigma_{\rm low}^2}} \exp\left(-{(\log_{10}\dot m_i - \mu_{\rm low})^2\over 2\sigma_{\rm low}^2}\right) ,
\end{multline}
where the product is over all values of $\dot m_i$ from the numerical distribution. It is numerically prudent to work with the logarithm of this quantity, $\ln {\cal L}$, to avoid numerical rounding errors. One then simply maximises this likelihood function, which we do using the {\tt emcee} MCMC code \cite{EMCEE}. 

\begin{figure}
    \centering
    \includegraphics[width=0.99\linewidth]{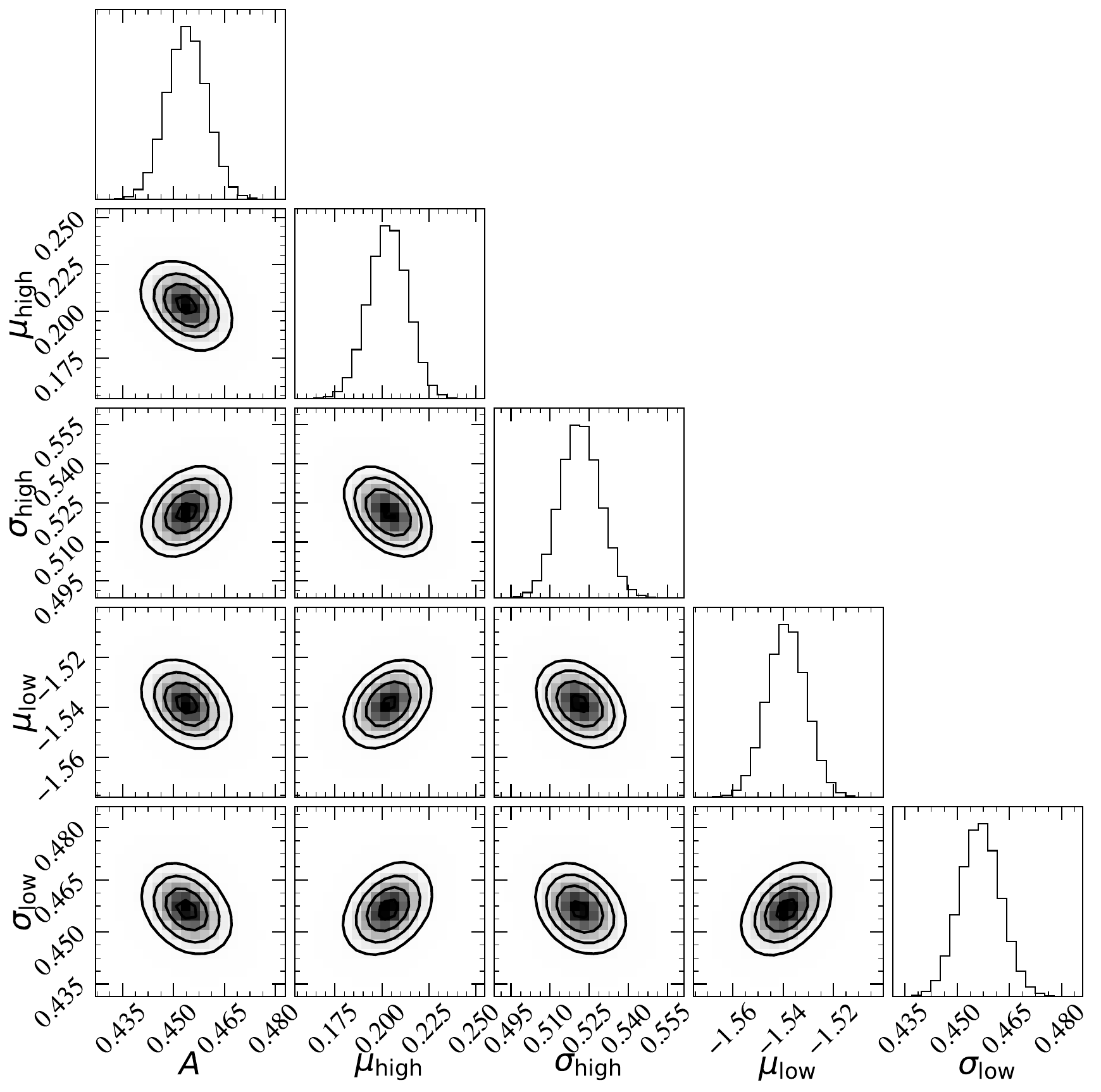}
    \caption{The posterior distributions of the parameters of a double Gaussian  fit to the numerical Eddington-normalised accretion rates at outflow launch (logarithm). The parameter $A$ (in effect the requirement for a second Gaussian) is constrained to be $A>0$ at extremely high $(\gg 5\sigma)$ significance.  The low-accretion rate transition scale is inferred to be $\dot m_{\rm low} = 10^{\mu_{\rm low}} \approx 0.028$, exactly as found for black hole X-ray binaries \citep{Vahdat2019}.}
    \label{fig:stat_test}
\end{figure}

The results of this numerical likelihood-maximisation are displayed in Figure \ref{fig:stat_test}. We note that the posterior distributions of the 5 fitted parameters of the double-Gaussian are precisely in keeping with the narrative of the paper. Namely, the parameter $A$ (in effect the requirement for a second Gaussian) is constrained to be $A>0$ at extremely high $(\gg 5\sigma)$ significance. Indeed, $A$ is very close to the value $5/11$, which would represent an exact split of prompt and delayed flares by accretion rate.  The low-accretion rate transition scale is inferred to be $\dot m_{\rm low} = 10^{\mu_{\rm low}} \approx 0.027$, exactly as found for black hole X-ray binaries \citep{Maccarone2003,Vahdat2019}, while the high-accretion rate transition scale is inferred to be $\dot m_{\rm high} = 10^{\mu_{\rm high}} \approx 1.6$. 

\subsection{Null hypothesis test}
It is important to determine that the bi-modal distribution of accretion rates we infer at launch is a result of robust accretion-outflow coupling, and not simply a coincidence of sampling evolving accretion rate profiles at random times in some fixed observing windows. To ascertain this we perform a null hypothesis test. 

\begin{figure}
    \centering
    \includegraphics[width=0.95\linewidth]{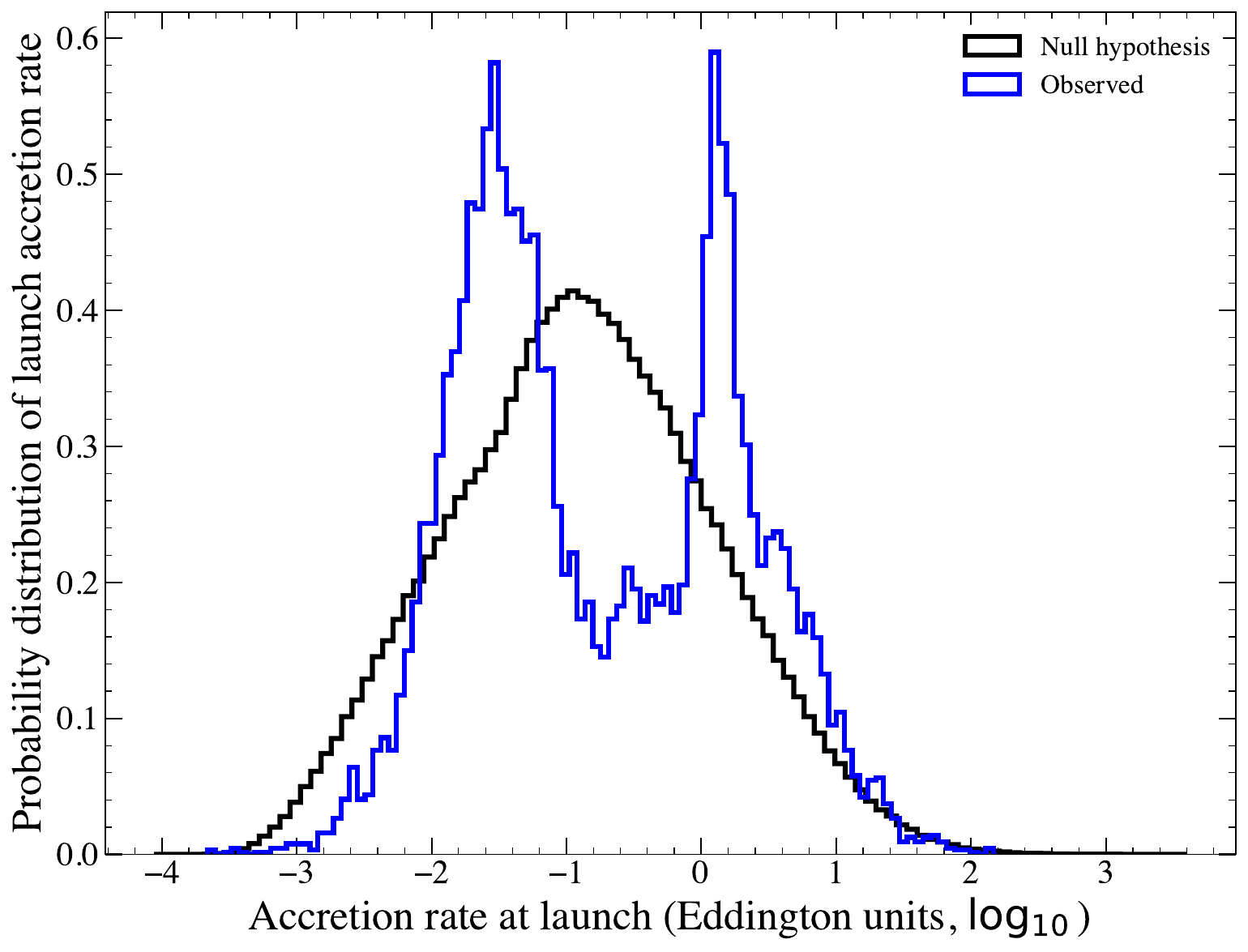}
    \caption{A comparison of the observed (blue) distribution of accretion rates at the time of outflow launch against the distribution of accretion rates under the null hypothesis that outflow launch time have no correlation with disk properties (black), and are instead log-uniformly distributed between $t_{\rm min}=10\, {\rm d}$ and $t_{\rm max}=7\, {\rm yr}$. This null hypothesis generates an accretion rate distribution well described by a single Gaussian centred at $\sim 0.15$, in tension with the observations.  }
    \label{fig:null}
\end{figure}

Following the population synthesis procedure outlined above, we now randomly sample ``launch times'' from a log-uniform distribution spanning 10 days to 7 years, plausibly bounding the upper and lower bounds that are observably feasible. This is intended to mimic the scenario when the observed launch time is random, or indeed not coupled to the accretion rate in the disk. By computing a population synthesis of the accretion rates in the flow at these times, we can compute the null-hypothesis distribution of $\dot m$, which we plot in Figure \ref{fig:null}. 

In Figure \ref{fig:null} we see that in this random launch time limit, the measured accretion rate distribution is well described by a single Gaussian centred at $\sim 0.15$, as opposed to the bi-modal distribution we observe. These two distributions are distinct according to a 2-sample KS test with $p$-value $<0.05$.

\subsection{Constraints on the outflow launch accretion rate for prompt and delayed flares}
\begin{figure}
    \centering
    \includegraphics[width=0.45\linewidth]{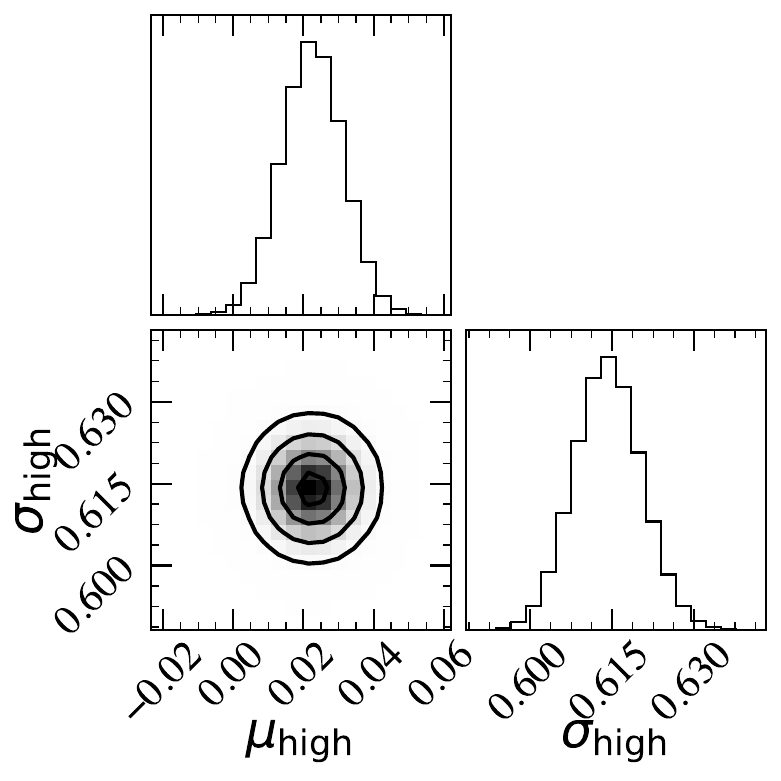}
    \includegraphics[width=0.45\linewidth]{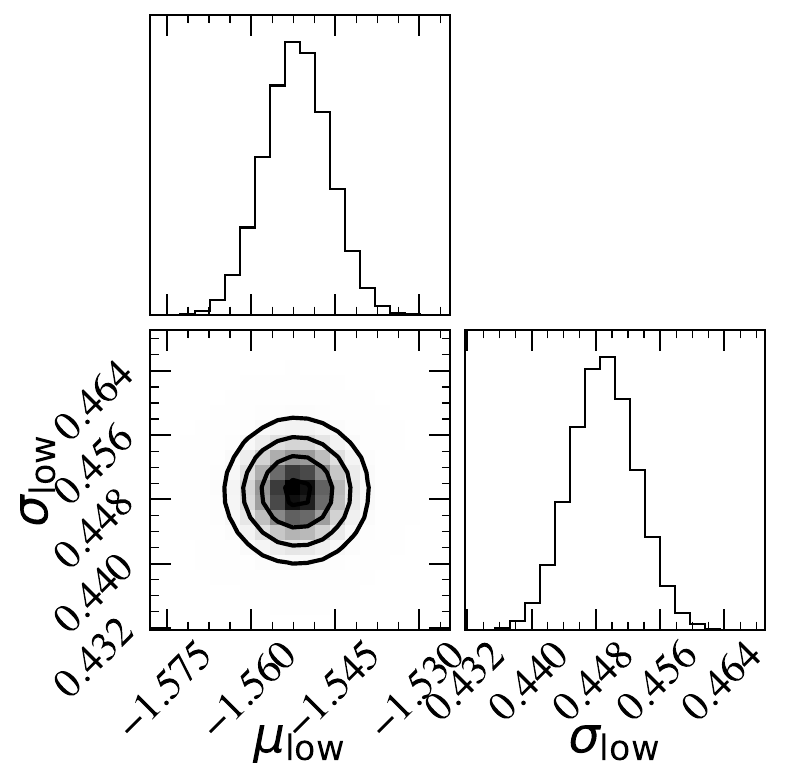}
    \caption{Single-Gaussian fits to the distribution of accretion rates at launch  for prompt (left) and delayed (right) flares. The prompt flares are consistent with $\dot m = 1$, and the delayed flares are consistent with $\dot m = 0.02$.  }
    \label{fig:single_gauss}
\end{figure}

We now split the radio outflow dates by the categorisation of ``prompt'' and   ``delayed'' flares, fitting each distribution by a single Gaussian. These single Gaussian fits are shown in Figure \ref{fig:single_gauss}. The prompt flares are again consistent with $\dot m = 1$, and the delayed flares are consistent with $\dot m = 0.02$. We see by comparison of Figures \ref{fig:stat_test} and \ref{fig:single_gauss} that the results (the preferred accretion rates at launch) are consistent regardless of if we split the population by radio type, or fit a double Gaussian to the entire population. 

\subsubsection{Accretion rate or bolometric luminosity?}
\begin{figure}
    \centering
    \includegraphics[width=0.95\linewidth]{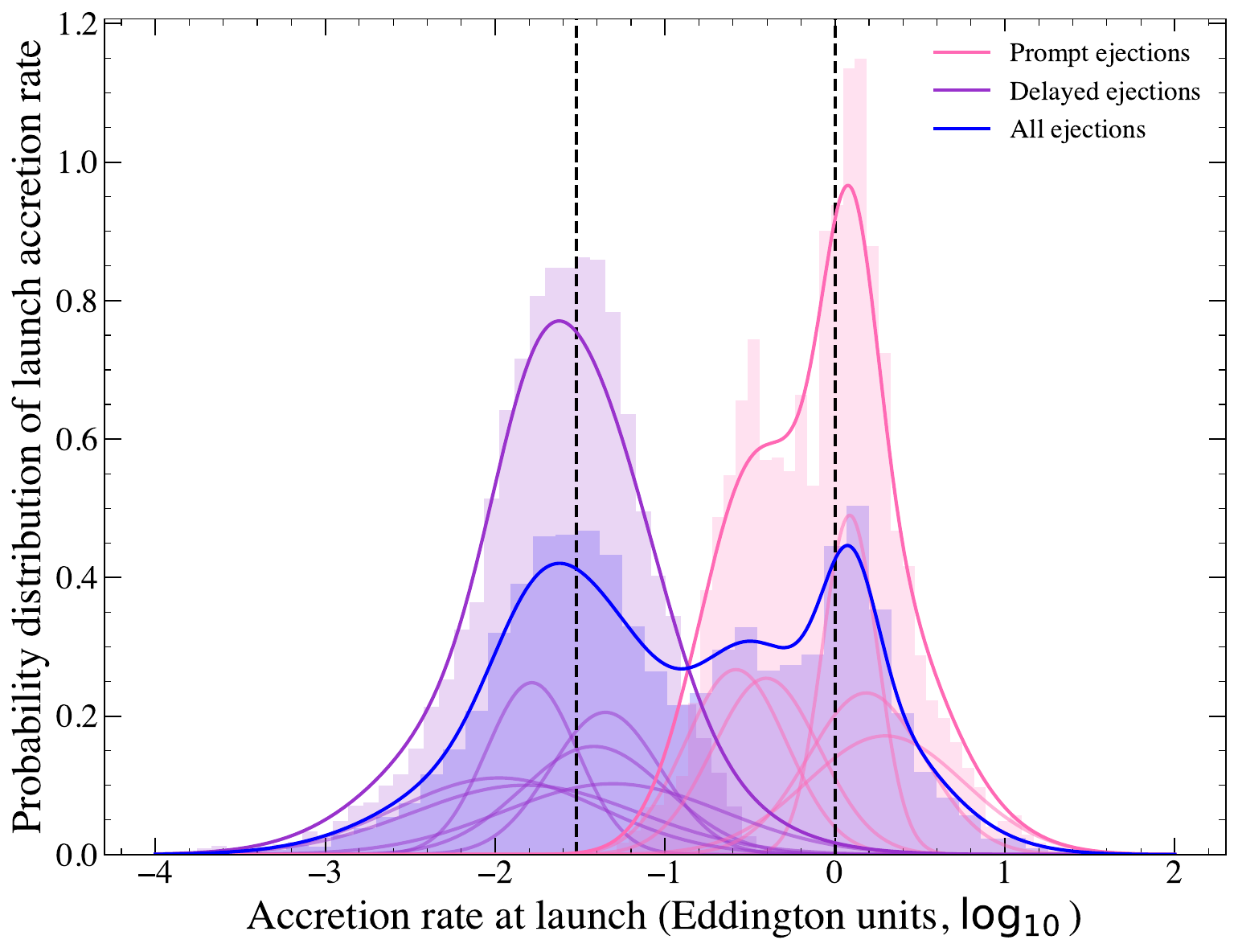}
    \caption{The same as Figure \ref{fig:solved} but now for the Eddington ratio of the bolometric luminosity instead of the mass accretion rate onto the black hole. The distribution remains bi-modal. Early time bolometric luminosities are typically lower than the associated accretion rate (for physical reasons), while at late times (delayed flares) the two measurements converge.  }
    \label{fig:lbol_dist}
\end{figure}

As discussed in early sections, in a time-dependent accretion model the Eddington ratios of the bolometric luminosity and mass accretion rate will generally not be equal at early times. In Figure \ref{fig:lbol_dist} we show that this choice does not impact our results. In Figure \ref{fig:lbol_dist} we reproduce Figure \ref{fig:solved}  (the Eddington ratio distribution at launch time), but now for the Eddington ratio of the bolometric luminosity instead of the mass accretion rate onto the black hole. The distribution remains bi-modal. Early time bolometric luminosities are typically lower than the associated accretion rate (for physical reasons), while at late times (delayed flares) the two measurements converge. 

\end{appendices}

\clearpage
\newpage 


\bibliography{sn-bibliography, andy}

\end{document}